\documentclass[letterpaper, 10 pt, conference]{ieeeconf}
\IEEEoverridecommandlockouts
\usepackage{cite}
\usepackage{amsmath,amssymb,amsfonts}

\let\proof\relax
\let\endproof\relax
\usepackage{amsthm}
\usepackage{comment}
\newtheorem{lem}{Lemma}
\newtheorem{thm}{Theorem}
\newtheorem{prop}{Proposition}
\newtheorem{cor}{Corollary}
\theoremstyle{definition}
\newtheorem{defn}{Definition}
\newtheorem{rem}{Remark}
\newtheorem{assumption}{Assumption}
\usepackage{booktabs}
\usepackage{multirow}

\usepackage{caption}  % Optional, but helpful for customization
\usepackage{subcaption} % Required for the subfigure environment

\usepackage{amsmath, amssymb, amsfonts}
\usepackage{xcolor}
\usepackage{url}
\newcommand{\boxend}{\hfill \ensuremath{\Box}}

\newcommand{\rev}[1]{#1}
\newenvironment{revenv}{}{}
\newcommand{\authnote}[1]{}

\newenvironment{blueenv}{}{}
\newcommand{\margin}[1]{}
\newcommand{\reza}[1]{#1}
\usepackage{algorithm}
\usepackage{algpseudocode}

\newcommand{\FedLQR}{\texttt{FedLQR}~}

\newcommand{\st}[1]{\textcolor{violet}{ST: #1}}
\newcommand{\added}[1]{#1}

\usepackage{bm}
\usepackage{dsfont}

\usepackage{mathtools}
\usepackage{array}

\usepackage{graphicx}
\usepackage{textcomp}
\def\BibTeX{{\rm B\kern-.05em{\sc i\kern-.025em b}\kern-.08em
    T\kern-.1667em\lower.7ex\hbox{E}\kern-.125emX}}
\begin{document}
\title{Resilient Federated LQR over Noisy and Lossy Uplinks \\via Scalar Communication
% ResiFedLQR: Resilient Federated Cooperative LQR
%via Directional Derivatives
}
\author{Mohammadreza Rostami, Shahriar Talebi, \emph{Member, IEEE} and Solmaz S. Kia, \emph{Senior Member, IEEE}
\thanks{This work is partially supported by the 2026 UCLA–UCI PI grant awarded by the Samueli Foundation (Corresponding author: M. Rostami).}
\thanks{M. Rostami and S. Kia are with the University of California, Irvine, CA, US, and S. Talebi is with the University of California, Los Angeles, CA, US. Emails: {\em\small mrostam2@uci.edu}, {\em\small s.talebi@ucla.edu} and {\em\small solmaz@uci.edu} }
}

\maketitle

\begin{abstract}
This paper studies model-free federated learning of a common LQR
controller for a fleet of similar systems that communicate with a server
over an uplink subject to limited bandwidth, packet dropout, and additive
noise. We propose \textsc{ResiFedLQR}, a scalar communication scheme in
which each agent sends one real number per packet, with redundancy,
instead of full gradient estimates. The scalars can be gradient
coordinates, projections of the gradient onto random directions
reproduced at the server from a shared seed, or the rollout costs
themselves. For the projection encoding, we prove that, with high
probability, every iterate stabilizes all agents and the fleet-average
cost converges linearly to a neighborhood of the optimum whose radius
separates heterogeneity, sampling, and communication effects, and we
derive a rule for choosing the number of transmissions from the channel
parameters and the fleet size. We then compare the encodings with
full-gradient transmission under the same uplink budget and identify the
signal-to-noise regime and power constraint under which each is
preferable. Numerical experiments on a heterogeneous fleet \reza{of BlueROV2
underwater vehicles} evaluate
stability and convergence under varying noise levels and dropout
probabilities.

\end{abstract}

% \begin{IEEEkeywords}
% Federated Learning, Linear Quadratic Regulator (LQR), Policy Optimization
% \end{IEEEkeywords}

\section{Introduction}\label{sec:intro}
\begin{figure}[t]
    \centering
        \includegraphics[width=0.9\linewidth]{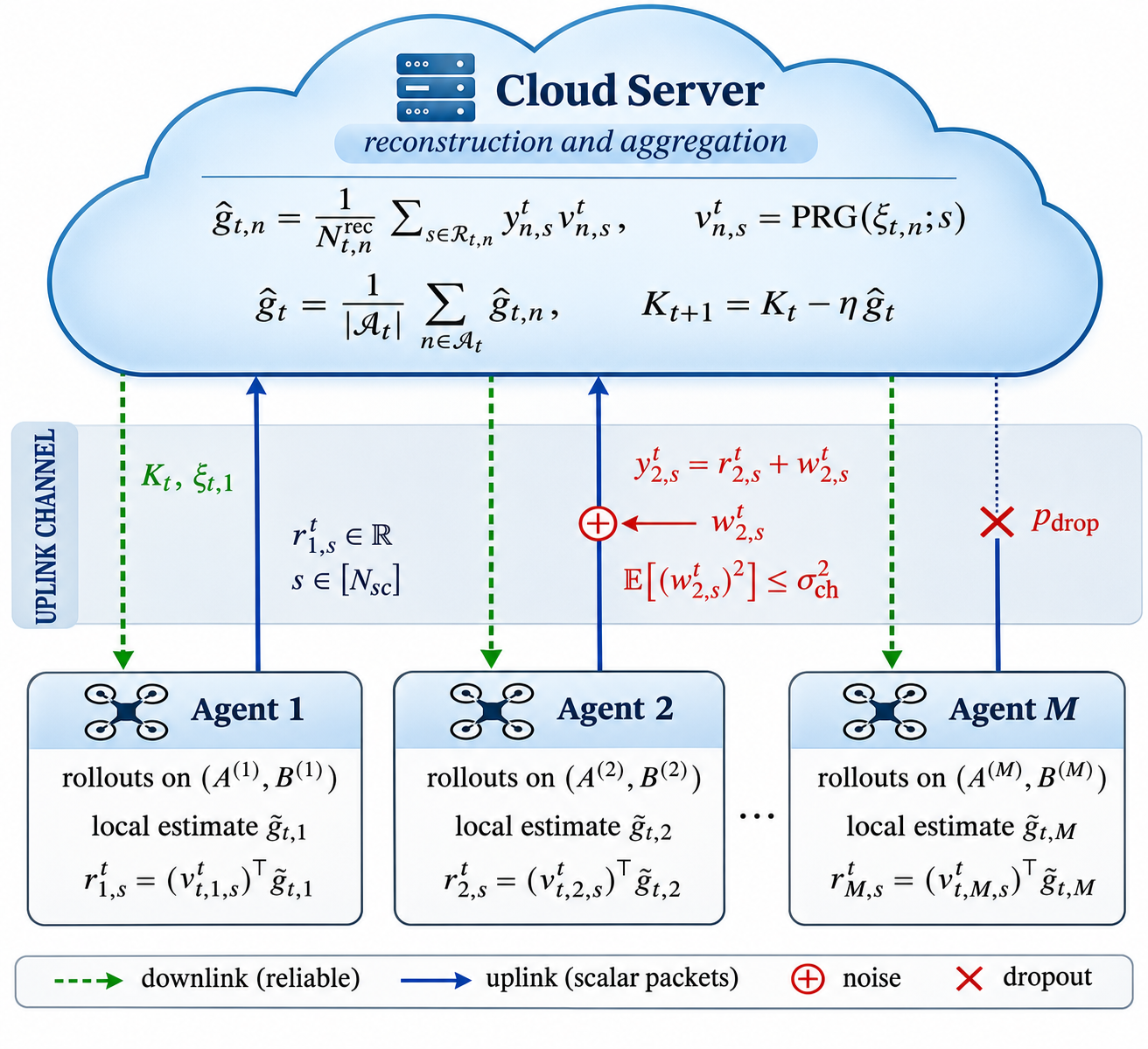}
    \caption{\small Federated LQR over an unreliable uplink. The server
    broadcasts the common gain $K_t$ over a reliable downlink. Each
    agent runs rollouts on its own unknown system, forms a local
    gradient estimate, and returns messages over an uplink subject to
    packet dropout and additive noise.}
    \label{fig:FL_LQR}
\end{figure}

Cooperative federated learning of linear quadratic
regulator (LQR) controllers is motivated by the cost of physical
experiments. Agents with similar dynamics, such as a fleet of small
drones, can pool rollout data through a central server to learn a
common state-feedback gain, reducing the number of rollouts required
per agent and accelerating
learning~\cite{ren_federated_2020,wang_model-free_2023}. In each
learning round, agents transmit local zeroth-order gradient estimates
to the server and receive an updated controller in return; see
Fig.~\ref{fig:FL_LQR}. Existing federated LQR
methods~\cite{wang_model-free_2023,alemzadeh_data-driven_2024} assume
reliable communication and transmit full gradient estimates, whose
dimension is $d=n_un_x$, number of elements of the control gain. In practice, however, uplink communication is
often the bottleneck: agents may operate with limited energy and
low-cost radios, leading to packet losses and additive channel noise.
Under such conditions, transmitting a full gradient is fragile: noise
distorts the descent direction, while a single packet loss removes an
agent's entire update. Because the learned controller is applied during
training, corrupted updates can even drive the common gain outside the
set of stabilizing controllers.

This article studies how to preserve closed-loop stability and learning
performance under noisy and lossy uplink communication, and quantifies
resilience as the range of noise levels, dropout probabilities, and
stepsizes over which learning remains stabilizing. To this end, we
consider a scalar communication architecture in which each packet
contains a single gradient-related scalar rather than an entire gradient
vector. Since the budget of one full-gradient transmission accommodates
$d$ scalar packets, communication resources can be used for redundancy,
making learning more resilient to packet losses and channel noise. We
investigate three scalar-message constructions: coordinate
transmission, random projection transmission, and direct transmission
of rollout costs. These considerations lead to three research questions:
\emph{(RQ1) Can scalar communication with redundancy preserve stabilizing
learning over unreliable channels? (RQ2) How much redundancy is required for
a given channel quality? (RQ3) And which scalar-message construction is
preferable under a given noise level and power constraint?}

\emph{Related Work:} Policy-gradient methods for LQR have been
extensively studied in both model-based and model-free settings, with
results on gradient domination, stability of the iterates, convergence,
and sample complexity~\cite{fazel_global_2018,malik_derivative-free_2019,
mohammadi_convergence_2021,bu_lqr_2019,hu_toward_2023,talebi_policy_2026},
including improved oracle complexity through two-point and
variance-reduced
estimators~\cite{malik_derivative-free_2019,toso2024oracle}. Cooperative
extensions include distributed policy iteration for homogeneous
networks~\cite{alemzadeh_data-driven_2024}, federated LQR tracking for
identical agents~\cite{ren_federated_2020}, and
\textsc{FedLQR}~\cite{wang_model-free_2023}, which accommodates
heterogeneous dynamics and quantifies both the sample-complexity benefit
of cooperation and the cost of heterogeneity. These works assume ideal
communication. In parallel, communication-efficient federated learning
has reduced communication overhead through local updates, quantization,
sparsification, and sign-based
compression~\cite{mcmahan_communication-efficient_2017,
konevcny_federated_2016,alistarh2017qsgd,
stich2018sparsified,bernstein2018signsgd}, while noisy and unreliable
wireless communication has been studied through over-the-air
aggregation, fading-channel models, and noisy-channel
analyses~\cite{yang2020ota,amiri2020fading,wei2022noisy}. However, these
works focus on optimization error and do not address the stability of a
physical system during learning. Random directional derivatives and
random projections have also been used in gradient-based learning,
zeroth-order optimization, and communication-efficient federated
learning~\cite{fournier2023can,silver2021learning,
rostami2024projected,nesterov2017random,rostami_fedscalar_2024, rostami2026scalar}. We
build on these ideas by using scalar messages as a communication
primitive for federated LQR over unreliable channels. A preliminary
conference version~\cite{MR-ST-SSK:26} introduced projection-based
scalar communication for federated LQR over an ideal channel and
analyzed the homogeneous-agent setting. The present article considers
heterogeneous agents communicating over a bandwidth-limited, noisy, and
lossy uplink, treats the choice of transmitted scalars as a design
variable, and compares multiple scalar encodings against full-gradient
communication.

\emph{Contributions:} We propose \textsc{ResiFedLQR}, a federated
model-free LQR algorithm in which agents communicate redundant scalar
messages over noisy and lossy uplinks. Our main contributions are:
\begin{itemize}
\item For projection-based communication, we establish high-probability
stability and linear convergence guarantees. Every iterate remains
stabilizing, and the fleet-average cost converges linearly to a
neighborhood whose size explicitly captures heterogeneity, sampling
error, communication noise, packet losses, and encoding effects.
\item We identify the key communication parameter governing stability
and convergence: the ratio between the policy dimension and the expected
number of scalar messages received by the server. This yields explicit
design rules that relate redundancy, fleet size, packet-drop
probability, and channel noise.
\item We compare coordinate, projection, and rollout-cost encodings to
one another and to full-gradient transmission under a common uplink
budget. All single-scalar encodings share the same protection against
losing an agent's update. Under a fixed receiver noise floor,
projections are more accurate than coordinates in mean square when the
per-coordinate signal-to-noise ratio is below approximately one, an
advantage that is tied to transmit energy, whereas their unbiasedness
under partial loss is not. Rollout-cost encoding removes local gradient
computation and the encoding error but amplifies the channel noise. The
analysis is supported by numerical stability-region studies.
\end{itemize}

\emph{Outline:} Section~\ref{sec:prob_def_object} formulates the
federated LQR problem and the uplink model. Section~\ref{sec:comm_model}
introduces the scalar encodings, and Section~\ref{sec:resifedlqr}
presents \textsc{ResiFedLQR}. Section~\ref{sec:analysis} establishes
stability and convergence, discusses the roles of heterogeneity, noise,
dropout, and redundancy, and compares the encodings.
Section~\ref{sec:numerics} reports numerical experiments, and
Section~\ref{sec:conclusion} concludes the article.

\emph{Notation:} $[M]:=\{1,\ldots,M\}$. For a matrix $X$, $\|X\|$ is
the Frobenius norm and $\|X\|_2$ the spectral norm; for a vector,
$\|\cdot\|$ is the Euclidean norm. $X\succ0$ ($X\succeq0$) means that
$X$ is symmetric positive (semi)definite, and $\sigma_{\min}(X)$ is its
smallest eigenvalue. A square matrix is Schur stable if its spectral
radius is less than one. A gain $K\in\mathbb{R}^{n_u\times n_x}$ is
identified with its vectorization in $\mathbb{R}^d$, $d:=n_un_x$,
whenever it appears in an inner product or as the argument of a
gradient; with this identification, $\|K\|$ is the same in both
representations. A scalar random variable $w$ is sub-Gaussian with
variance proxy $\sigma^2$ if $\mathbb{E}[e^{\lambda w}]\le e^{\lambda^2\sigma^2/2}$
for all $\lambda\in\mathbb{R}$. The main symbols are collected in Table~\ref{tab:nomenclature} in the
appendix. All proofs are given in the appendix.

\section{Problem Formulation and Objective}\label{sec:prob_def_object}

We consider a network of $M$ agents that cooperate through a central
server in a federated learning paradigm (Fig.~\ref{fig:FL_LQR}). Each
agent is governed by the discrete-time linear time-invariant (LTI)
dynamics
\begin{equation}\label{eq:dynamics}
  x^{(n)}_{k+1}=A^{(n)}x^{(n)}_k+B^{(n)}u^{(n)}_k,
  \qquad n\in[M],
\end{equation}
where $x^{(n)}_k\in\mathbb{R}^{n_x}$ and $u^{(n)}_k\in\mathbb{R}^{n_u}$
are the state and input of agent $n$ at time $k$. The dynamics
$(A^{(n)},B^{(n)})$ are unknown and may differ across agents, while all
agents share the same state and input dimensions.

\begin{assumption}[Initial-state distribution]\label{ass:init_state}
The initial state of every agent is drawn independently from a common
distribution $\mathcal D_0$ with
$\Sigma_0:=\mathbb E[x_0x_0^\top]\succ0$. We write
$\mu_0:=\lambda_{\min}(\Sigma_0)>0$. \hfill$\square$
\end{assumption}

Assumption~\ref{ass:init_state} is standard in policy optimization for
LQR~\cite{fazel_global_2018,malik_derivative-free_2019,mohammadi_convergence_2021,wang_model-free_2023}.
The cost depends on $\mathcal D_0$ only through $\Sigma_0$, and
$\Sigma_0\succ0$ ensures that finite cost certifies stability. Tail
conditions enter only through the accuracy of the zeroth-order oracle;
bounded-support~\cite{fazel_global_2018,malik_derivative-free_2019,wang_model-free_2023}
or sub-Gaussian~\cite{mohammadi_convergence_2021} initial-state
distributions are sufficient.

Each agent seeks a state-feedback controller minimizing the
infinite-horizon LQR cost
\begin{equation}\label{eq:individual_cost}
  J^{(n)}(K)
  =\mathbb E_{x^{(n)}_0\sim\mathcal D_0}\!\left[\,
   \sum\nolimits_{k=0}^{\infty}
   x^{(n)\top}_kQx^{(n)}_k+u^{(n)\top}_kRu^{(n)}_k\right],
\end{equation}
where $u^{(n)}_k=-Kx^{(n)}_k$ and $Q\succ0$, $R\succ0$ are shared across
the fleet. The agents have similar but nonidentical dynamics.

\begin{assumption}[Similarity of dynamics]\label{ass:similarity}
There exists a nominal model $(A,B)$ such that, for all $n\in[M]$,
\begin{equation}\label{eq:system_bound}
  \|A^{(n)}-A\|_2\le\epsilon_1,\qquad
  \|B^{(n)}-B\|_2\le\epsilon_2,
\end{equation}
for some heterogeneity parameters $\epsilon_1,\epsilon_2\ge0$.
\hfill$\square$
\end{assumption}

Assumption~\ref{ass:similarity} makes cooperation meaningful: the agents
solve closely related control problems, so rollout data collected by one
agent are informative for the others. Rather than learning separate
controllers, we therefore learn a common gain
$K\in\mathbb R^{n_u\times n_x}$ shared by the fleet. Every agent applies
$u_k^{(n)}=-Kx_k^{(n)}$, and the objective becomes
\begin{equation}\label{eq:avg_cost}
  J_{\mathrm{avg}}(K):=\frac1M\sum\nolimits_{n=1}^MJ^{(n)}(K).
\end{equation}
A shared gain need not be individually optimal for every agent.
However, in the homogeneous case $\epsilon_1=\epsilon_2=0$, all agents
share the same optimal gain and pooling rollouts yields a linear-in-$M$
speedup~\cite{ren_federated_2020}. Under small heterogeneity, the
minimizer of $J_{\mathrm{avg}}$ remains near-optimal for every agent,
while the per-agent sample complexity is reduced by approximately a
factor of $M$~\cite[Ths.~4--5]{wang_model-free_2023}. Cooperative
learning therefore spreads the cost of physical experimentation across
the fleet.

The key challenge is stability: a gain that stabilizes one agent may
destabilize another. Define
\begin{align}
  \mathcal S^{(n)}&:=\bigl\{K:A^{(n)}-B^{(n)}K
  \text{ is Schur stable}\bigr\},\nonumber\\
  \mathcal S&:=\bigcap\nolimits_{n=1}^{M}\mathcal S^{(n)},
  \label{eq:stablesets}
\end{align}
where $\mathcal S^{(n)}$ is the stabilizing set of agent $n$ and
$\mathcal S$ is the set of gains that stabilize the entire fleet. Each
$\mathcal S^{(n)}$ is open and generally
nonconvex~\cite{fazel_global_2018,bu_lqr_2019}, and $\mathcal S$ may be
empty. Since $Q\succ0$ and $\Sigma_0\succ0$, the
cost~\eqref{eq:individual_cost} is finite if and only if
$K\in\mathcal S^{(n)}$~\cite{fazel_global_2018}. Consequently,
\begin{equation}\label{eq:finite_iff_stable}
  J_{\mathrm{avg}}(K)<\infty\quad\Longleftrightarrow\quad K\in\mathcal S.
\end{equation}
Thus a bounded fleet-average cost certifies stability of every agent, a
property that underpins the analysis in Section~\ref{sec:analysis}.

\begin{assumption}[Initial stabilizing policy]\label{ass:init}
There exists a gain $K_0\in\mathcal S$ that stabilizes all agents
simultaneously. \hfill$\square$
\end{assumption}

Access to an initial stabilizing gain is standard in policy optimization
for LQR~\cite{fazel_global_2018,mohammadi_convergence_2021,hu_toward_2023,talebi_policy_2026,wang_model-free_2023}.
Requiring a common stabilizing gain is stronger but necessary for
meaningful cooperation: if $\mathcal S$ is empty, which can occur under
large heterogeneity~\cite[Props.~1--2]{wang_model-free_2023}, no common
controller can be deployed across the fleet. The gain $K_0$ may be
obtained from a conservative nominal design or offline synthesis.
Learning a stabilizing controller from
scratch~\cite{talebi_regularizability_2022} is beyond the scope of this
work.

The nominal federated LQR problem is
\begin{equation}\label{eq:fedlqr}
  \min_{K\in\mathcal S}\;J_{\mathrm{avg}}(K).
\end{equation}
Starting from $K_0$, the server broadcasts the current policy $K_t$ to
all agents at each round $t$. Since the dynamics are unknown, each agent
forms a local policy-gradient estimate from rollouts. Agent $n$ draws
$n_s$ perturbations $U_{t,n,i}$, $i\in[n_s]$, independently and
uniformly from the sphere $\{U:\|U\|=\mathsf r\}$, runs a rollout of
length $\tau$ under $K_t+U_{t,n,i}$ from an initial state drawn from
$\mathcal D_0$, and records the corresponding truncated cost
$\widehat J^{(n)}_{t,i}$. The local zeroth-order (ZO) estimate is
\begin{equation}\label{eq:zo_estimate}
  \tilde g_{t,n}=\frac{d}{n_s\mathsf r^2}
  \sum\nolimits_{i=1}^{n_s}\widehat J^{(n)}_{t,i}\,U_{t,n,i}.
\end{equation}
We use the one-point estimator for concreteness. The coordinate and
projection encodings developed below are independent of the chosen
zeroth-order oracle and apply equally to two-point and variance-reduced
estimators~\cite{malik_derivative-free_2019,toso2024oracle}. The server
aggregates the information received from the agents into a global
direction $\widehat g_t$ and updates the policy according to
\begin{equation}\label{eq:policy_update}
  K_{t+1}=K_t-\eta\,\widehat g_t,
\end{equation}
where $\eta>0$ is the stepsize.

The two communication directions are asymmetric. The downlink, through
which the server broadcasts $K_t$, is assumed reliable. The uplink,
however, is constrained by bandwidth, transmission power, and hardware
limitations, and is therefore subject to packet losses and additive
channel noise. When a full gradient estimate
$\tilde g_{t,n}\in\mathbb R^d$ is transmitted as a single packet, a
packet loss removes an entire agent contribution and channel noise
corrupts all $d$ entries simultaneously. This motivates the scalar
communication architecture developed in the sequel.

\begin{assumption}[Uplink dropout and channel noise]\label{assump:noise}
Each uplink packet is delivered independently with probability
$1-p_{\rm drop}$ and lost with probability $p_{\rm drop}\in[0,1)$. For
$N$ packets transmitted by agent $n$ in round $t$,
$\mathcal R_{t,n}(N)\subseteq[N]$ denotes the set of delivered packets.
A delivered packet with payload $m$ is received as
\begin{equation}\label{eq:channel}
  y=m+w,
\end{equation}
where $w$ is a zero-mean sub-Gaussian random variable with variance
proxy $\sigma_{\rm ch}^2$; for a vector payload, this holds for each
entry. The noise is independent across entries, packets, agents, and
rounds, and is independent of the dropout process and of all rollout
and encoding randomness. \hfill$\square$
\end{assumption}

\begin{rem}[Scope of the noise model]\label{rem:noise_model}
The noise level $\sigma_{\rm ch}$ is fixed per received real number and
does not depend on the payload magnitude. Under this model, larger
payloads enjoy larger signal-to-noise ratios. If transmit power is
normalized across symbols, the comparisons of
Section~\ref{sec:comparison} change accordingly; this is discussed in
Remark~\ref{rem:power}. \hfill$\square$
\end{rem}

Under Assumption~\ref{assump:noise}, the server receives only a noisy
and randomly thinned version of the agents' information. Consequently,
the update direction $\widehat g_t$ differs from the ideal federated
direction $\frac1M\sum_{n=1}^M\tilde g_{t,n}$, and excessive
communication error can drive the iterates out of $\mathcal S$. Our
objective is therefore to solve~\eqref{eq:fedlqr} while preserving
closed-loop stability under noisy and lossy uplink communication. In
particular, we seek communication schemes and aggregation rules such
that (1) every iterate $K_t$ remains in $\mathcal S$; (2) $J_{\mathrm{avg}}(K_t)$ converges linearly to a neighborhood of
its minimum whose radius explicitly separates heterogeneity, sampling,
and communication effects; (3) the resulting \emph{resilience region}, namely the set of noise
levels $\sigma_{\rm ch}$, dropout probabilities $p_{\rm drop}$, and
stepsizes $\eta$ for which (1) and (2) hold, can be characterized and
compared across communication architectures under the same uplink
budget.

%%%%Keep it here for if reviewers want proof for (6)
%The cost~\eqref{eq:individual_cost} is finite exactly on $\mathcal{S}^{(n)}$. If $A^{(n)}-B^{(n)}K$ is Schur stable, the closed-loop state and input decay geometrically from every initial condition, so the series in~\eqref{eq:individual_cost} converges. If it is not, it has an eigenvalue $\lambda$ with $|\lambda|\ge1$ and a unit left eigenvector $v$, i.e., $v^{*}(A^{(n)}-B^{(n)}K)=\lambda v^{*}$ and $\|v\|=1$, and the component of the state along $v$ does not decay: $|v^{*}x^{(n)}_k|=|\lambda|^k\,|v^{*}x^{(n)}_0|\ge|v^{*}x^{(n)}_0|$ for all $k$. Since $\Sigma_0\succ0$, this component is excited with positive expected energy, $\mathbb{E}|v^{*}x^{(n)}_0|^2=v^{*}\Sigma_0v>0$, and since $Q\succ0$, it is penalized at every step by at least $\lambda_{\min}(Q)\,|v^{*}x^{(n)}_k|^2$; hence the series diverges and $J^{(n)}(K)=+\infty$. Because the costs are nonnegative, \begin{equation}\label{eq:finite_iff_stable}   J_{\mathrm{avg}}(K)<\infty\quad\Longleftrightarrow\quad K\in\mathcal{S}. \end{equation}

\section{Scalar Message Constructions}\label{sec:comm_model}

To enlarge the resilience region of Section~\ref{sec:prob_def_object},
we replace the $d$-dimensional estimate $\tilde g_{t,n}$ by scalar
gradient-related messages, so that the uplink budget is spent on
redundancy rather than payload size. This section introduces the
corresponding encodings and decoders.

\subsection{Random Projections}\label{sec:random_proj}

Let the directions $v\in\mathbb R^d$ be drawn from an isotropic
distribution $\mathcal D$ satisfying
\begin{equation}\label{eq:isotropy}
  \mathbb E_{v\sim\mathcal D}[vv^\top]=I.
\end{equation}
For a local estimate $\tilde g_{t,n}$, define the scalar projection
$\zeta:=v^\top\tilde g_{t,n}$. The corresponding rank-one reconstruction
\begin{equation}\label{eq:rankone}
  \mathsf d:=\zeta v=vv^\top\tilde g_{t,n}
\end{equation}
is unbiased. Among isotropic directions, the worst-case reconstruction
variance is minimized when
\begin{equation}\label{eq:fixed_norm}
  \|v\|^2=d\qquad\text{almost surely},
\end{equation}
in which case\footnote{For any $v$ with $\mathbb E[vv^\top]=I$,
$\mathbb E\|vv^\top z-z\|^2
=\mathbb E[(v^\top z)^2\|v\|^2]-\|z\|^2$, which equals $(d-1)\|z\|^2$
when $\|v\|^2=d$ almost surely and $(d+1)\|z\|^2$ for Gaussian
$v\sim\mathcal N(0,I)$. More generally,
$\max_{\|z\|=1}\mathbb E[(v^\top z)^2\|v\|^2]
=\lambda_{\max}(\mathbb E[\|v\|^2vv^\top])
\ge\mathbb E\|v\|^4/d\ge d$, with equality when $\|v\|^2=d$ almost
surely.}
\begin{equation}\label{eq:encoding_variance}
  \mathbb E_v\|vv^\top\tilde g_{t,n}-\tilde g_{t,n}\|^2
  =(d-1)\|\tilde g_{t,n}\|^2.
\end{equation}
Thus a single projection provides a noisy but unbiased estimate of
$\tilde g_{t,n}$, while averaging $N$ independent reconstructions
reduces the reconstruction variance by a factor of $N$.

Both Rademacher directions, $v\in\{-1,+1\}^d$ with independent
equiprobable entries (cf.~\eqref{eq:rademacher_moments}), and directions
drawn uniformly from the sphere of radius $\sqrt d$
satisfy~\eqref{eq:isotropy}--\eqref{eq:fixed_norm}, and both can be
reproduced exactly from an integer seed. The scaled basis vectors
$\{\sqrt d\,e_i\}_{i\in[d]}$, drawn uniformly, also
satisfy~\eqref{eq:isotropy}--\eqref{eq:fixed_norm}; the corresponding
projection is a randomly selected coordinate of $\tilde g_{t,n}$ scaled
by $\sqrt d$, so that the analysis of Section~\ref{sec:analysis} covers
random coordinate sampling as a special case. We write
$\mathrm{PRG}_{\mathcal D}$ for the seeded generator of directions with
distribution $\mathcal D$ and use Rademacher directions unless stated
otherwise.

The key observation is that the directions themselves need not be
transmitted. At round $t$, the server broadcasts a seed $\xi_{t,n}$
together with $K_t$, and both agent and server generate the same
directions
\(
  v_{t,n,s}=\mathrm{PRG}_{\mathcal D}(\xi_{t,n};s),~ s\in[N_{\rm sc}].
\)
Agent $n$ therefore transmits only the scalar projections
\begin{equation}\label{eq:scalar_measurement_clean}
  r^t_{n,s}:=v_{t,n,s}^\top\tilde g_{t,n},
  \qquad s\in[N_{\rm sc}],
\end{equation}
one scalar per packet. Under Assumption~\ref{assump:noise}, the server
receives $y^t_{n,s}=r^t_{n,s}+w^t_{n,s}$ for
$s\in\mathcal R_{t,n}:=\mathcal R_{t,n}(N_{\rm sc})$, regenerates the
corresponding directions, and forms
\begin{equation}\label{eq:decoded_direction}
  \widehat g_{t,n}
  :=\frac{1}{N^{\rm rec}_{t,n}}\sum\nolimits_{s\in\mathcal R_{t,n}}
  y^t_{n,s}\,v_{t,n,s},
  \quad N^{\rm rec}_{t,n}:=|\mathcal R_{t,n}|,
\end{equation}
whenever $N^{\rm rec}_{t,n}\!>\!0$. By~\eqref{eq:rankone},
$\widehat g_{t,n}$ averages $N^{\rm rec}_{t,n}$ rank-one reconstructions
of $\tilde g_{t,n}$ together with their channel-noise~terms.

\subsection{Rollout Costs and Coordinates}

\begin{blueenv}
The scalar message need not be formed from a local gradient estimate:
the server can reconstruct a descent direction directly from rollout
costs when the zeroth-order perturbations are generated from the shared
seed. Let $U_{t,n,s}=\frac{\mathsf r}{\sqrt d}v_{t,n,s}$, $s\in[n_s]$,
where $v_{t,n,s}$ is uniform on the sphere of radius $\sqrt d$. Then
$\|U_{t,n,s}\|=\mathsf r$ and~\eqref{eq:zo_estimate} becomes
\[
  \tilde g_{t,n}=\frac{\sqrt d}{\mathsf r\,n_s}
  \sum\nolimits_{s=1}^{n_s}\widehat J^{(n)}_{t,s}\,v_{t,n,s},
\]
so that the rollout costs alone suffice for the server to form the
gradient estimate. Alternatively, the agent can transmit the coordinates
of $\tilde g_{t,n}$ themselves, each in its own packet, which involves
no encoding error. The three constructions are summarized next.

\begin{defn}[Scalar encodings]\label{def:gl_ol}
At round $t$, agent $n$ transmits $N_{\rm pk}$ scalars $r^t_{n,s}$, one
per packet, according to one of the following schemes:(1)  \emph{Gradient-level (GL) encoding:} $N_{\rm pk}=N_{\rm sc}$ and
$r^t_{n,s}=v_{t,n,s}^{\top}\tilde g_{t,n}$, as
in~\eqref{eq:scalar_measurement_clean}. (2) \emph{Oracle-level (OL) encoding:} $N_{\rm pk}=n_s$ and
\begin{equation}\label{eq:oracle_scalar}
  r^t_{n,s}=\widehat J^{(n)}_{t,s},\qquad s\in[n_s],
\end{equation}
where the rollouts use the seeded perturbations
$U_{t,n,s}=\frac{\mathsf r}{\sqrt d}v_{t,n,s}$. (3) \emph{Coordinate-level (CL) encoding:} $N_{\rm pk}=dN_{\rm F}$ and
each coordinate is transmitted $N_{\rm F}$ times,
\begin{equation}\label{eq:coord_scalar}
  r^t_{n,s}=[\tilde g_{t,n}]_{i_s},~
  i_s:=((s-1)\bmod d)+1,~ s\in[dN_{\rm F}].
\end{equation}
\hfill$\square$
\end{defn}

For GL and OL encoding, the server decodes
\begin{equation}\label{eq:unified_decode}
  \widehat g_{t,n}
  :=\frac{c}{N^{\rm rec}_{t,n}}\sum\nolimits_{s\in\mathcal R_{t,n}}
  y^t_{n,s}\,v_{t,n,s},
\end{equation}
where $\mathcal R_{t,n}:=\mathcal R_{t,n}(N_{\rm pk})$, $c=1$ for GL,
which recovers~\eqref{eq:decoded_direction}, and
$c=\sqrt d/\mathsf r$ for OL. For CL encoding, the server sets
$[\widehat g_{t,n}]_i$ to the average of the received copies of
coordinate $i$, that is, of $y^t_{n,s}$ over
$\big|\{s\in\mathcal R_{t,n}:i_s=i\}\big|$, whenever this set is nonempty.
\end{blueenv}

\subsection{Consequences for Resilience}

The scalar architecture has two consequences that are central to
resilience. First, each packet carries a single scalar rather than a
$d$-dimensional gradient. Consequently, the communication budget of one
full-gradient transmission can be redistributed across up to $d$
packets, creating redundancy that mitigates packet loss. Second, the
transmitted scalars differ in magnitude across constructions. A
projection has mean-square magnitude
$\mathbb E_v(v^\top\tilde g_{t,n})^2=\|\tilde g_{t,n}\|^2$ and yields an
unbiased rank-one estimate of $\tilde g_{t,n}$, whereas a coordinate has
mean-square magnitude $\|\tilde g_{t,n}\|^2/d$ on average and involves
no encoding error; this difference governs the comparison of the
encodings under channel noise.

Section~\ref{sec:analysis} builds on these constructions to address the
three research questions of Section~\ref{sec:intro}. It establishes
conditions under which learning remains stabilizing over noisy and lossy
channels, quantifies the redundancy that a given channel requires, and
compares the constructions with one another and with full-gradient
transmission under the same uplink budget.

\begin{algorithm}[t]\footnotesize
\caption{\small \textsc{ResiFedLQR}: Resilient Federated LQR via
Scalar Encoding over Noisy and Unreliable Uplink Channels}
\label{alg:nc_scalarfedlqr_inline}
\begin{algorithmic}[1]
\State \textbf{Input:} initial gain $K_0\in\mathcal{S}$, stepsize
$\eta$, horizon $T$, ZO parameters $(n_s,\tau,\rev{\mathsf r})$, mode
$\in\{\mathrm{GL},\mathrm{OL}\}$, direction distribution $\mathcal{D}$
satisfying~\eqref{eq:isotropy}--\eqref{eq:fixed_norm} (uniform on the
sphere of radius $\sqrt d$ if OL), scalar transmissions $N_{\rm sc}$ (GL)
\State $N_{\rm pk}\gets N_{\rm sc}$, $c\gets1$ if GL;\quad
$N_{\rm pk}\gets n_s$, $c\gets\sqrt d/\rev{\mathsf r}$ if OL
\Comment{mode setup}
\For{each round $t=0,1,\ldots,T-1$}
  \State Server broadcasts $K_t$ and seed $\xi_{t,n}\in\mathbb{Z}_+$
  to each agent $n\in[M]$
  \For{each agent $n\in[M]$ \textbf{in parallel}}
    \If{GL} compute $\tilde g_{t,n}$ at $K_t$ by~\eqref{eq:zo_estimate}
    \Comment{GL: local ZO estimate}
    \EndIf
    \For{$s=1,\ldots,N_{\rm pk}$}
      \State $v_{t,n,s}\gets\mathrm{PRG}_{\mathcal{D}}(\xi_{t,n};s)$
      \If{GL} $r^t_{n,s}\gets v_{t,n,s}^\top\tilde g_{t,n}$
      \Comment{GL: projection}
      \Else\ run a rollout of length $\tau$ under
      $K_t+\tfrac{\rev{\mathsf r}}{\sqrt d}\mathrm{mat}(v_{t,n,s})$;
      $r^t_{n,s}\gets\widehat J^{(n)}_{t,s}$
      \Comment{OL: rollout cost}
      \EndIf
      \State Transmit $r^t_{n,s}$ as one packet
    \EndFor
  \EndFor
  \State $\mathcal{A}_t\gets\varnothing$
  \For{each agent $n\in[M]$}
    \State $\mathcal{R}_{t,n}\gets$ indices of received packets;
    receive $y^t_{n,s}=r^t_{n,s}+w^t_{n,s}$, $s\in\mathcal{R}_{t,n}$
    \State $N^{\rm rec}_{t,n}\gets|\mathcal{R}_{t,n}|$
    \If{$N^{\rm rec}_{t,n}>0$}
      \State Regenerate $v_{t,n,s}\gets\mathrm{PRG}_{\mathcal{D}}(\xi_{t,n};s)$
      for $s\in\mathcal{R}_{t,n}$
      \State $\widehat g_{t,n}\gets\frac{c}{N^{\rm rec}_{t,n}}
      \sum_{s\in\mathcal{R}_{t,n}}y^t_{n,s}\,v_{t,n,s}$
      \Comment{GL and OL via $c$}
      \State $\mathcal{A}_t\gets\mathcal{A}_t\cup\{n\}$
    \EndIf
  \EndFor
  \If{$\mathcal{A}_t\neq\varnothing$}
    \State $\widehat g_t\gets\frac{1}{|\mathcal{A}_t|}
    \sum_{n\in\mathcal{A}_t}\widehat g_{t,n}$,\quad
    $K_{t+1}\gets K_t-\eta\,\mathrm{mat}(\widehat g_t)$
  \Else
    \State $K_{t+1}\gets K_t$
  \EndIf
\EndFor
\State \textbf{Output:} $K_T$
\end{algorithmic}
\end{algorithm}
%%%%%%%%%%%%%%%%%%%%%%%%%%%%%
%%%%%%%%%%%%%%%%%%%%%%%%%%%%%
\section{\textsc{ResiFedLQR}: Resilient Scalar Federated LQR}
\label{sec:resifedlqr}

Retransmission is a standard means of mitigating the impairments of
Assumption~\ref{assump:noise}: averaging repeated copies attenuates
channel noise, and redundancy compensates for packet losses. Under
bundled full-gradient transmission, each copy costs $d$ real numbers,
whereas a scalar packet costs one; the budget of $N_{\rm F}$
full-gradient transmissions therefore accommodates up to $dN_{\rm F}$
scalar packets. \textsc{ResiFedLQR}
(Algorithm~\ref{alg:nc_scalarfedlqr_inline}) uses this packet budget to
make the policy update resilient to both noise and dropout.

At round $t$, each agent transmits $N_{\rm pk}$ scalars according to
Definition~\ref{def:gl_ol}. Let
\begin{equation}\label{eq:active_set}
  \mathcal A_t:=\{n\in[M]:N^{\rm rec}_{t,n}>0\}
\end{equation}
denote the set of \emph{active} agents, i.e., those from which at least
one scalar is received. If $\mathcal A_t\neq\varnothing$, the server
decodes $\widehat g_{t,n}$ as in Section~\ref{sec:comm_model} for each
$n\in\mathcal A_t$, forms
\begin{equation}\label{eq:resi_decode}
  \widehat g_t:=\frac{1}{|\mathcal A_t|}
  \sum\nolimits_{n\in\mathcal A_t}\widehat g_{t,n},
\end{equation}
and applies~\eqref{eq:policy_update}; otherwise, it sets $K_{t+1}=K_t$.
Under CL encoding, the average in~\eqref{eq:resi_decode} is taken
coordinate by coordinate over the agents that delivered the
corresponding coordinate. Thus $\widehat g_t$ averages, over a random
subset of agents, directions decoded from random subsets of noisy
scalars; Section~\ref{sec:analysis} bounds its deviation from
$\nabla J_{\rm avg}(K_t)$.

\begin{rem}[Communication budget and resilience]\label{rem:budget}
For GL encoding, $N_{\rm sc}$ is a design parameter with
\begin{equation}\label{eq:nsc_range_problem_formulation}
  N_{\rm F}\le N_{\rm sc}\le dN_{\rm F}.
\end{equation}
At $N_{\rm sc}=N_{\rm F}$, the packet count is unchanged and the uplink
cost decreases by a factor of $d$; at $N_{\rm sc}=dN_{\rm F}$, the costs
coincide and the agent sends $d$ times as many packets. Under
Assumption~\ref{assump:noise},
$N^{\rm rec}_{t,n}\sim\mathrm{Bin}(N_{\rm pk},1-p_{\rm drop})$ and
$|\mathcal A_t|\sim\mathrm{Bin}(M,1-p_{\rm drop}^{N_{\rm pk}})$, where
$\mathrm{Bin}(N,p)$ denotes the binomial distribution with $N$ trials
and success probability $p$. Increasing $N_{\rm sc}$ therefore raises
both the expected number of received scalars per agent,
$(1-p_{\rm drop})N_{\rm sc}$, and the probability
$1-p_{\rm drop}^{N_{\rm sc}}$ that an agent is active, while the
averaging in~\eqref{eq:unified_decode} attenuates the channel noise.
Section~\ref{sec:analysis} gives a lower bound on $N_{\rm sc}$ under
which every iterate remains stabilizing and the cost decreases
(see~\eqref{eq:nsc_rule}). For OL encoding, $N_{\rm pk}=n_s$ is fixed by
the number of rollouts (see Remark~\ref{rem:gl_vs_ol}). \hfill$\square$
\end{rem}

Section~\ref{sec:analysis} addresses the three research questions by
characterizing how the number of received scalars, the dropout
probability, the channel noise, and the encoding choice determine the
resilience region of \textsc{ResiFedLQR}.

\section{Stability and Convergence Analysis}\label{sec:analysis}

This section answers the three research questions of
Section~\ref{sec:intro}. Sections~\ref{sec:error_decomp}--\ref{sec:descent}
show that, for a suitable stepsize, the iterates of
Algorithm~\ref{alg:nc_scalarfedlqr_inline} remain in a sublevel set of
$J_{\rm avg}$ contained in $\mathcal S$ and converge linearly to a
neighborhood of the optimum, which yields the certified resilience
region (RQ1). Section~\ref{sec:discussion} separates the contributions
of heterogeneity, sampling, and communication to the residual error and
derives the required redundancy (RQ2). Section~\ref{sec:comparison}
compares the encodings with one another and with full-gradient
transmission under the same uplink budget (RQ3). Throughout, $Q\succ0$,
$R\succ0$, and $\Sigma_0\succ0$ as in Section~\ref{sec:prob_def_object}. All proofs are given in the appendix.

\subsection{Error Decomposition}\label{sec:error_decomp}

Unlike \textsc{FedLQR} over an ideal channel, the server observes
neither the local estimates $\tilde g_{t,n}$ nor all agents. Under GL
encoding, \eqref{eq:decoded_direction} gives, for $n\in\mathcal{A}_t$,
\begin{align}
\label{eq:vectorized_client_decode_dropout}
  \widehat g_{t,n}
  =\underbrace{\tfrac{1}{N_{t,n}^{\rm rec}}\sum\nolimits_{s\in\mathcal R_{t,n}}
    v_{t,n,s}v_{t,n,s}^\top\tilde g_{t,n}}_{\bar g_{t,n}}
  \nonumber \\ +\underbrace{\tfrac{1}{N_{t,n}^{\rm rec}}\sum\nolimits_{s\in\mathcal R_{t,n}}
    w^t_{n,s}v_{t,n,s}}_{e^{\rm ch}_{t,n}} .
\end{align}
Averaging over $\mathcal{A}_t$ as in~\eqref{eq:resi_decode} yields
$\widehat g_t=\bar g_t+e_t^{\rm ch}$ with
$\bar g_t:=\tfrac{1}{|\mathcal{A}_t|}\sum_{n\in\mathcal{A}_t}\bar g_{t,n}$
and $e_t^{\rm ch}:=\tfrac{1}{|\mathcal{A}_t|}\sum_{n\in\mathcal{A}_t}e^{\rm ch}_{t,n}$.
Let
\begin{equation}\label{eq::zo_diff}
  g_t:=\nabla J_{\rm avg}(K_t),~~
  \tilde g_t:=\tfrac{1}{M}\sum_{n=1}^M\tilde g_{t,n},~~
  \tilde g_t^{\mathcal{A}}:=\tfrac{1}{|\mathcal{A}_t|}\sum_{n\in\mathcal{A}_t}\tilde g_{t,n}
\end{equation}
denote the exact gradient and the full and active-agent averages of the
local ZO estimates. The error of the update direction decomposes as
\begin{equation}\label{eq:total_error_resifed_dropout}
  e_t^{\rm tot}:=\widehat g_t-g_t
  =\underbrace{(\bar g_t-\tilde g_t^{\mathcal{A}})}_{e_t^{\rm proj}}
  +\underbrace{(\tilde g_t^{\mathcal{A}}-\tilde g_t)}_{e_t^{\rm drop}}
  +\underbrace{(\tilde g_t-g_t)}_{e_t^{\rm ZO}}
  +e_t^{\rm ch}.
\end{equation}
The terms $e_t^{\rm proj}$, $e_t^{\rm drop}$, and $e_t^{\rm ch}$ are
caused by the communication layer: projection onto finitely many random
directions, averaging over the active agents only, and channel noise.
They vanish under ideal gradient exchange, whereas $e_t^{\rm ZO}$ is
present even then. We group the communication-layer errors as
$e_t^{\rm rec}:=\widehat g_t-\tilde g_t=e_t^{\rm proj}+e_t^{\rm drop}+e_t^{\rm ch}$,
so that
\begin{equation}\label{eq:total_error_resifed}
  e_t^{\rm tot}=e_t^{\rm rec}+e_t^{\rm ZO}.
\end{equation}
Rounds with $\mathcal{A}_t=\varnothing$ leave $K_t$ unchanged, so the
results below concern active rounds.

\subsection{Properties of the Costs}

Let $c:=J_{\rm avg}(K_0)$ and
\begin{equation}\label{eq:sublevel}
  \mathcal{S}_c:=\{K\in\mathcal{S}: J_{\rm avg}(K)\le c\},
\end{equation}
the sublevel set of the initial gain, which is nonempty by
Assumption~\ref{ass:init}. For each agent, let $K^{(n)}_\star$ be the
minimizer of $J^{(n)}$ over $\mathcal{S}^{(n)}$,
$J^{(n)}_\star:=J^{(n)}(K^{(n)}_\star)$, and
$\Sigma^{(n)}_\star:=\sum_{k\ge0}(A^{(n)}-B^{(n)}K^{(n)}_\star)^k
\Sigma_0(A^{(n)}-B^{(n)}K^{(n)}_\star)^{k\top}$ the closed-loop state
covariance under $K^{(n)}_\star$. The following lemma collects known
properties of the single-agent cost.

\begin{lem}[Single-agent LQR properties]\label{lem:single_agent}
Fix $n\in[M]$.
\begin{enumerate}
\item[(a)] (\emph{Coercivity}) $J^{(n)}$ is real-analytic on the open set
$\mathcal{S}^{(n)}$, and $J^{(n)}(K)\to\infty$ as
$K\to\partial\mathcal{S}^{(n)}$ or $\|K\|\to\infty$
\cite[Lemma~3.7]{bu_lqr_2019}.
\item[(b)] (\emph{PL condition}) Let $\mu_n\!:=\!\frac{2\mu_0^2\,\sigma_{\min}(R)}{\|\Sigma^{(n)}_\star\|_2}$. For every $K\!\in\!\mathcal{S}^{(n)}$,
\begin{equation}\label{eq:individual_pl}
  \tfrac12\|\nabla J^{(n)}(K)\|^2
  \ge\mu_n\bigl(J^{(n)}(K)-J^{(n)}_\star\bigr),
\end{equation}
\cite[Lemma~11]{fazel_global_2018}, \cite[Lemma~2]{wang_model-free_2023}.
\item[(c)] (\emph{Gradient sensitivity to the model}) For any
$(A',B')$ with cost $J'$ and any $K$ stabilizing both
$(A^{(n)},B^{(n)})$ and $(A',B')$,
\begin{align}\label{eq:grad_sens}
  \|\nabla J^{(n)}(K)-&\nabla J'(K)\|
  \le{}\|A^{(n)}-A'\|_2\,h^{(n)}_{1,\rm het}(K)\nonumber\\
  &+\|B^{(n)}-B'\|_2\,h^{(n)}_{2,\rm het}(K),
\end{align}
where $h^{(n)}_{1,\rm het},h^{(n)}_{2,\rm het}$ are positive polynomials
in $\|K\|$, $\|A^{(n)}-B^{(n)}K\|_2$, $J^{(n)}(K)$, and the problem data
\cite[Lemma~3 and App.~E]{wang_model-free_2023}. \hfill$\square$
\end{enumerate}
\end{lem}

To transfer these properties to the fleet, let
$\operatorname{dist}(K,\mathcal{X}):=\inf_{K'\in\mathcal{X}}\|K-K'\|$,
$\mu_c:=\min_{n\in[M]}\mu_n$, and define
\begin{equation}\label{eq:rc_Scplus}
  r_c:=\tfrac12\!\inf_{K\in\mathcal{S}_c,\,K'\notin\mathcal{S}}\|K-K'\|,
  ~
  \mathcal{S}_c^+:=\{K\!:\!\operatorname{dist}(K,\mathcal{S}_c)\le r_c\},
\end{equation}
\begin{equation}\label{eq:constants}
  \bar L_c:=\sup_{K\in\mathcal{S}_c^+}\|\nabla^2J_{\rm avg}(K)\|_2,\quad
  G_c:=\sup_{K\in\mathcal{S}_c}\|\nabla J_{\rm avg}(K)\|,
\end{equation}
\begin{equation}\label{eq:hbar}
  \bar h_{q,c}:=\max_{n\in[M]}\sup_{K\in\mathcal{S}_c}h^{(n)}_{q,\rm het}(K),
  \quad
  b_{\rm het}:=\bigl(\epsilon_1\bar h_{1,c}+\epsilon_2\bar h_{2,c}\bigr)^2,
\end{equation}
where $q\in\{1,2\}$.

\begin{lem}[Fleet-level properties on $\mathcal{S}_c$]\label{lem:fleet}
Let Assumptions~\ref{ass:similarity} and~\ref{ass:init} hold. Then:
\begin{enumerate}
\item[(a)] $\mathcal{S}_c$ is compact and $\mathcal{S}_c\subseteq\mathcal{S}$;
\item[(b)] $r_c>0$, and $\mathcal{S}_c^+$ is compact with
$\mathcal{S}_c^+\subseteq\mathcal{S}$;
\item[(c)] $\bar L_c<\infty$ and $G_c<\infty$;
\item[(d)] $\bar h_{1,c},\bar h_{2,c}<\infty$, and for all $i,j\in[M]$
and $K\in\mathcal{S}_c$,
\begin{equation}\label{eq:grad_het}
  \|\nabla J^{(i)}(K)-\nabla J^{(j)}(K)\|\le2\sqrt{b_{\rm het}} .
\end{equation}
\end{enumerate}
\end{lem}

% \begin{proof}
% (a) By Lemma~\ref{lem:single_agent}(a) and the nonnegativity of the
% costs, $J_{\rm avg}$ is continuous on $\mathcal{S}$ and diverges as
% $K\to\partial\mathcal{S}$ or $\|K\|\to\infty$; hence $\mathcal{S}_c$ is
% closed, bounded, and contained in $\mathcal{S}$.
% (b) The complement of $\mathcal{S}$ is closed and disjoint from the
% compact set $\mathcal{S}_c$, so $r_c>0$, and every
% $K\in\mathcal{S}_c^+$ is at distance at least $r_c$ from it.
% (c) $\nabla J_{\rm avg}$ and $\nabla^2J_{\rm avg}$ are continuous on
% $\mathcal{S}$, hence bounded on the compact sets
% $\mathcal{S}_c\subseteq\mathcal{S}_c^+$.
% (d) The arguments of $h^{(n)}_{q,\rm het}$ are continuous on
% $\mathcal{S}_c$, with $J^{(n)}(K)\le MJ_{\rm avg}(K)\le Mc$, so the
% suprema in~\eqref{eq:hbar} are finite by (a). By~\eqref{eq:system_bound},
% $\|A^{(i)}-A^{(j)}\|_2\le2\epsilon_1$ and
% $\|B^{(i)}-B^{(j)}\|_2\le2\epsilon_2$. Applying~\eqref{eq:grad_sens}
% with $n=i$ and $(A',B')=(A^{(j)},B^{(j)})$, admissible since
% $K\in\mathcal{S}^{(i)}\cap\mathcal{S}^{(j)}$, gives~\eqref{eq:grad_het}.
% \end{proof}

The proof is given in the appendix. Heterogeneity enters the analysis only through $b_{\rm het}$, which
vanishes when $\epsilon_1=\epsilon_2=0$; by~\eqref{eq:grad_het}, the
average cost satisfies a PL inequality up to this bias.

\begin{lem}[Collective PL condition]\label{lem:collective_pl}
Under Assumptions~\ref{ass:similarity} and~\ref{ass:init}, for every
$K\in\mathcal{S}_c$,
\begin{equation}\label{eq:collective_pl}
  \tfrac12\|\nabla J_{\rm avg}(K)\|^2
  \ge \mu_c\bigl(J_{\rm avg}(K)-J^\star_{\rm avg}\bigr)-b_{\rm het},
\end{equation}
where $J^\star_{\rm avg}:=\min_{K\in\mathcal{S}}J_{\rm avg}(K)$.\boxend
\end{lem}
The proof is given in the appendix.
% \begin{proof}
% Let $g_n:=\nabla J^{(n)}(K)$ and $g:=\frac1M\sum_ng_n=\nabla J_{\rm avg}(K)$.
% Expanding the squares gives
% \[
%   \tfrac12\|g\|^2
%   =\frac{1}{2M}\sum\nolimits_{n=1}^M\|g_n\|^2
%    -\frac{1}{4M^2}\sum\nolimits_{i,j=1}^M\|g_i-g_j\|^2 .
% \]
% By~\eqref{eq:individual_pl} and $\mu_n\ge\mu_c$, the first term is at
% least $\mu_c\bigl(J_{\rm avg}(K)-\frac1M\sum_nJ^{(n)}_\star\bigr)
% \ge\mu_c\bigl(J_{\rm avg}(K)-J^\star_{\rm avg}\bigr)$, since
% $J^{(n)}_\star\le J^{(n)}(K')$ for every $K'\in\mathcal{S}$.
% By~\eqref{eq:grad_het}, the second term is at most
% $\frac{M(M-1)}{4M^2}\cdot4b_{\rm het}\le b_{\rm het}$.
% \end{proof}

\subsection{Zeroth-Order and Heterogeneity Errors}

For $\varepsilon>0$, define
$\mathcal{E}^{\rm ZO}_{t,n}:=\{\|\tilde g_{t,n}-\nabla J^{(n)}(K_t)\|\le\varepsilon\}$
and $\mathcal{E}^{\rm ZO}_t:=\bigcap_{n=1}^M\mathcal{E}^{\rm ZO}_{t,n}$.
The only property of the local gradient oracle used in the analysis is
that, for every $\varepsilon>0$ and $\delta_{\rm ZO}\in(0,1)$, its
sampling parameters can be chosen, polynomially in the problem data,
$1/\varepsilon$, and $\log(M/\delta_{\rm ZO})$, such that
\begin{equation}\label{eq:zo_event}
  \mathbb{P}\bigl(\mathcal{E}^{\rm ZO}_t\,\big|\,K_t\bigr)\ge1-\delta_{\rm ZO}
  \qquad\text{for every }K_t\in\mathcal{S}_c .
\end{equation}
The one-point estimator~\eqref{eq:zo_estimate} satisfies
\eqref{eq:zo_event} with parameters $(n_s,\tau,\mathsf r)$, provided
$\mathsf r\le r_c$ so that every perturbed gain lies in
$\mathcal{S}_c^+\subseteq\mathcal{S}$; this follows
from~\cite[Lemma~4]{wang_model-free_2023} and the compactness of
$\mathcal{S}_c$ under the tail conditions discussed after
Assumption~\ref{ass:init_state}. The two-point estimator also
satisfies~\eqref{eq:zo_event}, with improved oracle
complexity~\cite{malik_derivative-free_2019,toso2024oracle}. We fix
such a choice for the rest of the paper. On $\mathcal{E}^{\rm ZO}_t$,
$\|e^{\rm ZO}_t\|\le\varepsilon$, and the local estimates remain close
to their average.

\begin{lem}[Bounded gradient heterogeneity]\label{lem:heterogeneity}
Let $K_t\in\mathcal{S}_c$ and $\Delta_{t,n}:=\tilde g_{t,n}-\tilde g_t$.
On $\mathcal{E}^{\rm ZO}_t$, for all $n\in[M]$,
\begin{equation}\label{eq:gamma_B}
  \|\Delta_{t,n}\|\le\Delta_{\max}:=2\sqrt{b_{\rm het}}+2\varepsilon .
\end{equation}
\end{lem}

% \begin{proof}
% Write $\Delta_{t,n}=\bigl(\nabla J^{(n)}(K_t)-\nabla J_{\rm avg}(K_t)\bigr)
% +\bigl(\tilde g_{t,n}-\nabla J^{(n)}(K_t)\bigr)-e^{\rm ZO}_t$. The first
% term is at most $2\sqrt{b_{\rm het}}$ by~\eqref{eq:grad_het} and the
% triangle inequality; each of the other two is at most $\varepsilon$ on
% $\mathcal{E}^{\rm ZO}_t$.
% \end{proof}

The proof is given in the appendix. Bounded gradient heterogeneity is usually imposed as an assumption in
federated optimization~\cite{rostami2023federated}; for LQR, it follows
from Assumption~\ref{ass:similarity}.

\subsection{Reconstruction Error under Noise and Dropout}

We analyze the GL encoding. Recall from Remark~\ref{rem:budget} that
$N^{\rm rec}_{t,n}\sim\mathrm{Bin}(N_{\rm sc},q_{\rm sc})$ and
$|\mathcal{A}_t|\sim\mathrm{Bin}(M,q_{\rm act})$, where
$q_{\rm sc}:=1-p_{\rm drop}$ and $q_{\rm act}:=1-p_{\rm drop}^{N_{\rm sc}}$
are the delivery probability of one scalar and the activation
probability of one agent. Let
\begin{equation}\label{eq:nu_theta}
  \nu:=q_{\rm sc}N_{\rm sc},\quad
  \theta:=\sqrt{\frac{d\,\lambda_T}{Mq_{\rm act}\,\nu}},\quad
  \lambda_T:=\log\frac{6dT}{\delta},
\end{equation}
for a horizon $T\ge1$ and confidence level $\delta\in(0,1)$. Here $\nu$
is the expected number of received scalars per agent. With high
probability, at least $Mq_{\rm act}/2$ agents are active and each
delivers at least $\nu/2$ scalars; $\theta$ is then the relative
accuracy with which the $d$-dimensional average can be reconstructed
from these scalars, and it governs the projection and channel-noise
terms below.

\begin{lem}[Reconstruction error in one round]\label{lem:hp_erec_resifed}
Let Assumptions~\ref{ass:similarity}, \ref{ass:init},
and~\ref{assump:noise} hold, consider GL encoding, and fix a round $t$
with $K_t\in\mathcal{S}_c$. Let $\bar\nu:=\max\{1,\nu/2\}$ and
$\mathcal{E}^{\rm rec}_t:=\{N^{\rm rec}_{t,n}\ge\bar\nu\ \forall n\in\mathcal{A}_t\}$.
Then $\mathbb{P}(\mathcal{E}^{\rm rec}_t)\ge1-Me^{-\nu/8}$, and there is
an event of probability at least $1-\delta/T$ on whose intersection with
$\mathcal{E}^{\rm ZO}_t\cap\mathcal{E}^{\rm rec}_t\cap\{|\mathcal{A}_t|>0\}$
\begin{align}\label{eq:hp_erec_resifed}
  \|e_t^{\rm rec}\|
  \le\;&C\Bigl[\sqrt{\tfrac{d\lambda_T}{|\mathcal{A}_t|\,\nu}}
      +\tfrac{d\lambda_T}{|\mathcal{A}_t|\,\nu}\Bigr]
      \bigl(\|\tilde g_t\|+\Delta_{\max}+\sigma_{\rm ch}\bigr)
  \nonumber\\
  &+C\Bigl[\sqrt{\tfrac{M(1-q_{\rm act})\lambda_T}{|\mathcal{A}_t|^2}}
      +\tfrac{\lambda_T}{|\mathcal{A}_t|}\Bigr]\Delta_{\max},
\end{align}
where $C=2$. \hfill$\square$
\end{lem}
The proof is given in the appendix. The first line of~\eqref{eq:hp_erec_resifed} is the error of
reconstructing a $d$-dimensional vector from about $|\mathcal{A}_t|\nu$
noisy scalar projections; it scales with the size of the projected
vectors, $\|\tilde g_t\|+\Delta_{\max}$, plus the noise level. The
second line bounds $e^{\rm drop}_t$, the error of averaging over a
random subset of agents, and scales with their dissimilarity
$\Delta_{\max}$. Using $\|\tilde g_t\|\le\|g_t\|+\varepsilon$ on
$\mathcal{E}^{\rm ZO}_t$, a Chernoff bound on $|\mathcal{A}_t|$, and a
union bound over the horizon, we obtain a bound that is uniform in $t$
and separates a part proportional to $\|g_t\|$ from an additive~part. Its relative constant $\bar a$ depends only on the number of received
scalars, whereas its additive constant $\bar b$ collects the sampling
error, the projection and channel-noise errors scaled by $\theta$, and
the dropout error.

\begin{cor}[Uniform relative-plus-additive bound]\label{cor:uniform}
Let Assumption~\ref{assump:noise} hold, assume $\theta\le1$, and define
\begin{subequations}\label{eq:ab_bar}
\begin{align}
  \bar a&:=4C\,\theta,\label{eq:a_bar}\\
  \bar b&:=\varepsilon+\bar a\,(\varepsilon+\Delta_{\max}+\sigma_{\rm ch})
   \nonumber\\
   &\quad+2C\Bigl[\sqrt{\tfrac{(1-q_{\rm act})\lambda_T}{Mq_{\rm act}^2}}
     +\tfrac{\lambda_T}{Mq_{\rm act}}\Bigr]\Delta_{\max} .\label{eq:b_bar}
\end{align}
\end{subequations}
Let $\mathcal{E}^{\rm ZO}_T:=\bigcap_{t<T}\mathcal{E}^{\rm ZO}_t$, and
let $\mathcal{E}_T$ be the event that, for every active round $t<T$ with
$K_t\in\mathcal{S}_c$,
\begin{equation}\label{eq:rel_add}
  \|e_t^{\rm tot}\|\le\bar a\,\|g_t\|+\bar b .
\end{equation}
Then
$\mathbb{P}(\mathcal{E}_T\cap\mathcal{E}^{\rm ZO}_T)
\ge1-\delta-T\bigl(e^{-Mq_{\rm act}/8}+Me^{-\nu/8}+\delta_{\rm ZO}\bigr)$.
\end{cor}
The proof is given in the appendix.
% \begin{proof}
% By the Chernoff lower tail,
% $\mathbb{P}(|\mathcal{A}_t|<Mq_{\rm act}/2)\le e^{-Mq_{\rm act}/8}$.
% On $\{|\mathcal{A}_t|\ge Mq_{\rm act}/2\}$, substituting
% $1/|\mathcal{A}_t|\le2/(Mq_{\rm act})$ into~\eqref{eq:hp_erec_resifed}
% bounds its first bracket by $\sqrt2\theta+2\theta^2\le4\theta$, using
% $\theta\le1$, and its second by
% $2\sqrt{(1-q_{\rm act})\lambda_T/(Mq_{\rm act}^2)}+2\lambda_T/(Mq_{\rm act})$.
% Then $\|\tilde g_t\|\le\|g_t\|+\varepsilon$ and
% $\|e_t^{\rm tot}\|\le\|e_t^{\rm rec}\|+\varepsilon$ give~\eqref{eq:rel_add}.
% For each round, the bound~\eqref{eq:hp_erec_resifed}, the Chernoff
% event, $\mathcal{E}^{\rm rec}_t$, and $\mathcal{E}^{\rm ZO}_t$ fail with
% probabilities at most $\delta/T$, $e^{-Mq_{\rm act}/8}$, $Me^{-\nu/8}$,
% and $\delta_{\rm ZO}$, respectively, the last by~\eqref{eq:zo_event}; a
% union bound over the $T$ rounds completes the proof.
% \end{proof}

\subsection{Descent, Stability, and Convergence}\label{sec:descent}

Since $\bar b\ge\varepsilon>0$, a purely relative error bound cannot
hold near a stationary point, and the descent lemma below accounts for
both relative and additive errors. Any $L\ge\bar L_c$ is a smoothness
constant of $J_{\rm avg}$ on $\mathcal{S}_c^+$; we use
$L_c:=\max\{\bar L_c,\mu_c\}$, so that $\mu_c\le L_c$, which is used only
to place the contraction factor of Theorem~\ref{thm:main} in
$[\tfrac34,1)$.

\begin{lem}[One-step descent with relative and additive error]
\label{lem:federated_descent}
Let $K\in\mathcal{S}_c$, $g:=\nabla J_{\rm avg}(K)$, and
$K^+:=K-\eta\widehat g$ with $\widehat g=g+e$,
$\|e\|\le a\|g\|+b$, $a\in[0,1)$, $b\ge0$. If $\eta\|\widehat g\|\le r_c$
and $0<\eta\le\frac{1-a}{2L_c(1+a)^2}$, then
\begin{equation}\label{eq:one_step_descent}
  J_{\rm avg}(K^+)\le J_{\rm avg}(K)
  -\frac{\eta(1-a)}{4}\|g\|^2+\frac{3\eta\,b^2}{2(1-a)} .
\end{equation}
\end{lem}

% \begin{proof}
% Since $\|K^+-K\|\le r_c$, the segment $[K,K^+]$ lies in
% $\mathcal{S}_c^+$, on which $\|\nabla^2J_{\rm avg}\|_2\le L_c$; hence
% $J_{\rm avg}(K^+)\le J_{\rm avg}(K)-\eta\langle g,\widehat g\rangle
% +\frac{L_c\eta^2}{2}\|\widehat g\|^2$. Using
% $\langle g,e\rangle\ge-a\|g\|^2-b\|g\|$ and
% $\|\widehat g\|^2\le\bigl((1+a)\|g\|+b\bigr)^2\le2(1+a)^2\|g\|^2+2b^2$,
% \begin{gather*}
%   J_{\rm avg}(K^+)\le J_{\rm avg}(K)
%   -\eta\bigl[(1-a)-L_c\eta(1+a)^2\bigr]\|g\|^2\\
%   +\eta b\|g\|+L_c\eta^2b^2 .
% \end{gather*}
% For the stated $\eta$, the bracket is at least $(1-a)/2$. Young's
% inequality gives
% $\eta b\|g\|\le\frac{\eta(1-a)}{4}\|g\|^2+\frac{\eta b^2}{1-a}$, and
% $L_c\eta^2b^2\le\frac{\eta(1-a)b^2}{2(1+a)^2}\le\frac{\eta b^2}{2(1-a)}$.
% \end{proof}

The proof is given in the appendix. Let $N^{\rm up}_t:=\sum_{t'<t}\mathbf{1}\{|\mathcal{A}_{t'}|>0\}$ denote
the number of active rounds before $t$, and
$F_t:=J_{\rm avg}(K_t)-J^\star_{\rm avg}$ the optimality gap.

\begin{thm}[Stability and linear convergence of \textsc{ResiFedLQR}]
\label{thm:main}
Let Assumptions~\rev{\ref{ass:init_state}}--\ref{assump:noise} hold,
consider GL encoding with the ZO parameters satisfying~\eqref{eq:zo_event},
and let $\bar a,\bar b$ be as in~\eqref{eq:ab_bar}. \rev{Suppose
$\theta<1/(4C)$, so that $\bar a<1$,} and
\begin{equation}\label{eq:certif}
  \bar F:=\frac{1}{\mu_c}\Bigl[b_{\rm het}+\frac{3\bar b^{\,2}}{(1-\bar a)^2}\Bigr]
  \le F_0 .
\end{equation}
Then, for any stepsize
\begin{equation}\label{eq:stepsize}
  0<\eta\le\eta^\star:=\min\Bigl\{\frac{1-\bar a}{2L_c(1+\bar a)^2},\;
  \frac{r_c}{(1+\bar a)G_c+\bar b}\Bigr\},
\end{equation}
with probability at least
$1-\delta-T(e^{-Mq_{\rm act}/8}+Me^{-\nu/8}+\delta_{\rm ZO})$, for all
$t\le T$:
\begin{enumerate}
\item[(i)] (\emph{Stability}) $K_t\in\mathcal{S}_c\subseteq\mathcal{S}$,
i.e., every iterate stabilizes every agent;
\item[(ii)] (\emph{Convergence})
$F_t\le\rho^{N^{\rm up}_t}F_0+\bigl(1-\rho^{N^{\rm up}_t}\bigr)\bar F$,
\end{enumerate}
where $\rho:=1-\eta\mu_c(1-\bar a)/2\in[\tfrac34,1)$.\boxend
\end{thm}

% \begin{proof}
% Work on $\mathcal{E}_T\cap\mathcal{E}^{\rm ZO}_T$ of
% Corollary~\ref{cor:uniform} and argue by induction on $t$ that
% $F_t\le F_0$, which, since $c=J_{\rm avg}(K_0)$
% and by~\eqref{eq:finite_iff_stable}, is equivalent to
% $K_t\in\mathcal{S}_c$. The claim holds at $t=0$. If round $t$ is
% inactive, $K_{t+1}=K_t$. Otherwise,~\eqref{eq:rel_add} holds, so
% $\|\widehat g_t\|\le(1+\bar a)G_c+\bar b$ and
% $\eta\|\widehat g_t\|\le r_c$ by~\eqref{eq:stepsize}.
% Lemma~\ref{lem:federated_descent} with $(a,b)=(\bar a,\bar b)$ and
% Lemma~\ref{lem:collective_pl}, $\|g_t\|^2\ge2\mu_cF_t-2b_{\rm het}$,~give
% \begin{multline*}
%   F_{t+1}\le\Bigl(1-\frac{\eta\mu_c(1-\bar a)}{2}\Bigr)F_t
%   +\frac{\eta(1-\bar a)}{2}b_{\rm het}\\
%   +\frac{3\eta\bar b^{\,2}}{2(1-\bar a)}
%   =\rho F_t+(1-\rho)\bar F .
% \end{multline*}
% Since $\eta\le\frac{1-\bar a}{2L_c(1+\bar a)^2}$ and $\mu_c\le L_c$,
% $1-\rho\le\frac14$. Hence $F_{t+1}\le\max\{F_t,\bar F\}\le F_0$
% by~\eqref{eq:certif}, which proves (i). For (ii),
% $F_{t+1}-\bar F\le\rho\,(F_t-\bar F)$ on active rounds and
% $F_{t+1}-\bar F=F_t-\bar F$ on inactive rounds; since $\rho>0$,
% induction gives $F_t-\bar F\le\rho^{N^{\rm up}_t}(F_0-\bar F)$, which
% rearranges to (ii).
% \end{proof}

The proof is given in the appendix. Theorem~\ref{thm:main} answers RQ1 for GL encoding: scalar
communication with redundancy preserves stabilizing learning whenever
$\bar a<1$, \eqref{eq:certif}, and~\eqref{eq:stepsize} hold. These
conditions define the \emph{certified resilience region}, a set of
$(\eta,N_{\rm sc},p_{\rm drop},\sigma_{\rm ch})$ contained in the
resilience region of Section~\ref{sec:prob_def_object}; its empirical
counterpart is reported in Section~\ref{sec:numerics}.

The guarantee is conservative. The constants in~\eqref{eq:ab_bar}
and~\eqref{eq:certif} arise from Bernstein-type and union bounds, so
that $\bar a<1$ requires $Mq_{\rm act}\nu>64\,d\,\lambda_T$, and $L_c$,
$G_c$, and $r_c$ are not computable in the model-free setting; as
in~\cite{fazel_global_2018,malik_derivative-free_2019,wang_model-free_2023},
the stepsize is tuned in practice. \rev{Moreover, the probability in
Theorem~\ref{thm:main} is informative only when
$Mq_{\rm act}\gtrsim8\log(T/\delta)$ and $\nu\gtrsim8\log(MT/\delta)$,
e.g., $Mq_{\rm act}\gtrsim80$ for $T=1000$ and $\delta=0.05$. The
theorem therefore describes the large-fleet regime; in
Section~\ref{sec:numerics}, with $M=10$, it is used only to predict how
stability and convergence depend on
$(\sigma_{\rm ch},p_{\rm drop},N_{\rm sc})$, not as a certificate.}

\subsection{Discussion of the Guarantees}\label{sec:discussion}

The radius $\bar F$ in~\eqref{eq:certif} collects three sources of
error with different costs of removal.
\begin{enumerate}
\item[1)] \emph{Heterogeneity} contributes $b_{\rm het}/\mu_c
=\mathcal{O}((\epsilon_1+\epsilon_2)^2)$, independent of $N_{\rm sc}$,
$n_s$, and $\eta$. It is the price of a common
policy~\cite[Th.~3]{wang_model-free_2023} and can be removed only by
local fine-tuning.
\item[2)] \emph{Zeroth-order sampling} contributes through
$\varepsilon$ in $\bar b$; reducing it requires more or longer rollouts,
the most expensive resource.
\item[3)] \emph{Communication}  contributes through the relative error
$\bar a=4C\theta$ and through three terms of $\bar b$
in~\eqref{eq:b_bar}: the projection and channel-noise term
$\bar a(\sigma_{\rm ch}+\Delta_{\max}+\varepsilon)$ and the two dropout
terms, proportional to
$\sqrt{(1-q_{\rm act})/(Mq_{\rm act}^2)}\,\Delta_{\max}$ and to
$\Delta_{\max}/(Mq_{\rm act})$. Additional scalar transmissions, often
the cheapest resource, reduce $\bar a$ and the first two terms: $\theta$
decreases as $1/\sqrt{N_{\rm sc}}$, since
$\theta^2=d\lambda_T/(Mq_{\rm act}q_{\rm sc}N_{\rm sc})$, and
$1-q_{\rm act}=p_{\rm drop}^{N_{\rm sc}}$ decreases exponentially in
$N_{\rm sc}$. The last term does not decrease with $N_{\rm sc}$, only
with the fleet size, as $1/M$. It is conservative: the dropout error
$e^{\rm drop}_t$ is the error of averaging over the active agents only,
so it is zero in every round in which all agents are active, and this
occurs with probability $q_{\rm act}^M$, which tends to one as
$N_{\rm sc}$ grows; the last term does not capture this because it
stems from the worst-case term of Bernstein's inequality. The term
$\theta\Delta_{\max}$ couples heterogeneity and communication but,
unlike $b_{\rm het}$, imposes no irreducible accuracy floor.
\end{enumerate}

\paragraph{Effect on stepsize and rate}
The contraction factor $\rho=1-\eta\mu_c(1-\bar a)/2$ depends on the
channel only through $\bar a=4C\theta$, i.e., through dropout and the
number of transmissions. Channel noise and zeroth-order error enter
through $\bar b$, hence through the radius $\bar F$ and the second
stepsize bound in~\eqref{eq:stepsize}. As $\theta\to0$ with the first
bound active, the rate approaches that of exact federated gradient
descent, $\rho\to1-\mu_c/(4L_c)$.

\paragraph{Required redundancy}
The following rule answers RQ2. Given a target relative accuracy
$\bar a\le a_0<1$ and a target $b_0$ for the projection and
channel-noise term and for each dropout term of $\bar b$, these targets
are met if
\begin{subequations}\label{eq:nsc_rule}
\begin{align}
&  N_{\rm sc}\ge\frac{16C^2 d\,\lambda_T}{Mq_{\rm sc}^2}
  \max\Bigl\{\frac{1}{a_0^2},\;
  \frac{(\sigma_{\rm ch}+\Delta_{\max}+\varepsilon)^2}{b_0^2}\Bigr\},
  \label{eq:nsc_rule_a}\\
&  p_{\rm drop}^{N_{\rm sc}}\le
  \frac{Mq_{\rm act}^2b_0^2}{4C^2\lambda_T\Delta_{\max}^2},
  \qquad
  Mq_{\rm act}\ge\frac{2C\lambda_T\Delta_{\max}}{b_0}.
  \label{eq:nsc_rule_b}
\end{align}
\end{subequations}
Condition~\eqref{eq:nsc_rule_a} asks for enough scalars per agent: the
first argument of the maximum keeps the relative error below $a_0$, and
the second keeps the projection and channel-noise term of $\bar b$
below $b_0$; it uses $q_{\rm act}\ge q_{\rm sc}$. The first condition
in~\eqref{eq:nsc_rule_b} asks that an agent rarely lose all of its
packets, which keeps the first dropout term below $b_0$. The second is
a requirement on the fleet size rather than on $N_{\rm sc}$ and keeps
the second dropout term below $b_0$. The required redundancy grows
linearly with the policy dimension $d$ and, when noise dominates, with
the noise variance $\sigma_{\rm ch}^2$, and decreases linearly with the
fleet size $M$. It fits within the budget of Remark~\ref{rem:budget}
whenever the right-hand side of~\eqref{eq:nsc_rule_a} is at most
$dN_{\rm F}$, a requirement that is independent of $d$.

\subsection{\rev{Comparison of Encodings}}\label{sec:comparison}

We now turn to RQ3. All encodings of Definition~\ref{def:gl_ol} use
one-real-number packets and differ only in what the scalar is. We first
quantify what such packets buy over bundled full-gradient packets, and
then compare the encodings with one another.

\paragraph{Full-gradient transmission at equal uplink budget}
Let \textsc{FedLQR} transmit $\tilde g_{t,n}$ $N_{\rm F}$ times per
round as $N_{\rm F}$ packets, so that
$N^{\rm rec}_{t,n}\sim\mathrm{Bin}(N_{\rm F},q_{\rm sc})$, and let the
server average the received copies and then the active agents. Define
$\nu_{\rm F}:=q_{\rm sc}N_{\rm F}$,
$q^{\rm F}_{\rm act}:=1-p_{\rm drop}^{N_{\rm F}}$, and
$\theta_{\rm F}:=\sqrt{d\lambda_T/(Mq^{\rm F}_{\rm act}\nu_{\rm F})}$.

Full-gradient packets involve no projection, and the projection error
is the only term of~\eqref{eq:hp_erec_resifed} that scales with
$\|\tilde g_t\|$; channel noise and dropout contribute additive errors
only. The relative constant is therefore zero for \textsc{FedLQR}, and
the condition $\theta<1/(4C)$ of Theorem~\ref{thm:main}, whose sole
role is to ensure $\bar a<1$, is not needed: $\theta_{\rm F}$ enters
only through the additive constant. Since the proof of
Theorem~\ref{thm:main} uses the update error only
through~\eqref{eq:rel_add}, its claims carry over.

\begin{prop}[Full-gradient transmission over the same channel]
\label{prop:fedlqr_channel}
Let the hypotheses of Theorem~\ref{thm:main} hold\rev{, except those on
$\theta$}. Then the \textsc{FedLQR} update direction
satisfies~\eqref{eq:rel_add} with $\bar a^{\rm F}=0$ and
\begin{equation}\label{eq:ab_fedlqr}
  \bar b^{\rm F}=\varepsilon+4C\theta_{\rm F}\,\sigma_{\rm ch}
  +2C\Bigl[\sqrt{\tfrac{(1-q^{\rm F}_{\rm act})\lambda_T}{M(q^{\rm F}_{\rm act})^2}}
   +\tfrac{\lambda_T}{Mq^{\rm F}_{\rm act}}\Bigr]\Delta_{\max},
\end{equation}
with probability at least
$1-\delta-T(e^{-Mq^{\rm F}_{\rm act}/8}+Me^{-\nu_{\rm F}/8}+\delta_{\rm ZO})$,
and Theorem~\ref{thm:main} holds for \textsc{FedLQR} with
$(\bar a,\bar b)$ replaced by $(\bar a^{\rm F},\bar b^{\rm F})$.\boxend
\end{prop}
The proof is given in the appendix.

Fix $N_{\rm sc}=dN_{\rm F}$, so that GL encoding and \textsc{FedLQR} use
the same number $dN_{\rm F}$ of channel uses per agent and round. Then
$\nu=d\nu_{\rm F}$ and
$q_{\rm act}=1-p_{\rm drop}^{dN_{\rm F}}\ge q^{\rm F}_{\rm act}$, hence
$\theta\le\theta_{\rm F}/\sqrt d$. Comparing the upper
bounds~\eqref{eq:ab_bar} and~\eqref{eq:ab_fedlqr}, and hence the
available guarantees rather than the realized errors:
\begin{enumerate}
\item[1)] the dropout terms are smaller for GL, since
$1-q_{\rm act}=p_{\rm drop}^{dN_{\rm F}}\le p_{\rm drop}^{N_{\rm F}}$;
\item[2)] the channel-noise term is $4C\theta\sigma_{\rm ch}$ for GL
against $4C\theta_{\rm F}\sigma_{\rm ch}$, smaller by a factor of at
least $\sqrt d$ \rev{under the fixed noise floor of~\eqref{eq:channel}
(see Remark~\ref{rem:power} for the role of transmit energy)};
\item[3)] GL incurs the relative error $\bar a=4C\theta$ and the term
$4C\theta(\Delta_{\max}+\varepsilon)$, both absent under full-gradient
transmission; since
$\theta^2=\lambda_T/(Mq_{\rm act}q_{\rm sc}N_{\rm F})$, they decrease
as $1/\sqrt{MN_{\rm F}}$, that is, as the fleet size or the budget
grows.
\end{enumerate}
\begin{revenv}
Setting aside the dropout terms, the additive constant $\bar b$ of GL
is therefore smaller than the constant $\bar b^{\rm F}$ of
\textsc{FedLQR} when
\begin{equation}\label{eq:fedlqr_crossover}
  (\sqrt d-1)\,\sigma_{\rm ch}>\Delta_{\max}+\varepsilon
  \quad\text{and}\quad \theta<1/(4C),
\end{equation}
where the second condition is the one under which
Theorem~\ref{thm:main} applies to GL. In this case, the certified
resilience region of GL can contain operating points at
which~\eqref{eq:certif} fails for \textsc{FedLQR}. When the channel
noise is small compared with $\Delta_{\max}+\varepsilon$, the
projection error dominates and \textsc{FedLQR} has the smaller certified
radius; over a noiseless channel ($\sigma_{\rm ch}=0$), projections
bring no advantage beyond the dropout terms of item~1).
\end{revenv}

\begin{revenv}
\paragraph{Projections versus coordinates}
Coordinates offer the same packet granularity as projections without
projection error. The next result is exact rather than a comparison of
bounds: it gives the mean-square error of GL and CL encoding for a
single agent and a single round, conditional on
$\tilde g:=\tilde g_{t,n}$, at the same number $dN_{\rm F}$ of packets.

\begin{prop}[Projection vs.\ coordinate encoding]\label{prop:elementwise}
Let Assumption~\ref{assump:noise} hold with
$\mathbb{E}[w^2]=\sigma_{\rm ch}^2$, and let each scheme send
$dN_{\rm F}$ packets. Then, the followings hold:\\
(a) GL encoding with $N_{\rm sc}=dN_{\rm F}$ and
decoder~\eqref{eq:decoded_direction} satisfies, conditional on
$N^{\rm rec}_{t,n}=k\ge1$,
\begin{equation}\label{eq:mse_scalar}
  \mathbb{E}\|\widehat g_{t,n}-\tilde g\|^2
  =\frac{(d-1)\|\tilde g\|^2+d\,\sigma_{\rm ch}^2}{k}.
\end{equation}
(b) CL encoding, in which coordinate $i$ is sent in $N_{\rm F}$ packets,
$k_i$ of them are received, and the server sets $\widehat g_i$ to the
average of the received copies if $k_i\ge1$ and to zero otherwise,
satisfies
\begin{equation}\label{eq:mse_elem}
  \mathbb{E}\|\widehat g-\tilde g\|^2
  =d\,\kappa_{\rm F}\,\sigma_{\rm ch}^2+p_{\rm drop}^{N_{\rm F}}\|\tilde g\|^2,
\end{equation}
where $\kappa_{\rm F}:=\mathbb{E}\bigl[\mathbf 1\{k_1\ge1\}/k_1\bigr]$ with
$k_1\sim\mathrm{Bin}(N_{\rm F},q_{\rm sc})$. The same expression holds
for bundled full-gradient transmission.\\
(c) Let GL return $\widehat g_{t,n}=0$ when no packet is received, and
let $\kappa_{\rm sc}:=\mathbb{E}\bigl[\mathbf 1\{k\ge1\}/k\bigr]$ with
$k\sim\mathrm{Bin}(dN_{\rm F},q_{\rm sc})$. Then GL has the smaller
mean-square error if and only if
\begin{equation}\label{eq:elem_crossover}
  d(\kappa_{\rm F}-\kappa_{\rm sc})\,\sigma_{\rm ch}^2
  >\bigl[(d-1)\kappa_{\rm sc}+p_{\rm drop}^{dN_{\rm F}}
  -p_{\rm drop}^{N_{\rm F}}\bigr]\|\tilde g\|^2 .
\end{equation}
Without dropout, this reads $\sigma_{\rm ch}^2>\|\tilde g\|^2/d$.\boxend
\end{prop}

% \begin{proof}
% (a) Conditional on the received set,
% $\widehat g_{t,n}-\tilde g=\frac1k\sum_s\bigl[(v_sv_s^\top-I)\tilde g+w_sv_s\bigr]$,
% a sum of $k$ independent mean-zero terms with second moments
% $(d-1)\|\tilde g\|^2$ by~\eqref{eq:encoding_variance} and
% $d\sigma_{\rm ch}^2$ by~\eqref{eq:fixed_norm}; the cross terms vanish
% because $w_s$ is independent of $v_s$ and has zero mean.
% (b) Coordinate $i$ has error $\frac1{k_i}\sum w$ with second moment
% $\sigma_{\rm ch}^2/k_i$ if $k_i\ge1$, and error $-\tilde g_i$, which
% occurs with probability $p_{\rm drop}^{N_{\rm F}}$, otherwise. Summing
% over $i$ gives~\eqref{eq:mse_elem}. For bundled packets, all coordinates
% share the same $k$, which does not change the expectation.
% (c) By (a) and the zero decoder, the mean-square error of GL is
% $\kappa_{\rm sc}\bigl[(d-1)\|\tilde g\|^2+d\sigma_{\rm ch}^2\bigr]
% +p_{\rm drop}^{dN_{\rm F}}\|\tilde g\|^2$; comparing
% with~\eqref{eq:mse_elem} gives~\eqref{eq:elem_crossover}.
% \end{proof}

The proof is given in the appendix. Proposition~\ref{prop:elementwise} is best read through the
\emph{per-coordinate signal-to-noise ratio}
$\mathrm{SNR}:=\|\tilde g\|/(\sqrt d\,\sigma_{\rm ch})$, the
root-mean-square magnitude of a coordinate relative to the channel
noise. Without dropout, \eqref{eq:elem_crossover} reads
$\mathrm{SNR}<1$: projections are more accurate than coordinates when
$\mathrm{SNR}<1$, and coordinates are more accurate when
$\mathrm{SNR}>1$. Dropout shifts this threshold, but it remains close
to one unless $p_{\rm drop}^{N_{\rm F}}$ is large. The proposition also
shows that the dropout protection in item~1) above is a property of
one-number packets rather than of the encoding: under CL, as under GL,
an agent's contribution is lost entirely only with
probability~$p_{\rm drop}^{dN_{\rm F}}$.

\paragraph{Oracle-level encoding}
Under OL encoding, the decoder~\eqref{eq:unified_decode} averages
$\widehat J^{(n)}_{t,s}v_{t,n,s}$ over the received packets. Since
reception is independent of the rollouts, with $k$ received packets this
is the one-point estimate~\eqref{eq:zo_estimate} built from $k$ rollouts
instead of $n_s$. Let $\nu_{\rm O}:=q_{\rm sc}n_s$,
$q^{\rm O}_{\rm act}:=1-p_{\rm drop}^{n_s}$,
$\theta_{\rm O}:=\sqrt{d\lambda_T/(Mq^{\rm O}_{\rm act}\nu_{\rm O})}$,
and let $\varepsilon_{\rm O}$ be an accuracy for
which~\eqref{eq:zo_event} holds when every local estimate uses at least
$\max\{1,\nu_{\rm O}/2\}$ rollouts.

\begin{cor}[Oracle-level encoding]\label{cor:ol}
Let the hypotheses of Theorem~\ref{thm:main} hold, except those on
$\theta$, with $\mathcal D$ uniform on the sphere of radius $\sqrt d$.
Then the OL update direction satisfies~\eqref{eq:rel_add} with
$\bar a^{\rm O}=0$ and
\begin{align}\label{eq:ab_ol}
  \bar b^{\rm O}=\varepsilon_{\rm O}
  &+4C\theta_{\rm O}\,\frac{\sqrt d}{\mathsf r}\,\sigma_{\rm ch}\nonumber\\
  &+4C\Bigl[\sqrt{\tfrac{(1-q^{\rm O}_{\rm act})\lambda_T}{M(q^{\rm O}_{\rm act})^2}}
   +\tfrac{\lambda_T}{Mq^{\rm O}_{\rm act}}\Bigr]\sqrt{b_{\rm het}},
\end{align}
with probability at least
$1-\delta-T(e^{-Mq^{\rm O}_{\rm act}/8}+Me^{-\nu_{\rm O}/8}+\delta_{\rm ZO})$,
and Theorem~\ref{thm:main} holds for OL encoding with $(\bar a,\bar b)$
replaced by $(0,\bar b^{\rm O})$. \boxend
\end{cor}
The proof is given in the appendix.
% \begin{proof}
% Let $\widehat g^{\,\circ}_{t,n}$ denote the decoded direction without
% channel noise. Then
% \begin{align*}
%   \widehat g_t-g_t
%   ={}&\tfrac{1}{|\mathcal A_t|}\textstyle\sum_{n\in\mathcal A_t}
%      \bigl(\widehat g^{\,\circ}_{t,n}-\nabla J^{(n)}(K_t)\bigr)\\
%   &+\tfrac{1}{|\mathcal A_t|}\textstyle\sum_{n\in\mathcal A_t}
%      \bigl(\nabla J^{(n)}(K_t)-g_t\bigr)+e^{\rm ch}_t .
% \end{align*}
% On $\mathcal E^{\rm rec}_t$, every active agent has at least
% $\max\{1,\nu_{\rm O}/2\}$ received rollouts, so the first term is at
% most $\varepsilon_{\rm O}$ on the corresponding oracle event. The second
% term is bounded as in Step~1 of the proof of
% Lemma~\ref{lem:hp_erec_resifed}, with $\Delta_{t,n}$ replaced by
% $\nabla J^{(n)}(K_t)-g_t$, whose norm is at most $2\sqrt{b_{\rm het}}$
% by~\eqref{eq:grad_het}. The third term is bounded as in Step~3, with
% the coefficients $c_{n,s}$ multiplied by $\sqrt d/\mathsf r$. The
% remaining steps follow the proof of Corollary~\ref{cor:uniform}.
% \end{proof}
\begin{rem}[Gradient-level vs.\ oracle-level encoding]\label{rem:gl_vs_ol}
By Corollary~\ref{cor:ol}, OL has no relative error, requires no
condition on $\theta$, and needs no local gradient computation; however,
its channel-noise term carries the factor $\sqrt d/\mathsf r$, and each
packet is the cost of one rollout, so a lost packet wastes a rollout
and repeating a cost attenuates channel noise but not the sampling
error $\varepsilon_{\rm O}$. GL decouples $N_{\rm sc}$ from $n_s$ and
does not amplify the noise, at the price of the encoding
error~\eqref{eq:encoding_variance}. OL is thus preferable over nearly
noiseless channels with inexpensive rollouts and GL otherwise;
\reza{in Fig.~\ref{fig:gl_ol}, at equal rollouts and equal packets, OL
loses stability at a noise level about $290$ times lower, against a
factor of $\sqrt d/(2\mathsf r)=260$ for the two-point oracle used
there.} Table~\ref{tab:gl_vs_ol} summarizes the three encodings.
\hfill$\square$
\end{rem}

\begin{table}[t]
\centering\scriptsize
\caption{Comparison of the scalar encodings of
Definition~\ref{def:gl_ol}.}
\label{tab:gl_vs_ol}
\setlength{\tabcolsep}{2pt}
\renewcommand{\arraystretch}{1.15}
\begin{tabular}{@{}p{0.23\columnwidth}p{0.25\columnwidth}p{0.26\columnwidth}p{0.22\columnwidth}@{}}
\toprule
 & GL & OL & CL\\
\midrule
Scalar sent & $v^\top\tilde g_{t,n}$ & rollout cost $\widehat J^{(n)}_{t,s}$ & $[\tilde g_{t,n}]_i$\\
Packets/round & $N_{\rm sc}$ (design choice) & $n_s$ (= rollouts) & $dN_{\rm F}$\\
Local computation & ZO estimate & none & ZO estimate\\
Encoding error & $(d{-}1)\|\tilde g_{t,n}\|^2/N^{\rm rec}_{t,n}$ & none & none\\
Channel-noise 2nd moment & $d\,\sigma_{\rm ch}^2/N^{\rm rec}_{t,n}$ & $d^2\sigma_{\rm ch}^2/(\rev{\mathsf r}^2N^{\rm rec}_{t,n})$ & $\approx d^2\sigma_{\rm ch}^2/N^{\rm rec}_{t,n}$\\
A lost packet removes & one projection & one rollout & one copy of one coordinate\\
Requirement on $\mathcal{D}$ & \eqref{eq:isotropy}--\eqref{eq:fixed_norm} & \rev{uniform on sphere of radius $\sqrt d$} & none\\
\bottomrule
\end{tabular}
\end{table}
\begin{figure}[t]
    \centering
    \includegraphics[width=0.7\linewidth]{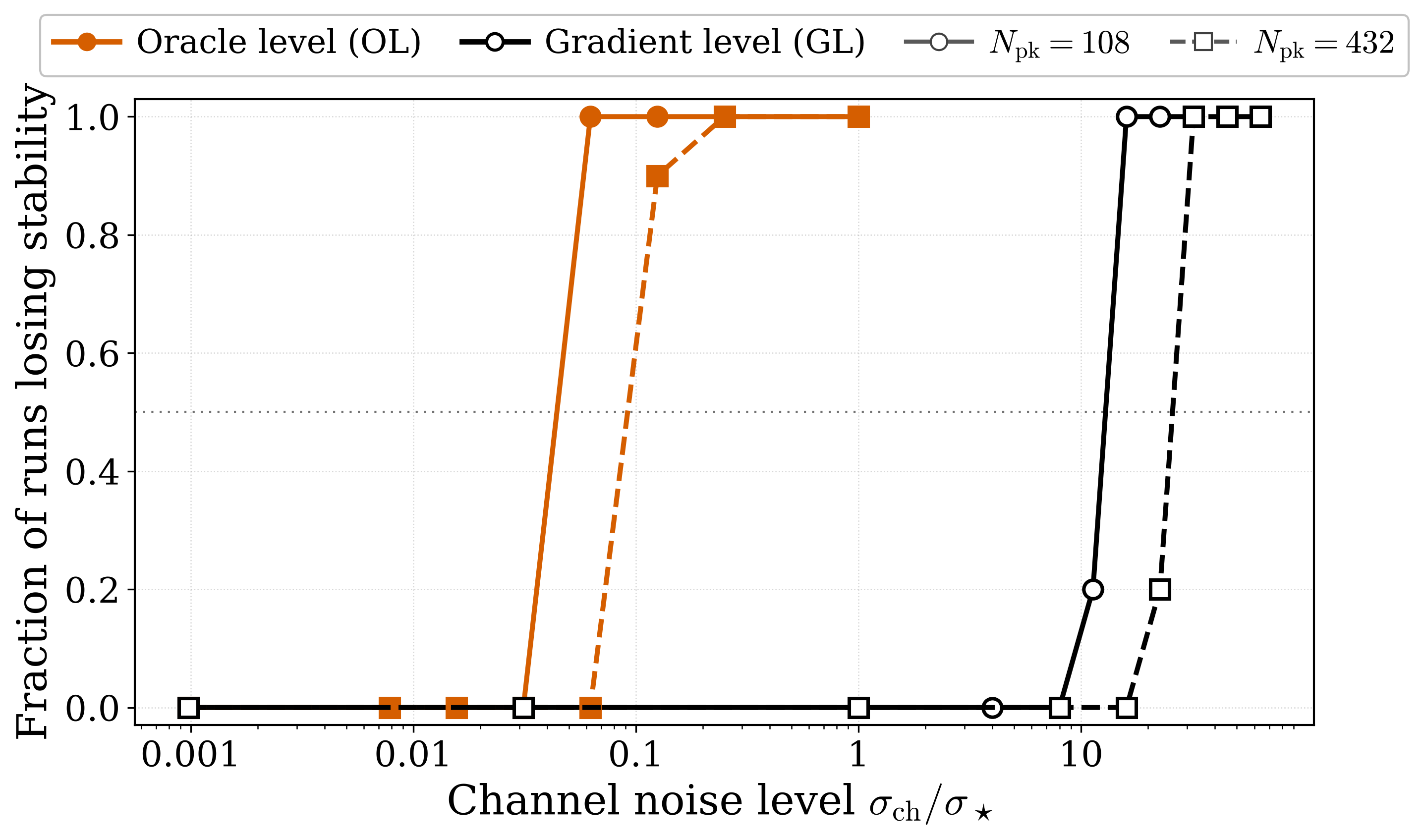}
    \caption{\small \reza{Fraction of realizations losing closed-loop
    stability versus channel noise for gradient-level (GL) and
    oracle-level (OL) encoding, at equal rollouts ($n_s=5$) and equal packets
   ($N_{\rm pk}\in\{108,432\}$ per agent and round), in the setting of
   Section~\ref{sec:numerics} ($d=108$,
   $\mathsf r=0.02$, $p_{\rm drop}=0.2$, $10$ realizations per noise level).}}
    \label{fig:gl_ol}
\end{figure}

\begin{rem}[Unbiasedness and the role of the noise model]\label{rem:power}
Any isotropic, fixed-norm encoding of Section~\ref{sec:random_proj},
including dense random projections and rescaled random coordinates,
yields a per-agent estimate $\widehat g_{t,n}$ that is unbiased for
every realized delivery pattern with at least one received packet, and
Theorem~\ref{thm:main} applies to all such encodings with the same
constants. The server can therefore average active agents directly,
without tracking which coordinates each agent delivered. Deterministic
coordinate transmission (CL) lacks this per-agent property: with the
zero-fill decoder of Proposition~\ref{prop:elementwise}(b), it is biased
under partial loss, and the bias is removed only by averaging each
coordinate over the agents that delivered it. The noise comparisons in~\eqref{eq:mse_scalar},
\eqref{eq:elem_crossover}, and item~2) above, and in
Section~\ref{sec:numerics}, are made at an equal number of packets under
the fixed noise floor of~\eqref{eq:channel}. In this setting a
projection carries mean-square signal $\|\tilde g_{t,n}\|^2$, whereas a
coordinate carries $\|\tilde g_{t,n}\|^2/d$ on average, so the noise
advantage of projections reflects a $d$-fold larger transmit energy
rather than a coding gain: at equal average transmit energy, CL scaled
by $\sqrt d$ attains the same noise term without projection error.
Comparisons under amplitude- or power-constrained uplinks are left for
future work. \hfill$\square$
\end{rem}
\end{revenv}

\paragraph{Answers to the research questions}
\emph{RQ1.} Scalar communication with redundancy preserves stabilizing
learning with high probability within the certified resilience region,
for isotropic encodings by Theorem~\ref{thm:main} and for rollout costs
by Corollary~\ref{cor:ol}. \emph{RQ2.} The required redundancy is given
by~\eqref{eq:nsc_rule}: it grows linearly with the policy dimension and,
when noise dominates, with the noise variance, and decreases linearly
with the fleet size. \emph{RQ3.} The analysis reveals that no encoding
dominates. The protection against dropout stems from packet granularity
and is common to all three. Among the gradient-based encodings,
projections are more accurate than coordinates only when the
per-coordinate signal-to-noise ratio is below approximately one under a
fixed noise floor, an effect of transmit energy rather than of coding
(Proposition~\ref{prop:elementwise}, Remark~\ref{rem:power}). Rollout
costs are preferable only over nearly noiseless channels with
inexpensive rollouts (Remark~\ref{rem:gl_vs_ol}).

\section{Numerical Results}\label{sec:numerics}

\begin{figure}[t]
    \centering
    \begin{subfigure}{0.32\linewidth}
        \centering
        \includegraphics[width=\linewidth]{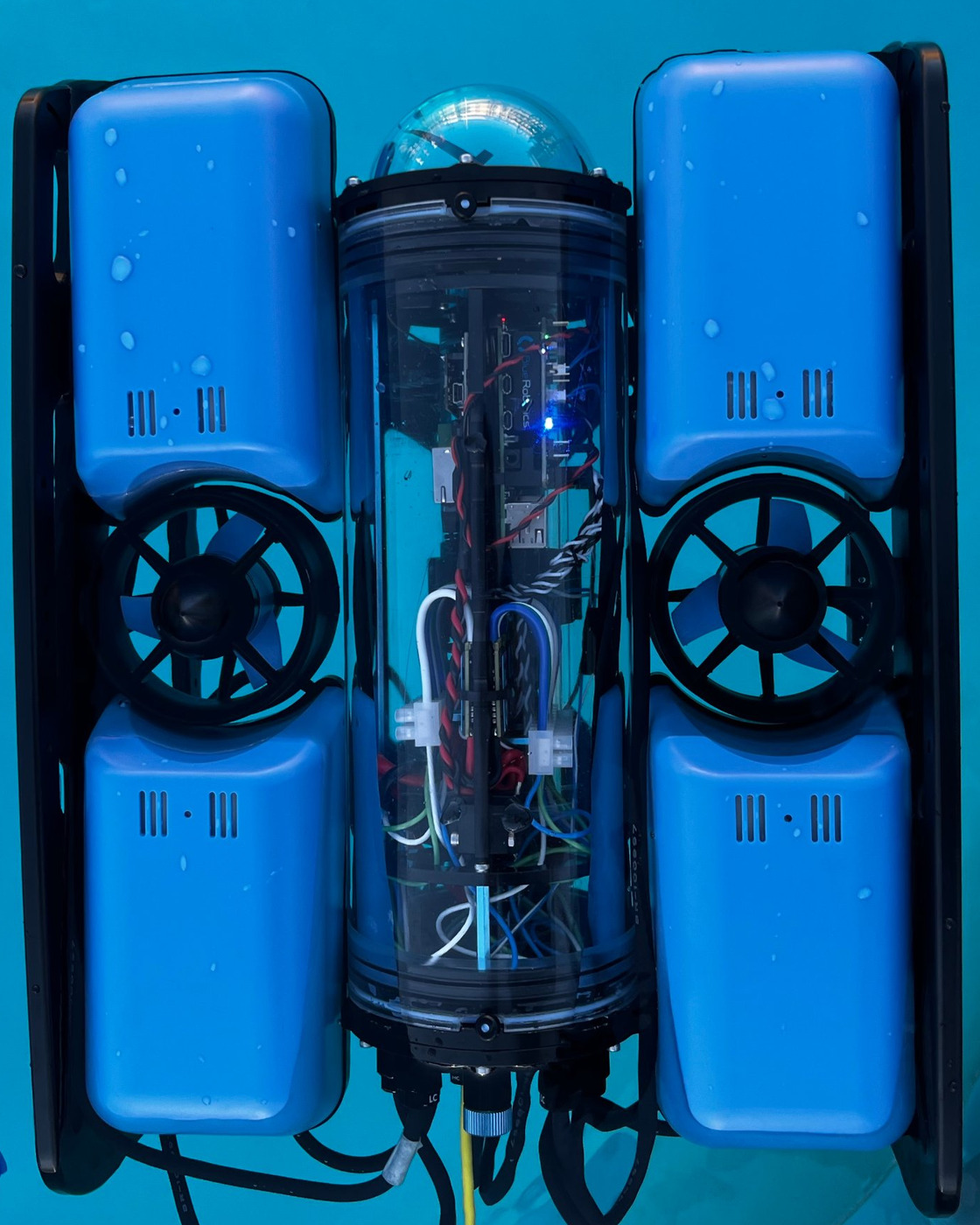}
        \caption{\reza{Vehicle}}
    \end{subfigure}
    \hfill
    \begin{subfigure}{0.64\linewidth}
        \centering
        \includegraphics[width=\linewidth]{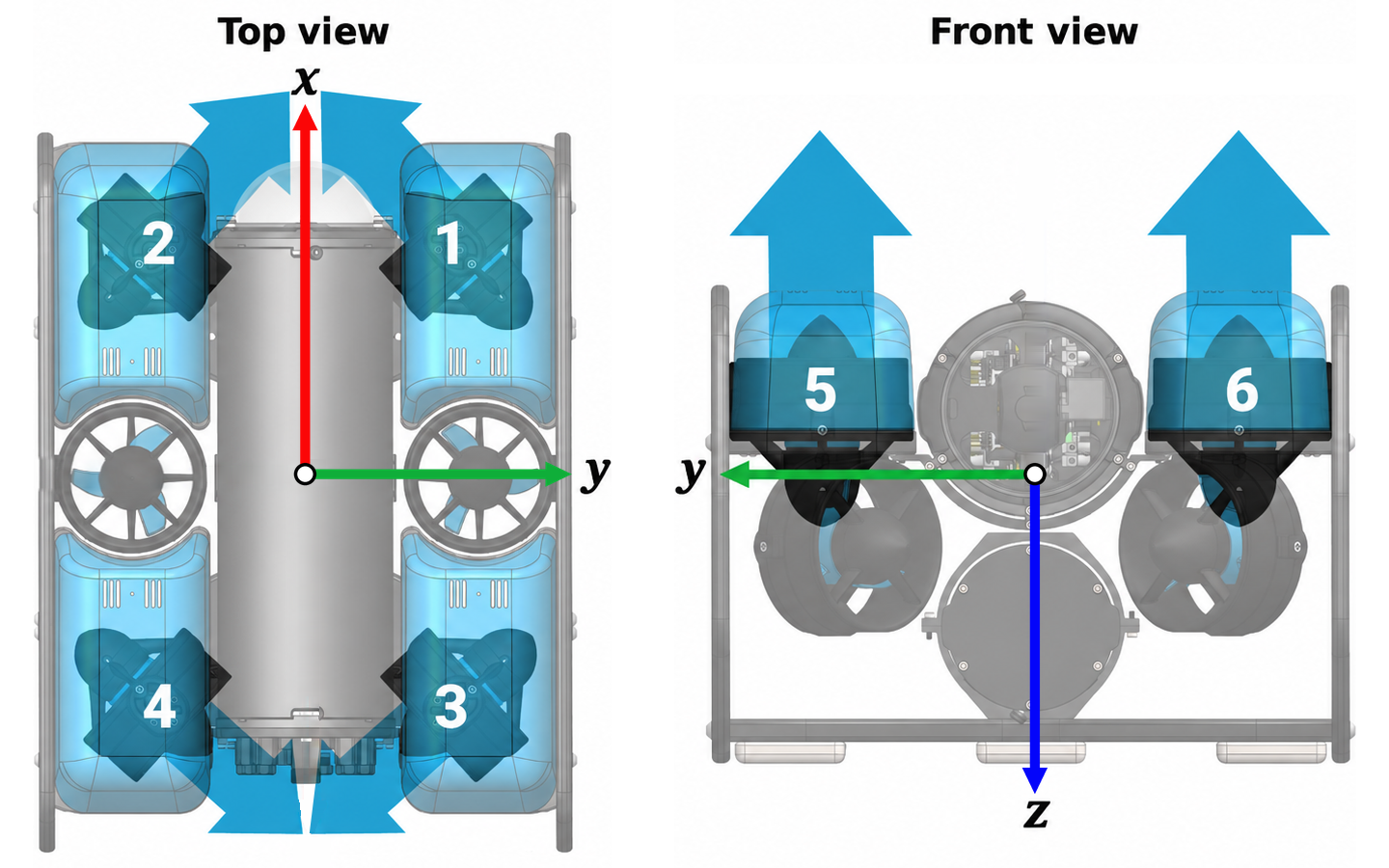}
        \caption{\reza{Thruster layout and body frame}}
    \end{subfigure}
    \caption{\small \reza{BlueROV2 in its standard configuration. (a) Top
    view of the vehicle. (b) Thrusters 1--4 are horizontal and vectored at
    $45^\circ$ (surge, sway, yaw), and thrusters 5 and 6 are vertical
    (heave, roll); the wide arrows show the thrust directions, and the body
    axes $x$, $y$, and $z$ point forward, to starboard, and down. The Heavy
    configuration used in the simulations has the same horizontal thrusters
    and four vertical thrusters, which also actuate pitch.}}
    \label{fig:rov}
\end{figure}

\reza{This section examines the three research questions numerically on a
fleet of BlueROV2 underwater vehicles (Fig.~\ref{fig:rov}) performing
model-free station-keeping control over a noisy and lossy uplink.} Unless stated
otherwise, \textsc{ResiFedLQR} is run with GL encoding and Rademacher
directions and is compared with bundled full-gradient \textsc{FedLQR} on
the same fleet and over the same channel, so that differences arise
solely from the communication mechanism.
Section~\ref{sec:numerical_stability_region} reports the empirical
resilience region (RQ1), Section~\ref{sec:numerical_noise_convergence}
the effect of redundancy (RQ2), and Section~\ref{sec:numerical_mse} the
comparison of encodings (RQ3). Section~\ref{sec:numerical_large_fleet}
repeats the comparison on \reza{a larger fleet}, and
Section~\ref{sec:numerical_het_dim} tests the dependence on
heterogeneity and policy dimension.

\subsection{\reza{BlueROV2 Fleet and Experimental Setup}}\label{sec:numerics_setup}

\emph{Vehicle model.} We consider the BlueROV2 in its Heavy
configuration, whose eight thrusters actuate all six degrees of
freedom, with the hydrodynamic coefficients identified
in~\cite{wu_bluerov2_2018}. \reza{Fig.~\ref{fig:rov} shows the vehicle in
its standard configuration, which has two vertical thrusters; the Heavy
configuration has four, mounted outboard, and thereby also actuates
pitch.} Following~\cite{fossen_handbook_2021}, let
$\mathsf q\in\mathbb{R}^6$ collect the position and Euler angles in the
North--East--Down frame and $\mathsf v\in\mathbb{R}^6$ the linear and
angular velocities in the body frame \reza{of Fig.~\ref{fig:rov}(b)} (the
usual symbols $\eta$ and $\nu$
are already used here for the stepsize and in~\eqref{eq:nu_theta}), and
let $\mathsf f\in\mathbb{R}^6$ be the wrench delivered by the thrusters.
Linearizing the equations of motion about hover, with the constant net
buoyancy compensated by feedforward,~gives
\begin{equation}\label{eq:rov_linear}
\begin{aligned}
  \dot{\mathsf q}&=\mathsf v,\qquad
  T_{\rm th}\dot{\mathsf f}=u-\mathsf f,\\
  \mathsf M\dot{\mathsf v}&=-\mathsf D\mathsf v-\mathsf G\mathsf q
  +\mathsf S\mathsf f,
\end{aligned}
\end{equation}
where $\mathsf M$ is the sum of the rigid-body and added-mass inertia
matrices, $\mathsf D$ is the linear damping matrix, $\mathsf G$ is the
linearized restoring matrix, and $u\in\mathbb{R}^6$ is the commanded
body wrench. The delivered wrench follows the command with the thruster
time constant $T_{\rm th}$, and both are normalized by the single-axis
wrench limits
$\mathsf S=\operatorname{diag}(113.7,113.7,160.8,35.1,19.3,30.4)$ (in N
and N\,m). The nominal parameters are the mass $11.5$~kg, weight
$112.8$~N, buoyancy $114.8$~N, moments of inertia $0.16$~kg\,m$^2$ about
each axis, center of gravity $z_g=0.02$~m below the center of buoyancy,
added masses $(5.5,12.7,14.57,0.12,0.12,0.12)$, linear damping
coefficients $(4.03,6.22,5.18,0.07,0.07,0.07)$, and $T_{\rm th}=0.1$~s.
The offset $z_g$ couples surge with pitch and sway with roll through
$\mathsf M$ and $\mathsf G$; heave and yaw are decoupled.
Discretizing~\eqref{eq:rov_linear} with a zero-order hold at the control
rate of $10$~Hz yields~\eqref{eq:dynamics} with state
$x=[\mathsf q^\top,\mathsf v^\top,\mathsf f^\top]^\top$, $n_x=18$,
$n_u=6$, and policy dimension $d=108$. The delivered wrench is part of
the state because the thruster time constant equals the control
period.

\reza{\emph{Fleet.} We consider $M=10$ vehicles. The first is the nominal
vehicle; for the others, the mass, the three moments of inertia, the
offset $z_g$, the six added masses, the six linear damping coefficients, and the thruster time constant are each scaled by an independent factor
drawn uniformly from $[1-\epsilon,1+\epsilon]$ with $\epsilon=0.2$, which
represents differences in
payload and buoyancy foam, drawn once and kept fixed across all
realizations and algorithms. Assumption~\ref{ass:similarity} holds with
$\epsilon_1\le1.99$ and $\epsilon_2\le0.78$, i.e., about $25\%$ and
$17\%$ of the norms of the nominal matrices. Every vehicle has four
open-loop eigenvalues at one, associated with position and yaw, and is
therefore not Schur stable.}

\reza{\emph{Cost and initial gain.} The cost matrices are
$Q=\operatorname{diag}(I_3,I_3,0.1I_6,0.01I_6)$, which weights position,
attitude, velocities, and delivered wrench, and $R=I_6$ on the
normalized commands. The initial state is drawn as
$x_0\sim\mathcal N(0,\Sigma_0)$ with standard deviations of $0.5$~m in
position, $0.1$~rad in attitude, $0.1$ in the velocities, and $0.05$ in
the delivered wrench. Both algorithms are initialized from the common
gain $K_0$ given by the LQR design for the nominal vehicle with input
weight $100R$, a cautious design that stabilizes every vehicle: the
closed-loop spectral radius is $0.950$ for the nominal vehicle and at
most $0.955$ over the fleet.}

\reza{\emph{Learning.} Unless otherwise stated, the horizon is $T=1000$
rounds and the stepsize is $\eta=2\times10^{-4}$. Each local ZO estimate
uses the two-point estimator with $n_s=5$ perturbations: for each
perturbation $U_{t,n,i}$, the agent runs two rollouts of length
$\tau=128$ ($12.8$~s), under $K_t+U_{t,n,i}$ and $K_t-U_{t,n,i}$ from the
same initial state, with costs $\widehat J^{(n),+}_{t,i}$ and
$\widehat J^{(n),-}_{t,i}$, and forms
$\tilde g_{t,n}=\frac{d}{2n_s\mathsf r^2}\sum_{i=1}^{n_s}
(\widehat J^{(n),+}_{t,i}-\widehat J^{(n),-}_{t,i})U_{t,n,i}$,
with smoothing radius $\mathsf r=0.02$. The one-point
estimator~\eqref{eq:zo_estimate} is too noisy at this dimension: over an
ideal channel it loses stability within a few rounds for
$\eta\ge2\times10^{-5}$ and makes no progress for
$\eta=2\times10^{-6}$. The rollouts are generated with the linear
models; as a check, a gain learned by \textsc{ResiFedLQR} was applied to
the nonlinear six-degree-of-freedom model
of~\cite{fossen_handbook_2021}, where it lowers the cost of a $15$~s
response from $13.4$ under $K_0$ to $6.8$, against $6.4$ for the gain
that minimizes $J_{\rm avg}$.}

\reza{\emph{Uplink.}} The uplink follows Assumption~\ref{assump:noise}:
each packet is
dropped independently with probability $p_{\rm drop}$, and every
received scalar or gradient entry is corrupted by independent zero-mean
Gaussian noise with standard deviation $\sigma_{\rm ch}$. \reza{Noise
levels are reported in units of
$\sigma_\star:=(\mathbb E\|\tilde g_{0,n}\|^2/d)^{1/2}=47.6$, the
root-mean-square coordinate of a local estimate at $K_0$: a reported
value $\sigma_{\rm ch}=1$ means that the noise added to each received
real number in~\eqref{eq:channel} has standard deviation
$\sigma_\star$.} For the paired comparisons we use
\reza{\[
(N_{\rm F},N_{\rm sc})\in\{(1,108),(2,216),(4,432)\},
\]}
which, since \reza{$d=108$}, correspond to equal uplink budgets
($N_{\rm sc}=dN_{\rm F}$, Remark~\ref{rem:budget}). When traffic is
reported in bits, each real number is represented with $32$ bits.
\reza{The per-coordinate signal-to-noise ratio at $K_0$ is
$\mathrm{SNR}=1$, $0.5$, $0.33$, and $0.25$ for
$\sigma_{\rm ch}=1,2,3,4$, and it decreases during learning, since the
local estimates shrink as the gain improves (their root-mean-square
coordinate falls from $47.6$ at $K_0$ to $0.16$ at the optimum). The
closed-loop experiments therefore operate at a per-coordinate
signal-to-noise ratio of at most one,} which by
Proposition~\ref{prop:elementwise} is the regime that favors projections
over coordinates. For these configurations, $\theta$
in~\eqref{eq:nu_theta} ranges from \reza{$0.72$ to $1.43$ at
$p_{\rm drop}=0.2$ and $\theta_{\rm F}/\theta\ge10.4$}; the condition
$\theta<1/(4C)$ of Theorem~\ref{thm:main} is thus not met, and these
experiments test the dependence predicted by the analysis rather than
its constants\reza{. Meeting this condition at $d=108$ requires more
than $64\,d\,\lambda_T\approx1.1\times10^{5}$ received scalars per
round, which is beyond the fleets simulated here.}
\subsection{Stabilizing Learning and the Empirical Resilience Region}
\label{sec:numerical_stability_region}

We first ask whether scalar communication with redundancy preserves
stabilizing learning (RQ1), and over which range of channel conditions
and stepsizes. We vary the stepsize $\eta$ and the dropout probability
$p_{\rm drop}$ at the equal budget
\reza{$(N_{\rm F},N_{\rm sc})=(2,216)$, over a grid of seven
logarithmically spaced stepsizes
$\eta\in[5\times10^{-5},8\times10^{-4}]$ and five equally spaced dropout
probabilities $p_{\rm drop}\in[0.1,0.8]$}. For each pair
$(\eta,p_{\rm drop})$ and noise level $\sigma_{\rm ch}$, we estimate
from \reza{$8$} realizations the probability $\widehat p_{\rm fail}$
that the learning trajectory loses closed-loop stability, i.e., that
$\max_{n}\rho(A^{(n)}-B^{(n)}K_t)\ge1$ for some $t\le T$.
\reza{Realization $i$ at a grid point uses the same random seed at every
noise level, so that the noise levels are compared on common
realizations. With $8$ realizations, $\widehat p_{\rm fail}=0$ certifies
$p_{\rm fail}\lesssim0.31$ at $95\%$ confidence.} For a method
$\mathcal M$, the empirical resilience region is
\begin{equation}\label{eq:empirical_stability_region}
\mathcal R_{\mathcal M}(\sigma_{\rm ch})
:=\bigl\{(\eta,p_{\rm drop}):
\widehat p^{\,\mathcal M}_{\rm fail}(\eta,p_{\rm drop},\sigma_{\rm ch})=0\bigr\},
\end{equation}
the empirical counterpart of the certified resilience region of
Section~\ref{sec:descent}. To remove sampling artifacts, a grid
point is counted in $\mathcal R_{\mathcal M}$ only if no failure is
observed at it or at any grid point with smaller stepsize and dropout
probability.

\begin{figure*}[t]
    \centering
    \includegraphics[width=\textwidth]{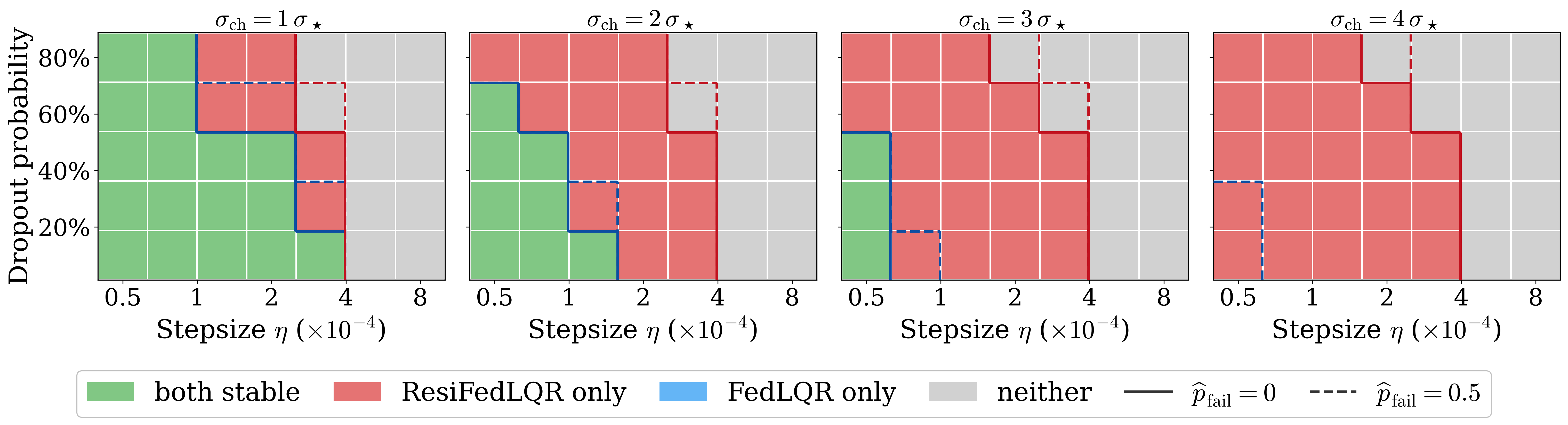}
    \caption{\small Empirical resilience
    regions~\eqref{eq:empirical_stability_region} over stepsize and
    dropout probability for $\sigma_{\rm ch}\in\{1,2,3,4\}$
    at \reza{$(N_{\rm F},N_{\rm sc})=(2,216)$}.}
    \label{fig:stability_regions_resifed}
\end{figure*}

\begin{figure*}[t]
    \centering
    \includegraphics[width=\textwidth]{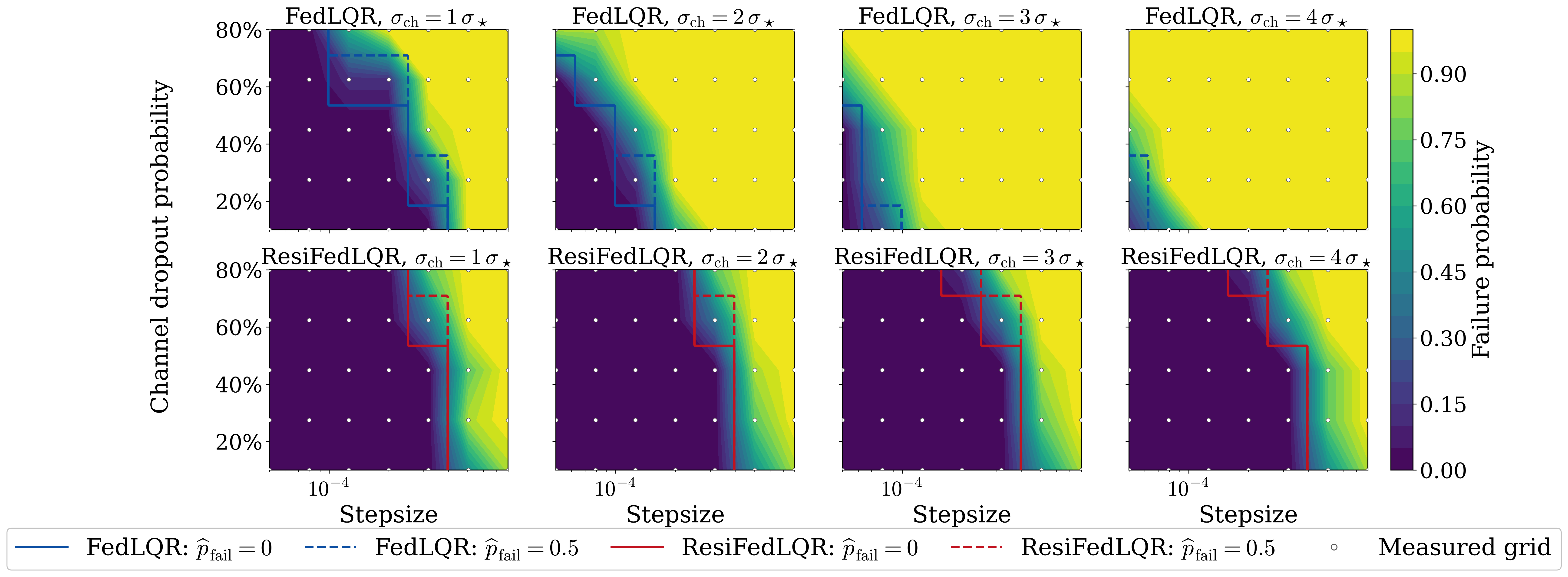}
    \caption{\small Empirical failure probability $\widehat p_{\rm fail}$
    over stepsize and dropout probability. Top: \textsc{FedLQR}; bottom:
    \textsc{ResiFedLQR}. Darker colors indicate smaller
    $\widehat p_{\rm fail}$; solid and dashed curves mark
    $\widehat p_{\rm fail}=0$ and $0.5$.}
    \label{fig:failure_heatmaps_resifed}
\end{figure*}

Fig.~\ref{fig:stability_regions_resifed} compares the regions of the
two algorithms. Green marks operating points at which both methods are
stable, red the additional region stabilized only by
\textsc{ResiFedLQR}, and gray points at which neither is; solid
boundaries correspond to $\widehat p_{\rm fail}=0$ and dashed boundaries
to $\widehat p_{\rm fail}=0.5$. No operating point is stabilized
only by \textsc{FedLQR}. Table~\ref{tab:stability_summary} gives the
coverage of the grid: the region of \textsc{FedLQR} shrinks from
\reza{$49\%$ at $\sigma_{\rm ch}=1$ to $23\%$ at $\sigma_{\rm ch}=2$ and
$9\%$ at $\sigma_{\rm ch}=3$, and is empty at $\sigma_{\rm ch}=4$,
whereas that of \textsc{ResiFedLQR} stays between $66\%$ and $63\%$.}
Fig.~\ref{fig:failure_heatmaps_resifed} reports the
full failure-probability landscape: as $\sigma_{\rm ch}$ increases, the
low-failure region of \textsc{FedLQR} contracts rapidly, whereas
\reza{that of \textsc{ResiFedLQR} is essentially unchanged. The reason
is that these noise levels are far below what \textsc{ResiFedLQR}
tolerates at this budget (about $\sigma_{\rm ch}=18$,
Section~\ref{sec:numerical_mse}): its failures occur at the largest
stepsizes and are caused by the stepsize, which at
$\eta=8\times10^{-4}$ makes both methods fail everywhere. Its region
does shrink at larger noise, to $57\%$ at $\sigma_{\rm ch}=8$ and $31\%$
at $\sigma_{\rm ch}=16$ (five realizations per point).}

\begin{table}[t]
\centering
\caption{Fraction of the \reza{$7\times5$} $(\eta,p_{\rm drop})$ grid
in the empirical resilience
region~\eqref{eq:empirical_stability_region}.}
\label{tab:stability_summary}
\begin{tabular}{lcccc}
\hline
$\sigma_{\rm ch}$ & 1 & 2 & 3 & 4\\
\hline
\textsc{FedLQR}     & \reza{49\%} & \reza{23\%} & \reza{9\%}  & \reza{0\%}\\
\textsc{ResiFedLQR} & \reza{66\%} & \reza{66\%} & \reza{63\%} & \reza{63\%}\\
\hline
\end{tabular}
\end{table}

These observations answer RQ1 affirmatively in the tested regime and
agree with the structure of Theorem~\ref{thm:main}: the
conditions~\eqref{eq:certif} and~\eqref{eq:stepsize} tighten as
$\sigma_{\rm ch}$ and $p_{\rm drop}$ grow, so that, for a fixed channel,
a larger resilience region corresponds to a larger admissible stepsize
and, for a fixed stepsize, to a larger tolerable dropout probability.
As shown in Section~\ref{sec:comparison}, two mechanisms contribute
to the enlargement relative to \textsc{FedLQR}: the protection against
losing an agent's contribution, which is shared by any scheme with
one-real-number packets, and the attenuation of channel noise, which
relies on the noise model~\eqref{eq:channel} and on the low
signal-to-noise ratio of these experiments.
Section~\ref{sec:numerical_mse} separates the two and shows that the
second dominates at moderate dropout.

\subsection{Effect of Redundancy}
\label{sec:numerical_noise_convergence}

\begin{figure*}[t]
    \centering
    \begin{subfigure}{0.24\textwidth}
        \centering
        \includegraphics[width=\linewidth]{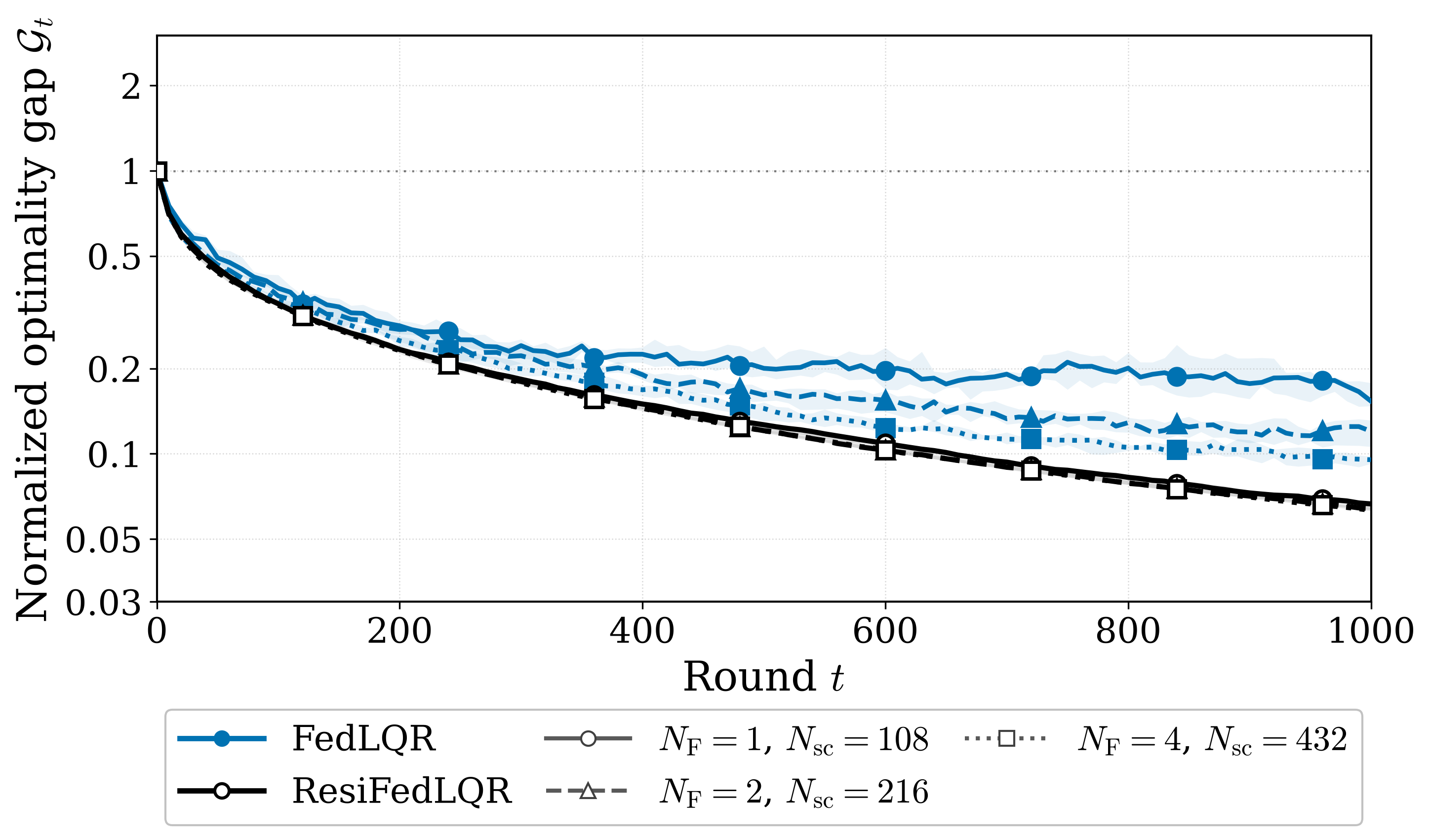}
        \caption{$\sigma_{\rm ch}=1$}
        \label{fig:gap_noise_1}
    \end{subfigure}
    \hfill
    \begin{subfigure}{0.24\textwidth}
        \centering
        \includegraphics[width=\linewidth]{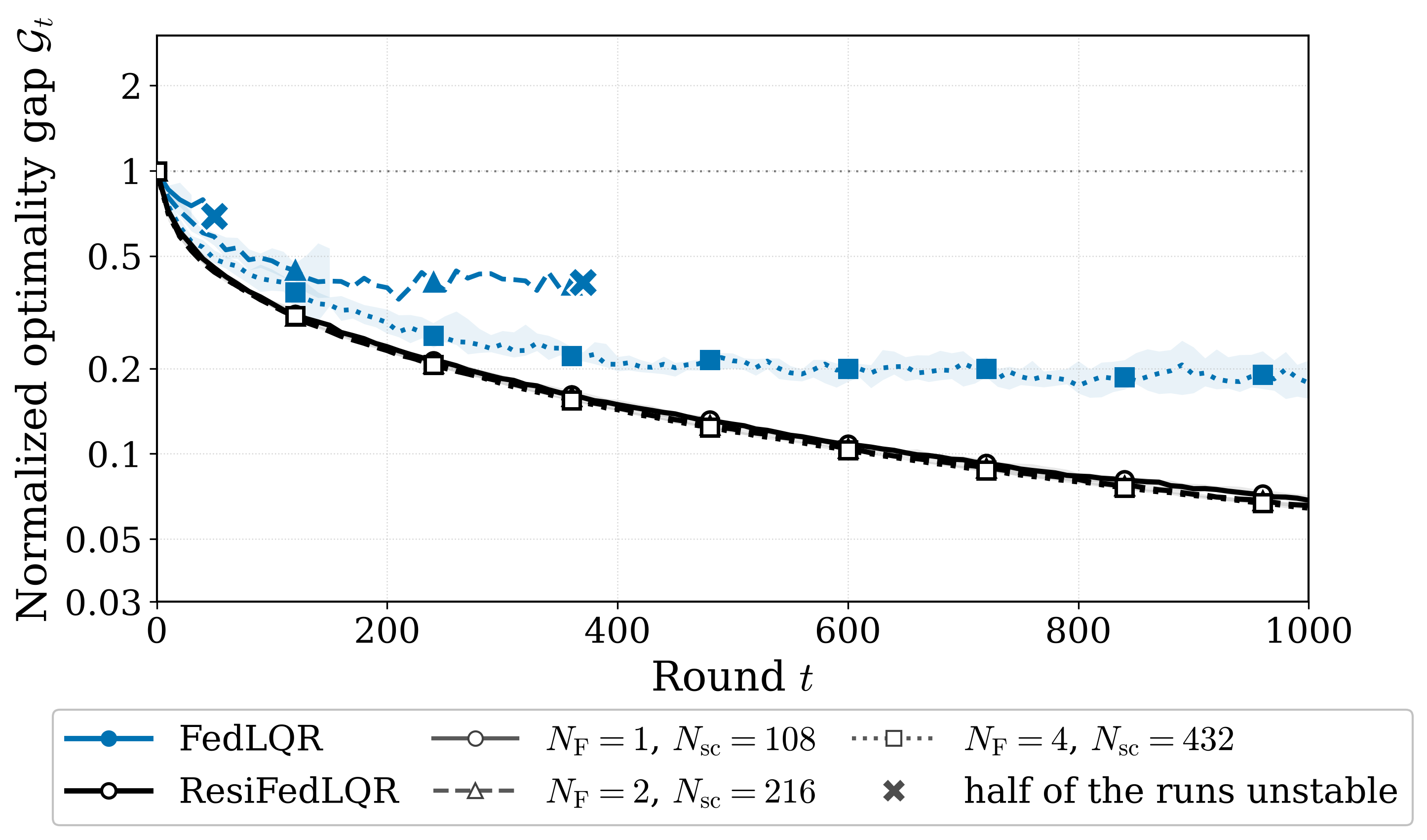}
        \caption{$\sigma_{\rm ch}=2$}
        \label{fig:gap_noise_2}
    \end{subfigure}
    \hfill
    \begin{subfigure}{0.24\textwidth}
        \centering
        \includegraphics[width=\linewidth]{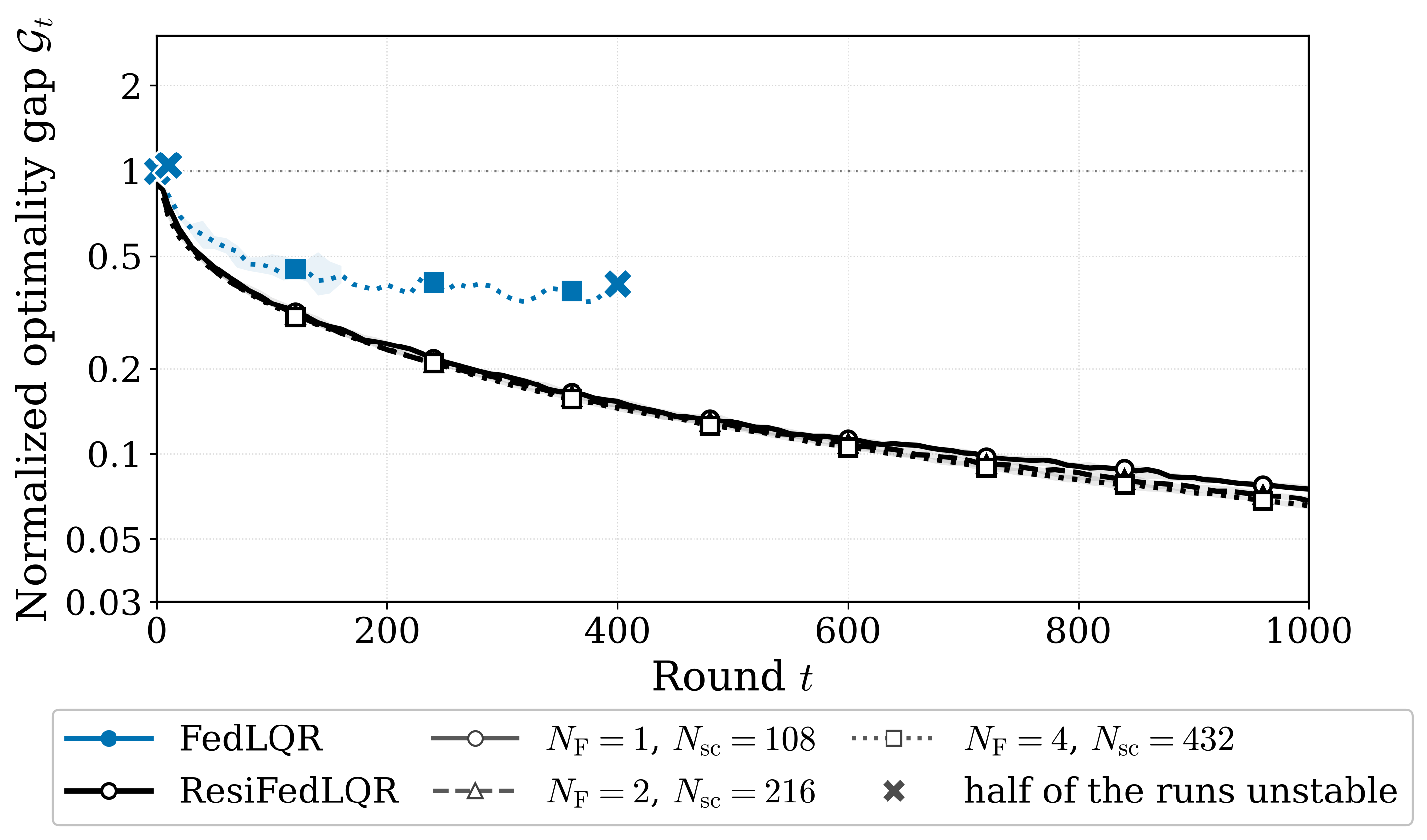}
        \caption{$\sigma_{\rm ch}=3$}
        \label{fig:gap_noise_3}
    \end{subfigure}
    \hfill
    \begin{subfigure}{0.24\textwidth}
        \centering
        \includegraphics[width=\linewidth]{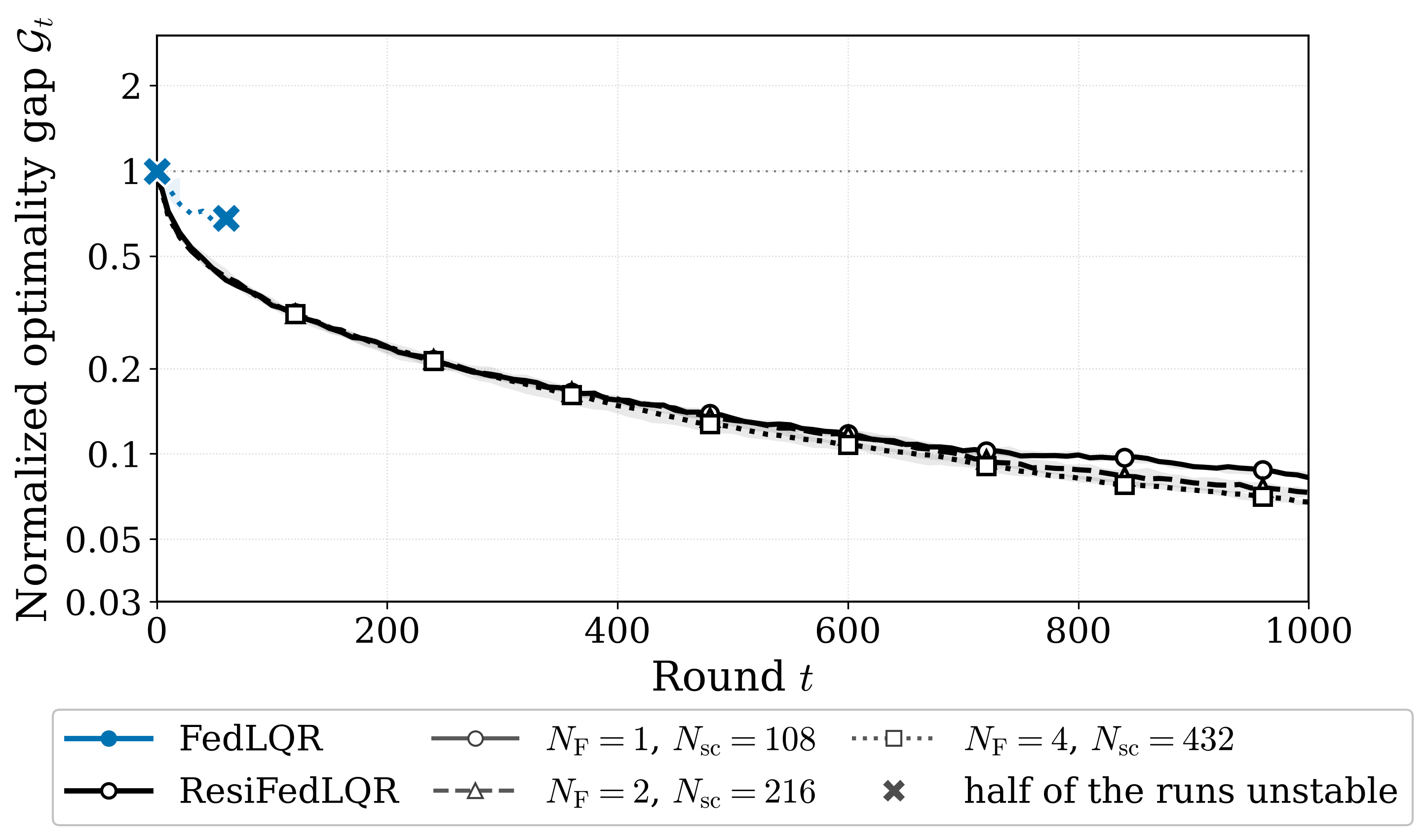}
        \caption{$\sigma_{\rm ch}=4$}
        \label{fig:gap_noise_4}
    \end{subfigure}
    \caption{\small Normalized optimality gap~\eqref{eq:numerical_normalized_gap}
    versus round under increasing uplink channel noise, at equal uplink
    budget $N_{\rm sc}=dN_{\rm F}$ and $p_{\rm drop}=0.2$. \reza{Lines
    are medians over $12$ realizations and shaded bands are
    interquartile ranges. A line ends with a cross at the last round at
    which fewer than half of the realizations have lost closed-loop
    stability;} the dotted line marks $\mathcal G_0$.}
    \label{fig:normalized_gap_noise_resifed}
\end{figure*}

We next examine how much redundancy a given channel requires (RQ2). We
measure performance by the normalized optimality~gap
\begin{equation}\label{eq:numerical_normalized_gap}
    \mathcal G_t:=\frac{J_{\rm avg}(K_t)-J_{\rm avg}(K^\star)}{J_{\rm avg}(K^\star)} ,
\end{equation}
\reza{where $J_{\rm avg}$ is the fleet-average cost~\eqref{eq:avg_cost},
evaluated exactly from the models of the ten vehicles, and $K^\star$ is
its minimizer; thus $\mathcal G_0=1.00$.}
Fig.~\ref{fig:normalized_gap_noise_resifed} compares \textsc{FedLQR}
and \textsc{ResiFedLQR} for $\sigma_{\rm ch}\in\{1,2,3,4\}$ at
$p_{\rm drop}=0.2$ \reza{for $N_{\rm F}\in\{1,2,4\}$
($N_{\rm sc}\in\{108,216,432\}$), and Table~\ref{tab:final_gap}
summarizes the three budgets. With $\eta=2\times10^{-4}$, the horizon
$T=1000$ covers the transient phase, in which the gap decreases from
$\mathcal G_0=1.00$ to about $0.07$;} the residual gap is examined over a
longer horizon in Section~\ref{sec:numerical_het_dim}.

\reza{\begin{table}[t]
\centering\scriptsize
\caption{Median of $\mathcal G_T$ over $12$ realizations at
$p_{\rm drop}=0.2$ ($\mathcal G_0=1.00$); in parentheses, the number of
realizations that lose closed-loop stability; $\infty$: more than half
of the realizations lose stability. F: \textsc{FedLQR};
R: \textsc{ResiFedLQR}.}
\label{tab:final_gap}
\setlength{\tabcolsep}{3pt}
\begin{tabular}{@{}c cc cc@{}}
\toprule
 & \multicolumn{2}{c}{$\sigma_{\rm ch}=1$} & \multicolumn{2}{c}{$\sigma_{\rm ch}=2$}\\
$(N_{\rm F},N_{\rm sc})$ & F & R & F & R\\
\midrule
$(1,108)$ & 0.15 (1) & 0.07 (0) & $\infty$ (12) & 0.07 (0)\\
$(2,216)$ & 0.12 (0) & 0.06 (0) & $\infty$ (11) & 0.07 (0)\\
$(4,432)$ & 0.10 (0) & 0.06 (0) & 0.18 (2) & 0.06 (0)\\
\midrule
 & \multicolumn{2}{c}{$\sigma_{\rm ch}=3$} & \multicolumn{2}{c}{$\sigma_{\rm ch}=4$}\\
$(N_{\rm F},N_{\rm sc})$ & F & R & F & R\\
\midrule
$(1,108)$ & $\infty$ (12) & 0.08 (0) & $\infty$ (12) & 0.08 (0)\\
$(2,216)$ & $\infty$ (12) & 0.07 (0) & $\infty$ (12) & 0.07 (0)\\
$(4,432)$ & $\infty$ (10) & 0.07 (0) & $\infty$ (12) & 0.07 (0)\\
\bottomrule
\end{tabular}
\end{table}}

\reza{The redundancy that \textsc{FedLQR} needs to keep learning stable
grows with the channel noise. At $\sigma_{\rm ch}=1$, one of its $12$
realizations loses stability with a single upload and none with
$N_{\rm F}\ge2$; at $\sigma_{\rm ch}=2$, it needs $N_{\rm F}=4$ to keep
its median gap finite; and for $\sigma_{\rm ch}\ge3$, at least $10$ of
its $12$ realizations lose stability at every budget. For
\textsc{ResiFedLQR}, no realization loses stability at any budget or
noise level, and its median gap ends between $0.06$ and $0.08$. Where
both methods descend, the median gap of \textsc{FedLQR} is $1.5$ to
$2.8$ times larger and its interquartile range $2.9$ to $23$ times
wider.} This monotone dependence of
the required redundancy on the noise level is the qualitative behavior
predicted by~\eqref{eq:nsc_rule}; since the constants of that rule are
not met at $M=10$, the experiment does not test its scaling
quantitatively.

\begin{figure}[t]
    \centering
    \includegraphics[width=\linewidth]{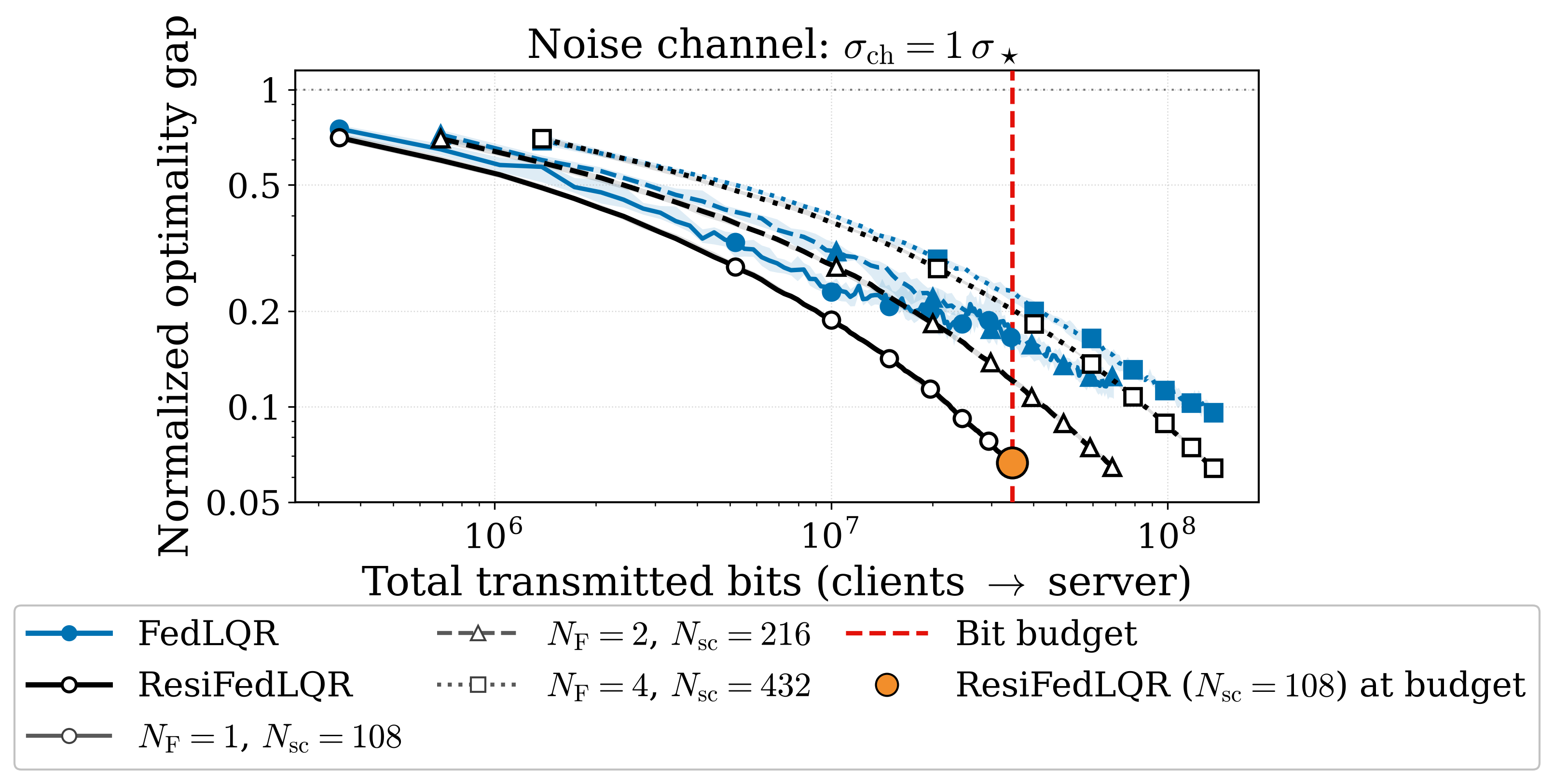}
        \caption{\small Normalized optimality
    gap~\eqref{eq:numerical_normalized_gap} versus cumulative uplink
    bits for $\sigma_{\rm ch}=1$ and $p_{\rm drop}=0.2$. Lines are
    medians over \reza{$12$} realizations \reza{and shaded bands are
    interquartile ranges}; the vertical line marks the
    budget of \reza{$3.46\times10^{7}$} bits and the dotted line marks
    $\mathcal G_0$.}
    \label{fig:communication_efficiency_resifed}
\end{figure}

Redundancy has a cost in traffic, so
Fig.~\ref{fig:communication_efficiency_resifed} plots
$\mathcal G_t$ against cumulative uplink bits for
$\sigma_{\rm ch}=1$ and $p_{\rm drop}=0.2$. The vertical line
marks a budget of \reza{$3.46\times10^{7}$ bits, the total uplink
traffic of the pair $(1,108)$ over the $T$ rounds. At this budget,
\textsc{ResiFedLQR} with $N_{\rm sc}=108$ reaches a median gap of
$0.066$, the lowest among all configurations of either method, against
$0.153$ for \textsc{FedLQR} with $N_{\rm F}=1$; configurations with more
transmissions per round have completed fewer rounds and remain between
$0.12$ and $0.23$.} Hence, over a mild channel, the smallest redundancy
is the most bit-efficient choice, and additional transmissions become
necessary only as the channel deteriorates.

\subsection{Comparison of Encodings}
\label{sec:numerical_mse}

\begin{figure}[t]
    \centering
    \includegraphics[width=0.8\linewidth]{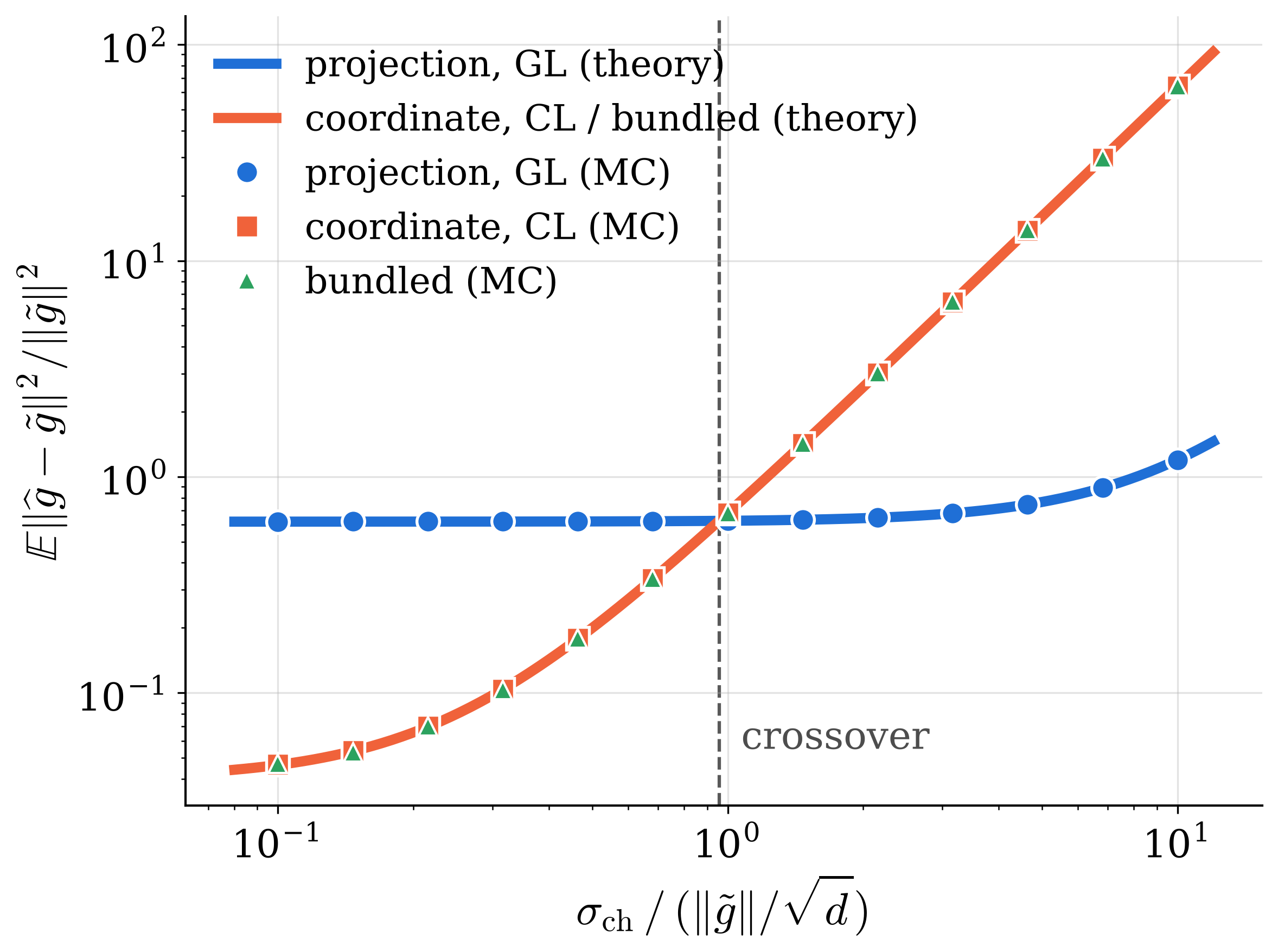}
    \caption{\small Normalized mean-square reconstruction error of one
    transmission round versus normalized channel noise, for
    \reza{$d=108$},
    $N_{\rm F}=2$, and $p_{\rm drop}=0.2$ at equal budget. Lines:
    \eqref{eq:mse_scalar} and~\eqref{eq:mse_elem}; markers: Monte Carlo
    estimates from $4\times10^{4}$ realizations. The dashed line marks
    the crossover.}
    \label{fig:mse_check}
\end{figure}

We finally turn to the choice of the scalar message (RQ3). To isolate
the encoding from the learning dynamics, we first verify
Proposition~\ref{prop:elementwise} for a single transmission round. For
a fixed vector $\tilde g\in\mathbb{R}^d$ with \reza{$d=108$},
$N_{\rm F}=2$, and
$p_{\rm drop}=0.2$, we estimate the mean-square reconstruction error of
GL encoding (\reza{$N_{\rm sc}=dN_{\rm F}=216$} Rademacher projections),
CL encoding, and bundled full-gradient transmission from $4\times10^{4}$
independent channel realizations per noise level; all three decoders
return zero when no packet is received. Fig.~\ref{fig:mse_check} shows
the error, normalized by $\|\tilde g\|^2$, against the channel noise
normalized by the root-mean-square coordinate magnitude
$\|\tilde g\|/\sqrt d$, i.e., against the inverse per-coordinate
signal-to-noise ratio. The empirical errors agree
with~\eqref{eq:mse_scalar} and~\eqref{eq:mse_elem}, evaluated with the
exact binomial expectations, to within \reza{$2\%$}, and CL and bundled
transmission coincide, as stated in Proposition~\ref{prop:elementwise}(b).
The curves cross at $\sigma_{\rm ch}=0.96\,\|\tilde g\|/\sqrt d$, the
value given by~\eqref{eq:elem_crossover}. Below the crossover, the
projection error dominates and GL saturates at a normalized error of
\reza{$0.62$}, against $0.05$ for CL; above it, GL is more accurate by a
factor that approaches $d$, reaching \reza{$53$} at
$\sigma_{\rm ch}=10\,\|\tilde g\|/\sqrt d$.

\begin{figure}[t]
    \centering
    \begin{subfigure}{\linewidth}
        \centering
        \includegraphics[width=0.7\linewidth]{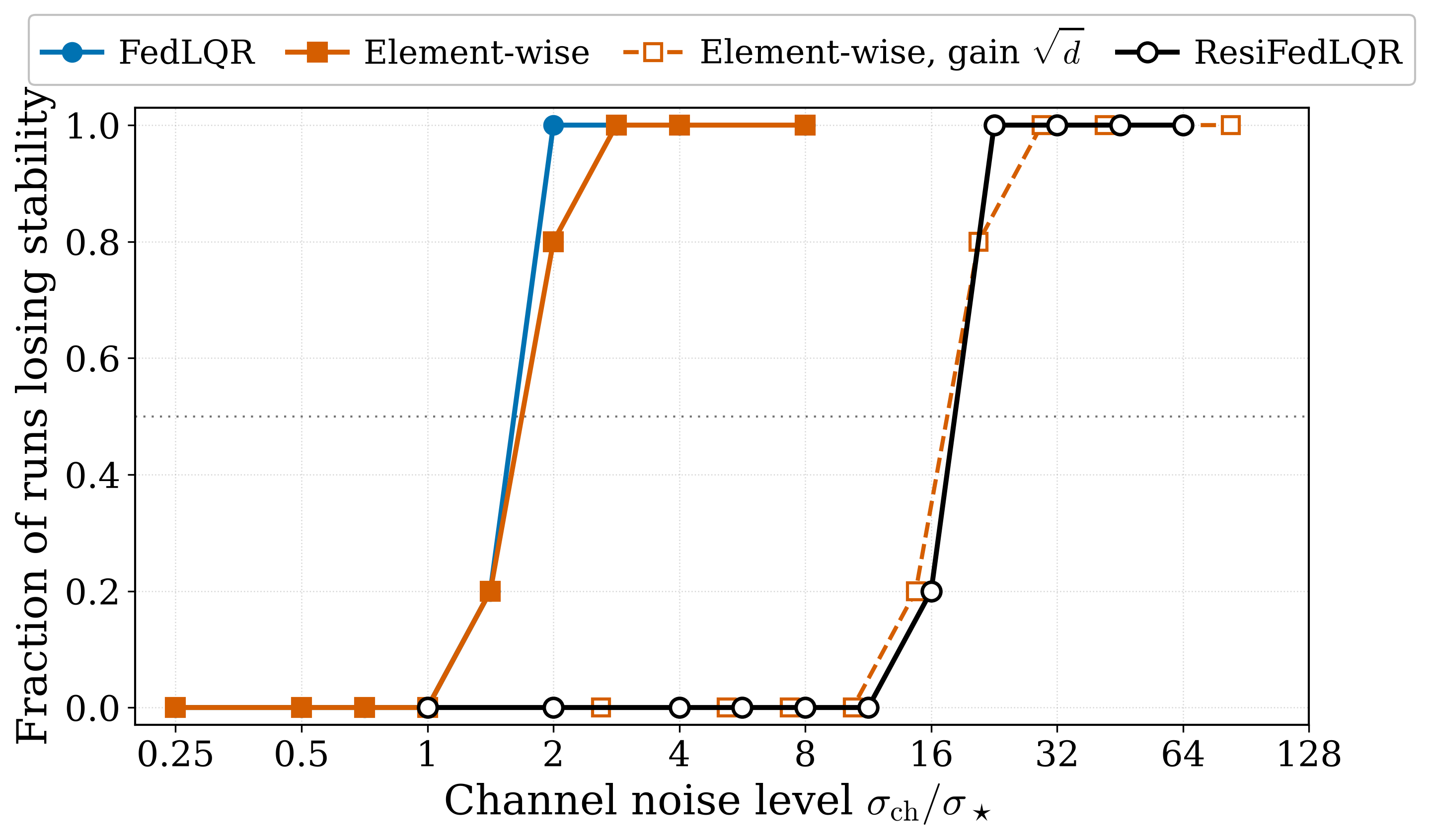}
        \caption{$M=10$}
    \end{subfigure}
    \par\vspace{1mm}
    \begin{subfigure}{\linewidth}
        \centering
        \includegraphics[width=0.7\linewidth]{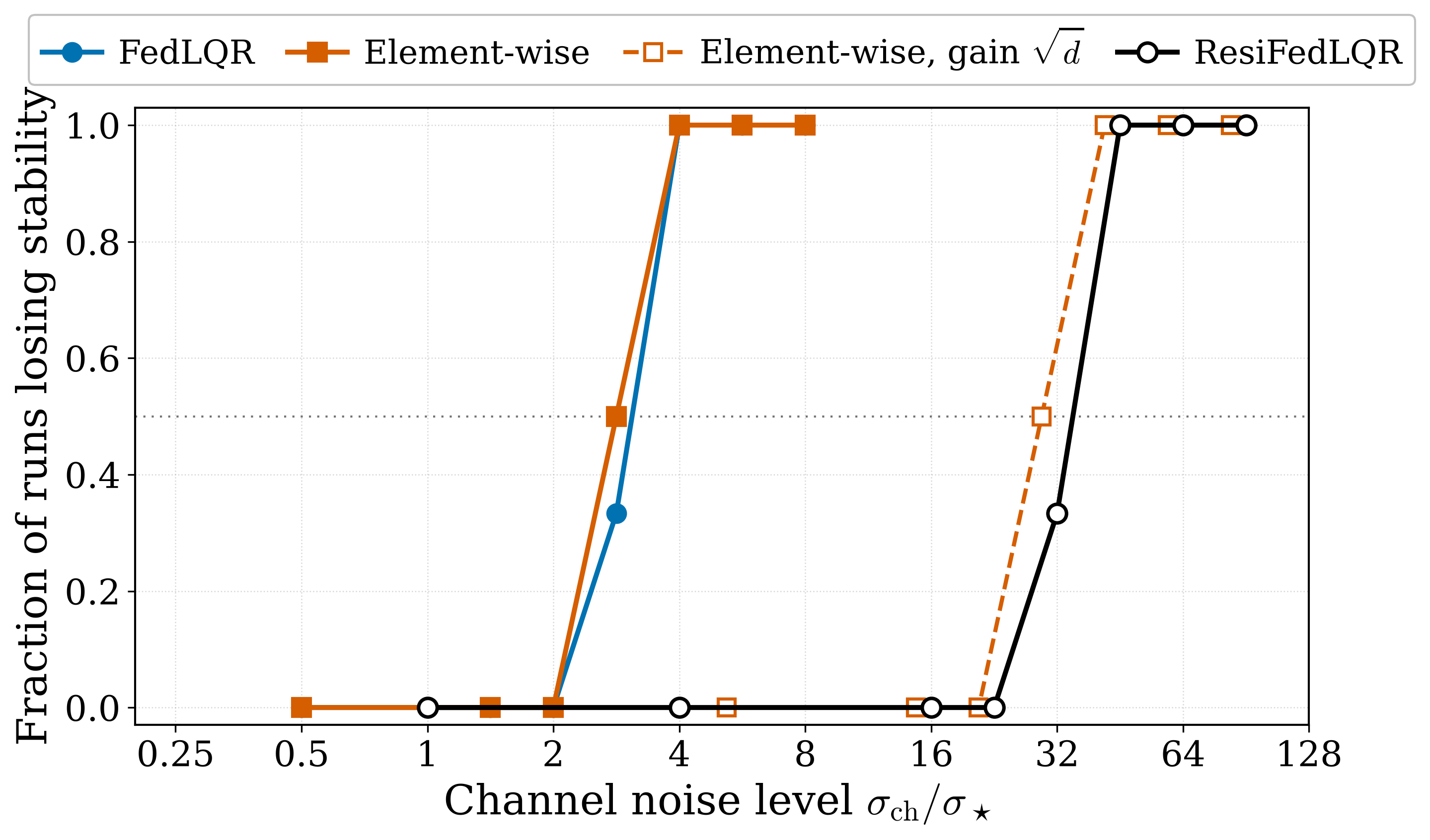}
        \caption{\reza{$M=40$}}
    \end{subfigure}
    \caption{\small Fraction of realizations losing closed-loop
    stability versus channel noise for bundled full-gradient
    transmission (\textsc{FedLQR}), CL encoding (element-wise) without
    and with a transmit gain of $\sqrt d$, and GL encoding
    (\textsc{ResiFedLQR}), at \reza{$(N_{\rm F},N_{\rm sc})=(2,216)$,
    $p_{\rm drop}=0.2$, $\eta=2\times10^{-4}$, and $T=1000$, with $10$
    (a) and $6$ (b) realizations per noise level.} (a) Fleet of
    Section~\ref{sec:numerics_setup} ($M=10$). (b) The same fleet
    extended to \reza{$M=40$} agents.}
    \label{fig:encoding_failure}
\end{figure}

In closed loop, Fig.~\ref{fig:encoding_failure}(a) compares four
schemes on the fleet of Section~\ref{sec:numerics_setup} at
\reza{$(N_{\rm F},N_{\rm sc})=(2,216)$} and $p_{\rm drop}=0.2$: bundled
full-gradient transmission, CL encoding with each coordinate averaged
over the agents that delivered it, CL encoding with a transmit gain of
$\sqrt d$, which has the same average transmit energy as GL, and GL
encoding. With \reza{$10$} realizations per noise level, half of the
realizations lose stability at \reza{$\sigma_{\rm ch}=1.61$ for bundled
transmission, $1.68$ for CL, $17.5$ for CL with gain $\sqrt d$, and
$18.2$ for GL.} Three observations follow. First, CL and bundled
transmission fail at the same noise level: at this dropout probability,
failures are driven by channel noise, and one-number packets alone do
not help. Second, GL tolerates about \reza{eleven} times more noise than
either, in agreement with the factor \reza{$\sqrt d=10.4$} of
Section~\ref{sec:comparison}.
Third, CL with gain $\sqrt d$ matches GL to within the sampling error.
At the low signal-to-noise ratio of these experiments, the advantage of
projections over coordinates is therefore an effect of transmit energy,
as stated in Remark~\ref{rem:power}, and it is not specific to
projections.

Finally, Fig.~\ref{fig:gl_ol} compares GL and OL encoding at equal
rollouts and equal packets \reza{($N_{\rm pk}\in\{108,432\}$)}: the two
perform alike over a nearly
noiseless channel, and OL loses stability at a noise level about
\reza{$290$ times lower ($\sigma_{\rm ch}=0.044$ and $0.092$ against
$12.9$ and $25.8$), in line with the factor
$\sqrt d/(2\mathsf r)=260$ by which the OL decoder amplifies the channel
noise for the two-point oracle (Corollary~\ref{cor:ol}; the gain is
halved because an OL packet carries the difference of two costs). OL
needs only the $n_s=5$ packets, whereas GL needs a number of projections
comparable to $d$: with $5$ packets per agent, GL loses stability in all
ten realizations even at negligible noise.} These results support the
answer to RQ3 given in
Section~\ref{sec:comparison}: no encoding dominates, the noise tolerance
of projections is matched by coordinates at equal transmit energy, and
rollout costs are preferable only over nearly noiseless channels.

\subsection{\reza{A Larger Fleet}}
\label{sec:numerical_large_fleet}

The fleet of Section~\ref{sec:numerics_setup} is too small for the
conditions of Theorem~\ref{thm:main}\reza{, and, as noted there, a fleet
that meets them at $d=108$ is beyond what we simulate. We therefore test
the predicted dependence on the fleet size, by repeating the comparison
with $M=40$ agents, generated as in Section~\ref{sec:numerics_setup},
at the same budget $(N_{\rm F},N_{\rm sc})=(2,216)$, for which
$\theta=0.51$ at $p_{\rm drop}=0.2$.}

Fig.~\ref{fig:encoding_failure}(b) shows the fraction of realizations
that lose stability. Half of the realizations fail at
\reza{$\sigma_{\rm ch}=3.08$ for bundled transmission, $2.83$ for CL,
$29.4$ for CL with gain $\sqrt d$, and $34.9$ for GL.} The ordering and
the conclusions of Section~\ref{sec:numerical_mse} are unchanged: GL
tolerates between \reza{$11$ and $12$} times more noise than bundled and
CL transmission, and CL at equal transmit energy matches it \reza{to
within the sampling error. Relative to $M=10$, the tolerated noise is
$1.9$ times larger for bundled transmission and GL and $1.7$ times
larger for CL, close to the factor $\sqrt 4=2$ predicted by
$\theta\propto1/\sqrt M$.}
Table~\ref{tab:large_fleet_region} reports the number of failing
realizations over a grid of stepsizes and dropout probabilities at
\reza{$\sigma_{\rm ch}=3$. GL encoding shows no failure at any of the
$16$ grid points, against $7$ for bundled transmission; with only $3$
realizations per point, however, no observed failure certifies no more
than $p_{\rm fail}\lesssim0.63$ at $95\%$ confidence, so this table is
indicative.} The enlargement of the empirical resilience region
observed in Section~\ref{sec:numerical_stability_region} therefore
persists \reza{for the larger fleet.}

\reza{\begin{table}[t]
\centering\scriptsize
\caption{Number of realizations, out of $3$, losing closed-loop
stability for $M=40$, $(N_{\rm F},N_{\rm sc})=(2,216)$, and
$\sigma_{\rm ch}=3$. F: bundled full-gradient transmission; G: GL
encoding.}
\label{tab:large_fleet_region}
\setlength{\tabcolsep}{3pt}
\begin{tabular}{@{}c cc cc cc cc@{}}
\toprule
 & \multicolumn{2}{c}{$p_{\rm drop}=0.1$} & \multicolumn{2}{c}{$0.3$} & \multicolumn{2}{c}{$0.5$} & \multicolumn{2}{c}{$0.7$}\\
$\eta$ & F & G & F & G & F & G & F & G\\
\midrule
$5\times10^{-5}$ & 0 & 0 & 0 & 0 & 0 & 0 & 0 & 0\\
$1\times10^{-4}$ & 0 & 0 & 0 & 0 & 0 & 0 & 3 & 0\\
$2\times10^{-4}$ & 2 & 0 & 1 & 0 & 3 & 0 & 3 & 0\\
$4\times10^{-4}$ & 3 & 0 & 3 & 0 & 3 & 0 & 3 & 0\\
\bottomrule
\end{tabular}
\end{table}}

\subsection{Effect of Heterogeneity and Policy Dimension}
\label{sec:numerical_het_dim}

We close with two experiments that test the predicted dependence
of the guarantees on the heterogeneity level and on the policy
dimension, with \reza{$8$ and $10$ realizations per configuration,
respectively,} and $p_{\rm drop}=0.2$.

\begin{figure}[t]
    \centering
    \begin{subfigure}{\linewidth}
        \centering
        \includegraphics[width=0.7\linewidth]{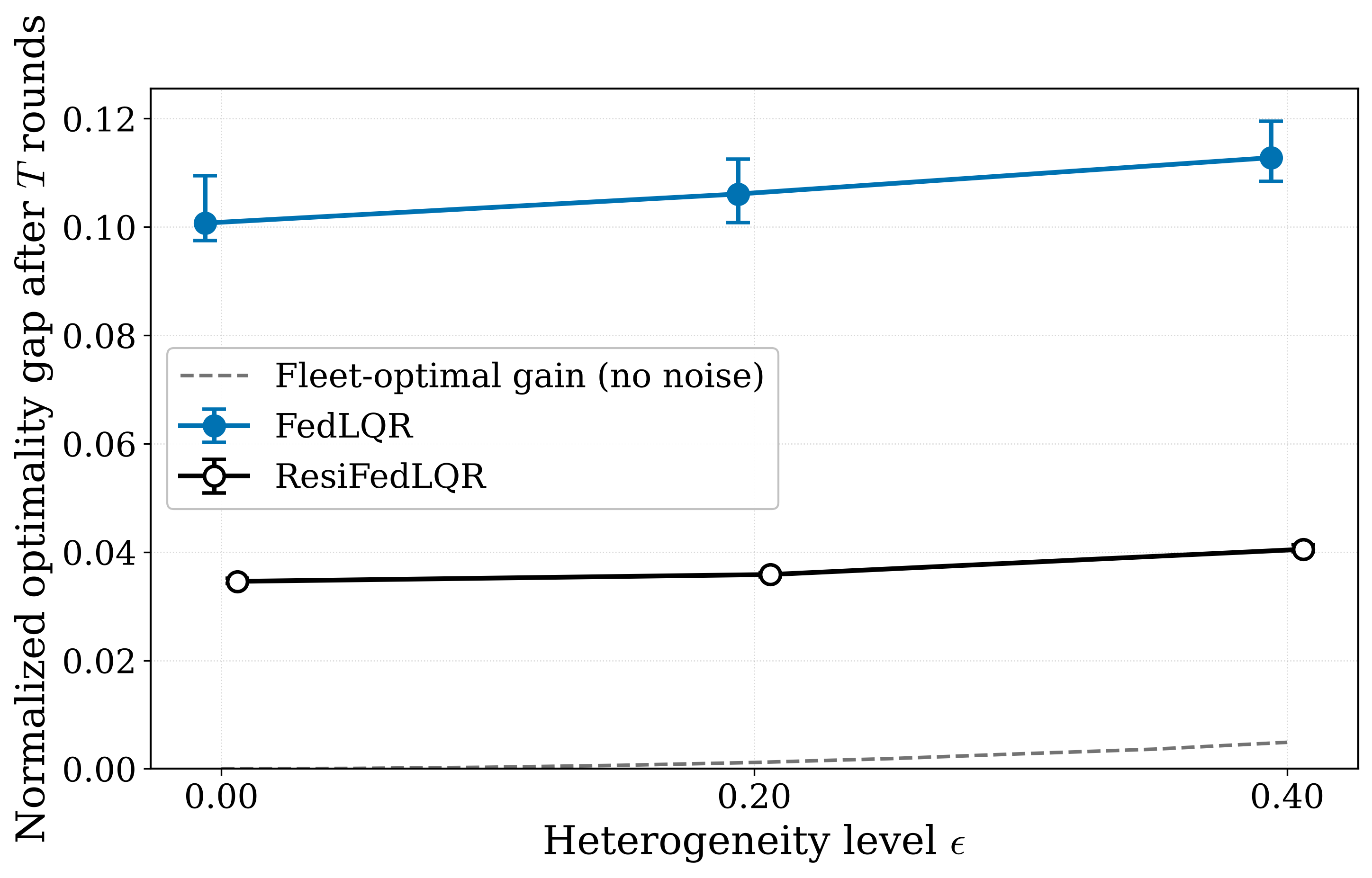}
        \caption{Heterogeneity}
    \end{subfigure}
    \par\vspace{1mm}
    \begin{subfigure}{\linewidth}
        \centering
        \includegraphics[width=0.7\linewidth]{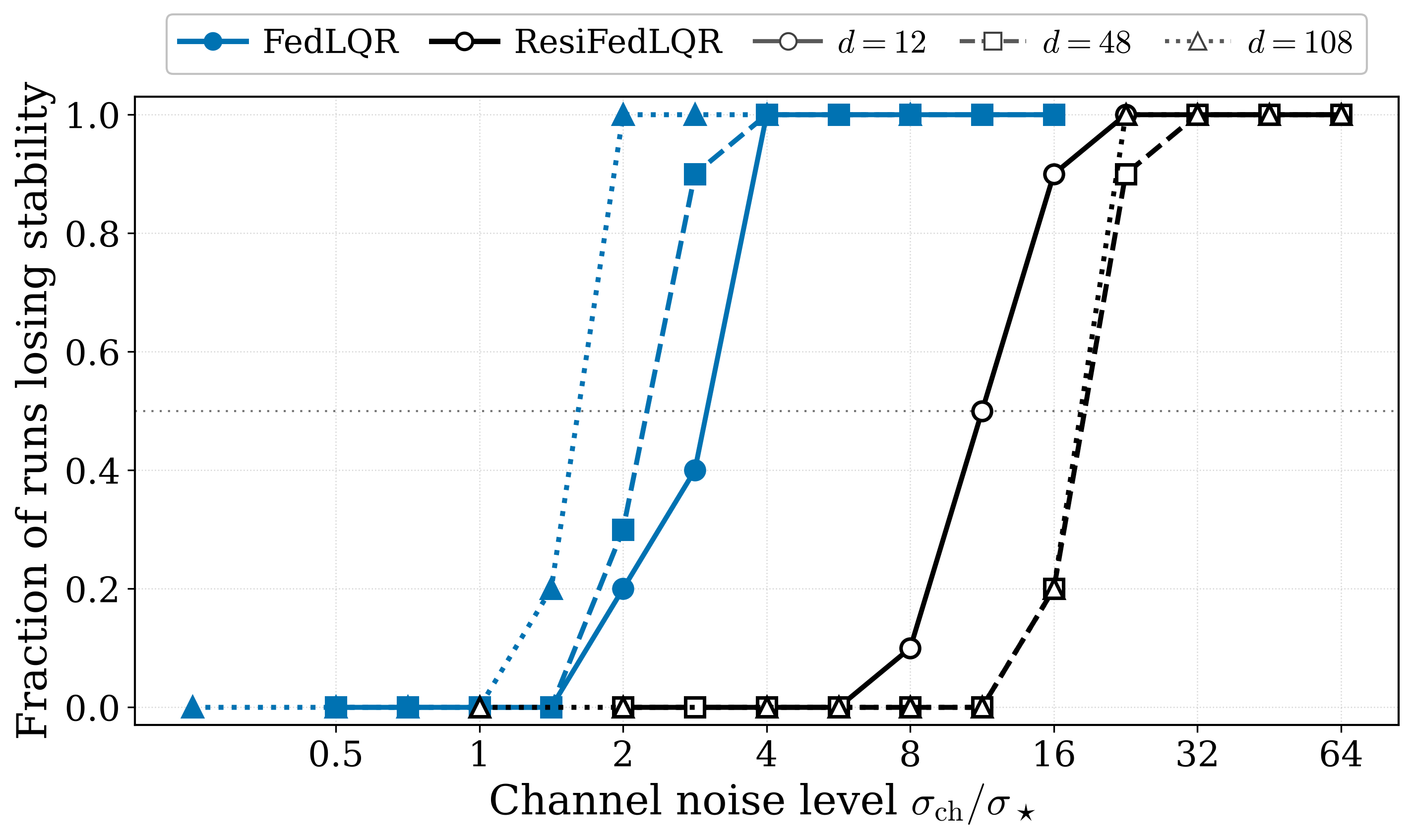}
        \caption{Policy dimension}
    \end{subfigure}
    \caption{\small (a) Gap after \reza{$T=2000$} rounds versus
    heterogeneity level \reza{$\epsilon$}: median and interquartile
    range over \reza{$8$} realizations; dashed: gain minimizing
    $J_{\rm avg}$.
    (b) Fraction of realizations losing stability versus channel noise
    for \reza{$d\in\{12,48,108\}$} at equal budget.}
    \label{fig:het_dim}
\end{figure}

\reza{\emph{Heterogeneity.} We vary the parameter spread
$\epsilon\in\{0,0.2,0.4\}$, keeping the same underlying random draws, at
$\sigma_{\rm ch}=1$ and $(N_{\rm F},N_{\rm sc})=(2,216)$, and run
$T=2000$ rounds with $\eta=2\times10^{-4}$. Fig.~\ref{fig:het_dim}(a)
shows the gap averaged over the last $200$ rounds, measured here
relative to the average of the agents' individually optimal costs, so
that the gain that minimizes $J_{\rm avg}$ has a gap of its own. The
dashed curve is that gap ($0$, $0.001$, and $0.005$): it is the price of
a common policy, which no communication scheme removes. It is small for
this vehicle: at $\epsilon=0.2$, the common gain keeps every agent
within $0.3\%$ of its own optimal cost. The residual of
\textsc{ResiFedLQR} rises with it ($0.035$, $0.036$, and $0.041$), as
predicted by the heterogeneity term of $\bar F$ in~\eqref{eq:certif};
the remaining offset, present even for identical agents, is due to
sampling and channel noise. The residual of \textsc{FedLQR} is about
three times larger ($0.101$, $0.106$, and $0.113$), and its
interquartile range is about ten times wider. No realization lost
stability.}

\emph{Policy dimension.} By Section~\ref{sec:comparison}, at equal
budget the channel-noise term of the update error is of order $\sqrt d$
times smaller for GL encoding than for \textsc{FedLQR}
under~\eqref{eq:channel}, so \textsc{ResiFedLQR} should tolerate about
$\sqrt d$ times more noise. We test this on three systems of increasing
dimension \reza{obtained from the same vehicle: the hover
model~\eqref{eq:rov_linear} splits exactly into decoupled parts, and we
consider the surge--pitch part ($6$ states, $2$ inputs, $d=12$), the
surge--pitch and sway--roll parts together ($12$ states, $4$ inputs,
$d=48$), and the full vehicle ($d=108$), with the same fleet, cost
weights, stepsize, and noise unit, $N_{\rm F}=2$ ($N_{\rm sc}=2d$), and
$T=1000$.} Fig.~\ref{fig:het_dim}(b) shows the fraction of realizations
that lose closed-loop stability as a function of $\sigma_{\rm ch}$. For
\reza{$d=12$, $48$, and $108$, half of the \textsc{FedLQR} realizations
fail at $\sigma_{\rm ch}=3.00$, $2.24$, and $1.61$, whereas
\textsc{ResiFedLQR} reaches that point only at $11.3$, $18.6$, and
$18.2$, i.e., at $3.8$, $8.3$, and $11.3$ times more noise, close to the
predicted $\sqrt d=3.5$, $6.9$, and $10.4$.}
By Remark~\ref{rem:power}, this scaling reflects the larger
transmit energy of projections under a fixed noise floor; for
\reza{$d=108$},
Fig.~\ref{fig:encoding_failure}(a) confirms that CL encoding with a gain
of $\sqrt d$ exhibits the same tolerance.

\section{Conclusion}
\label{sec:conclusion}
This paper investigated scalar communication for federated model-free
LQR over unreliable uplinks. Within the unified \textsc{ResiFedLQR}
framework, we studied projection-, coordinate-, and rollout-cost-based
scalar messages sent with redundancy. For isotropic encodings and rollout
costs, we established high-probability stability of every iterate and
linear convergence to a neighborhood of the optimum, which together
define a certified resilience region and a rule for the redundancy a
given channel requires, and we quantified the effects of redundancy,
channel noise, dropout, and heterogeneity. No encoding dominates:
projections outperform coordinates only at low signal-to-noise ratio, an
effect of transmit energy, and rollout costs avoid local computation but
amplify noise. Numerical studies \reza{on a fleet of BlueROV2 underwater
vehicles} corroborated these predictions in closed loop: at equal uplink
budget and moderate dropout, projections tolerated substantially more
channel noise than bundled full-gradient
transmission, coordinates matched them when sent at equal transmit
energy, and coordinates without this gain performed no better than
bundled transmission. Future work includes guarantees for moderate fleet
sizes and power-constrained uplink models.

\bibliographystyle{ieeetr}
\bibliography{references, bib/Reference, bib/alias}

\appendix
%\section{Proofs of Lemma~\ref{lem:hp_erec_resifed} and Proposition~\ref{prop:fedlqr_channel}}\label{app:proofs}

\subsection{Nomenclature}
Table~\ref{tab:nomenclature} collects the main symbols.

\begin{table}[h]
\centering\scriptsize
\caption{Nomenclature.}
\label{tab:nomenclature}
\setlength{\tabcolsep}{3pt}
\renewcommand{\arraystretch}{1.1}
\begin{tabular}{@{}l p{0.74\columnwidth}@{}}
\toprule
\multicolumn{2}{@{}l}{\emph{Problem}}\\
$M$, $d$ & number of agents; policy dimension $d=n_un_x$\\
$J^{(n)}$, $J_{\rm avg}$ & agent and fleet-average costs~\eqref{eq:individual_cost}, \eqref{eq:avg_cost}\\
$\mathcal S$, $\mathcal S_c$ & gains stabilizing all agents~\eqref{eq:stablesets}; sublevel set of $K_0$~\eqref{eq:sublevel}\\
$\epsilon_1,\epsilon_2$ & heterogeneity bounds~\eqref{eq:system_bound}\\
$F_t$, $\bar F$ & optimality gap $J_{\rm avg}(K_t)-J^\star_{\rm avg}$; certified radius~\eqref{eq:certif}\\
$T$, $\delta$ & horizon; confidence level\\
\midrule
\multicolumn{2}{@{}l}{\emph{Gradient oracle}}\\
$n_s$, $\tau$, $\mathsf r$ & rollouts per estimate; rollout length; smoothing radius\\
$\tilde g_{t,n}$ & local zeroth-order estimate~\eqref{eq:zo_estimate}\\
$\varepsilon$, $\delta_{\rm ZO}$ & oracle accuracy and failure probability~\eqref{eq:zo_event}\\
\midrule
\multicolumn{2}{@{}l}{\emph{Uplink}}\\
$p_{\rm drop}$, $\sigma_{\rm ch}$ & dropout probability; channel-noise level (Assumption~\ref{assump:noise})\\
$N_{\rm F}$, $N_{\rm sc}$ & number of full-gradient and scalar transmissions per agent and round\\
$N^{\rm rec}_{t,n}$ & number of scalars received from agent $n$ in round $t$\\
$\mathcal A_t$ & active agents, $N^{\rm rec}_{t,n}>0$~\eqref{eq:active_set}\\
$N^{\rm up}_t$ & number of rounds before $t$ with $\mathcal A_t\neq\varnothing$\\
$q_{\rm sc}$ & delivery probability of one scalar, $1-p_{\rm drop}$\\
$q_{\rm act}$ & probability that an agent is active, $1-p_{\rm drop}^{N_{\rm sc}}$\\
$\nu$ & expected scalars received per agent, $q_{\rm sc}N_{\rm sc}$\\
$\lambda_T$, $\theta$ & $\log(6dT/\delta)$; $\sqrt{d\lambda_T/(Mq_{\rm act}\nu)}$~\eqref{eq:nu_theta}\\
\midrule
\multicolumn{2}{@{}l}{\emph{Analysis}}\\
$\mu_c$ & PL constant, $\min_n\mu_n$ (Lemma~\ref{lem:single_agent})\\
$L_c$, $G_c$, $r_c$ & smoothness and gradient bounds~\eqref{eq:constants}; margin to instability~\eqref{eq:rc_Scplus}\\
$b_{\rm het}$, $\Delta_{\max}$ & heterogeneity bias~\eqref{eq:hbar}; dissimilarity bound~\eqref{eq:gamma_B}\\
$C$ & constant of Lemma~\ref{lem:hp_erec_resifed}, $C=2$\\
$\bar a$, $\bar b$ & relative and additive error constants~\eqref{eq:ab_bar}\\
$\eta^\star$, $\rho$ & stepsize bound~\eqref{eq:stepsize}; contraction factor (Theorem~\ref{thm:main})\\
\bottomrule
\end{tabular}
\end{table}

\subsection{Proof of Lemma~\ref{lem:fleet}}

(a) By Lemma~\ref{lem:single_agent}(a) and the nonnegativity of the
costs, $J_{\rm avg}$ is continuous on $\mathcal{S}$ and diverges as
$K\to\partial\mathcal{S}$ or $\|K\|\to\infty$; hence $\mathcal{S}_c$ is
closed, bounded, and contained in $\mathcal{S}$.
(b) The complement of $\mathcal{S}$ is closed and disjoint from the
compact set $\mathcal{S}_c$, so $r_c>0$, and every
$K\in\mathcal{S}_c^+$ is at distance at least $r_c$ from it.
(c) $\nabla J_{\rm avg}$ and $\nabla^2J_{\rm avg}$ are continuous on
$\mathcal{S}$, hence bounded on the compact sets
$\mathcal{S}_c\subseteq\mathcal{S}_c^+$.
(d) The arguments of $h^{(n)}_{q,\rm het}$ are continuous on
$\mathcal{S}_c$, with $J^{(n)}(K)\le MJ_{\rm avg}(K)\le Mc$, so the
suprema in~\eqref{eq:hbar} are finite by (a). By~\eqref{eq:system_bound},
$\|A^{(i)}-A^{(j)}\|_2\le2\epsilon_1$ and
$\|B^{(i)}-B^{(j)}\|_2\le2\epsilon_2$. Applying~\eqref{eq:grad_sens}
with $n=i$ and $(A',B')=(A^{(j)},B^{(j)})$, admissible since
$K\in\mathcal{S}^{(i)}\cap\mathcal{S}^{(j)}$, gives~\eqref{eq:grad_het}.
\hfill$\blacksquare$

\subsection{Proof of Lemma~\ref{lem:collective_pl}}

Let $g_n:=\nabla J^{(n)}(K)$ and $g:=\frac1M\sum_ng_n=\nabla J_{\rm avg}(K)$.
Expanding the squares gives
\[
  \tfrac12\|g\|^2
  =\frac{1}{2M}\sum\nolimits_{n=1}^M\|g_n\|^2
   -\frac{1}{4M^2}\sum\nolimits_{i,j=1}^M\|g_i-g_j\|^2 .
\]
By~\eqref{eq:individual_pl} and $\mu_n\ge\mu_c$, the first term is at
least $\mu_c\bigl(J_{\rm avg}(K)-\frac1M\sum_nJ^{(n)}_\star\bigr)
\ge\mu_c\bigl(J_{\rm avg}(K)-J^\star_{\rm avg}\bigr)$, since
$J^{(n)}_\star\le J^{(n)}(K')$ for every $K'\in\mathcal{S}$.
By~\eqref{eq:grad_het}, the second term is at most
$\frac{M(M-1)}{4M^2}\cdot4b_{\rm het}\le b_{\rm het}$.
\hfill$\blacksquare$

\subsection{Proof of Lemma~\ref{lem:heterogeneity}}

Write $\Delta_{t,n}=\bigl(\nabla J^{(n)}(K_t)-\nabla J_{\rm avg}(K_t)\bigr)
+\bigl(\tilde g_{t,n}-\nabla J^{(n)}(K_t)\bigr)-e^{\rm ZO}_t$. The first
term is at most $2\sqrt{b_{\rm het}}$ by~\eqref{eq:grad_het} and the
triangle inequality; each of the other two is at most $\varepsilon$ on
$\mathcal{E}^{\rm ZO}_t$.
\hfill$\blacksquare$
\subsection{Concentration Inequalities}

We use two standard results. The first is the matrix Bernstein
inequality~\cite[Th.~1.6]{tropp_user-friendly_2012}, specialized to
$d\times1$ matrices.

\begin{lem}[Vector Bernstein inequality]\label{lem:bernstein}
Let $Y_1,\ldots,Y_N\in\mathbb{R}^d$ be independent random vectors with
$\mathbb{E}[Y_i]=0$ and $\|Y_i\|\le R$ almost surely, and let
$V\ge\sum\nolimits_{i=1}^N\mathbb{E}\|Y_i\|^2$. Then, for every $\delta\in(0,1)$,
with probability at least $1-\delta$,
\begin{equation}\label{eq:bernstein}
  \Bigl\|\sum\nolimits_{i=1}^NY_i\Bigr\|
  \le\sqrt{2V\log\tfrac{d+1}{\delta}}+\tfrac{2R}{3}\log\tfrac{d+1}{\delta}.
\end{equation}
\end{lem}

\begin{proof}
For $d\times1$ matrices the variance statistic
of~\cite[Th.~1.6]{tropp_user-friendly_2012} is
$\max\{\|\sum_i\mathbb{E}[Y_iY_i^\top]\|_2,\sum_i\mathbb{E}\|Y_i\|^2\}
=\sum_i\mathbb{E}\|Y_i\|^2\le V$, since the first matrix is positive
semidefinite with trace equal to the second quantity. The theorem then
gives $\mathbb{P}(\|\sum_iY_i\|\ge u)\le(d+1)\exp\bigl(-\tfrac{u^2/2}{V+Ru/3}\bigr)$
for $u\ge0$. Setting $\ell:=\log\frac{d+1}{\delta}$ and
$u:=\sqrt{2V\ell}+\frac{2R\ell}{3}$, one checks that
$u^2\ge2\ell V+\frac{2\ell R}{3}u$, so the right-hand side is at most
$\delta$.
\end{proof}

The second concerns linear combinations of sub-Gaussian scalars (see
Section~\ref{sec:prob_def_object} for the definition).

\begin{lem}[Sub-Gaussian linear combinations]\label{lem:subgauss}
Let $w_1,\ldots,w_N$ be independent, mean-zero, sub-Gaussian scalars
with variance proxy $\sigma^2$, and let $c_1,\ldots,c_N\in\mathbb{R}^d$
be deterministic. Then, for every $\delta\in(0,1)$, with probability at
least $1-\delta$,
\begin{equation}\label{eq:subgauss}
  \Bigl\|\sum\nolimits_{i=1}^Nc_iw_i\Bigr\|
  \le\sigma\sqrt{2\log\tfrac{2d}{\delta}\;\sum\nolimits_{i=1}^N\|c_i\|^2}.
\end{equation}
The same bound holds for $\sum_{i=1}^N c_i\,\omega_i$ with
$\omega_i\in\mathbb{R}^d$ having independent, mean-zero, sub-Gaussian
entries with variance proxy $\sigma^2$ and $c_i\in\mathbb{R}$
deterministic, with $\sum_i\|c_i\|^2$ replaced by $d\sum_ic_i^2$.
\end{lem}

\begin{proof}
The $j$th coordinate $S_j:=\sum_ic_{ij}w_i$ is sub-Gaussian with
variance proxy $\sigma^2\sum_ic_{ij}^2$, so
$\mathbb{P}\bigl(|S_j|\ge\sigma\sqrt{2(\sum_ic_{ij}^2)\log\frac{2d}{\delta}}\bigr)\le\delta/d$.
A union bound over $j\in[d]$ and $\|S\|^2=\sum_jS_j^2$
give~\eqref{eq:subgauss}. In the second case $S_j=\sum_ic_i(\omega_i)_j$
has variance proxy $\sigma^2\sum_ic_i^2$ for each $j$, and the same
argument applies.
\end{proof}

\subsection{Proof of Lemma~\ref{lem:hp_erec_resifed}}

Throughout the proof, $t$ is fixed and omitted from subscripts where no
confusion arises. Let $\chi_{n,s}\in\{0,1\}$ indicate delivery of the
$s$th packet of agent $n$, so that $\mathbb{P}(\chi_{n,s}=1)=q_{\rm sc}$
and $N^{\rm rec}_n=\sum_{s=1}^{N_{\rm sc}}\chi_{n,s}$, and let
$I_n:=\mathbf{1}\{N^{\rm rec}_n>0\}$, so that
$|\mathcal{A}_t|=\sum_{n=1}^MI_n$ and $\mathbb{P}(I_n=1)=q_{\rm act}$.
By Assumption~\ref{assump:noise}, the indicators $\{\chi_{n,s}\}$ are
independent across $n$ and $s$, and the directions $\{v_{n,s}\}$, the
noises $\{w_{n,s}\}$, the indicators, and the local estimates
$\{\tilde g_n\}$ are mutually independent. \rev{The directions $v_{n,s}$
are taken to be independent draws from $\mathcal D$; for a seeded
pseudorandom generator, this is the standard idealization.}

\emph{Decomposition.} With the convention $I_n/N^{\rm rec}_n=0$ when
$N^{\rm rec}_n=0$, \eqref{eq:unified_decode} with $c=1$
and~\eqref{eq:resi_decode} give
\begin{equation}\label{eq:ghat_pf}
  \widehat g_t=\frac{1}{|\mathcal{A}_t|}\sum_{n=1}^M\frac{I_n}{N^{\rm rec}_n}
  \sum_{s=1}^{N_{\rm sc}}\chi_{n,s}\bigl(v_{n,s}^\top\tilde g_n+w_{n,s}\bigr)v_{n,s}.
\end{equation}
Since $(v_{n,s}^\top\tilde g_n)v_{n,s}=\tilde g_n+(v_{n,s}v_{n,s}^\top-I)\tilde g_n$
and $\sum_s\chi_{n,s}/N^{\rm rec}_n=1$ when $I_n=1$,
\begin{equation}\label{eq:erec_split}
  e^{\rm rec}_t=\widehat g_t-\tilde g_t
  =e^{\rm drop}_t+e^{\rm proj}_t+e^{\rm ch}_t,
\end{equation}
where, with $\Delta_n:=\tilde g_n-\tilde g_t$,
\begin{subequations}\label{eq:three_terms}
\begin{align}
  e^{\rm drop}_t&=\frac{1}{|\mathcal{A}_t|}\sum\nolimits_{n=1}^M(I_n-q_{\rm act})\Delta_n,\label{eq:edrop_pf}\\
  e^{\rm proj}_t&=\frac{1}{|\mathcal{A}_t|}\sum_{n=1}^M\frac{I_n}{N^{\rm rec}_n}
   \sum_{s=1}^{N_{\rm sc}}\chi_{n,s}\,(v_{n,s}v_{n,s}^\top-I)\tilde g_n,\label{eq:eproj_pf}\\
  e^{\rm ch}_t&=\frac{1}{|\mathcal{A}_t|}\sum_{n=1}^M\frac{I_n}{N^{\rm rec}_n}
   \sum_{s=1}^{N_{\rm sc}}\chi_{n,s}\,w_{n,s}v_{n,s}.\label{eq:ech_pf}
\end{align}
\end{subequations}
Here~\eqref{eq:edrop_pf} follows from
$\frac{1}{|\mathcal{A}_t|}\sum_nI_n\tilde g_n-\tilde g_t
=\frac{1}{|\mathcal{A}_t|}\sum_nI_n\Delta_n$ and $\sum_n\Delta_n=0$;
the three terms coincide with those
in~\eqref{eq:total_error_resifed_dropout}. We bound each on an event of
probability at least $1-\delta/(3T)$ and conclude by a union bound.
Note that $\log\frac{3T(d+1)}{\delta}\le\lambda_T$ since $d+1\le2d$.

\emph{Facts about the directions.} By~\eqref{eq:fixed_norm},
$vv^\top-I$ has the eigenvalue $d-1$ (eigenvector $v$) and the
eigenvalue $-1$ with multiplicity $d-1$, so $\|vv^\top-I\|_2=d-1$.
By~\eqref{eq:isotropy}, $\mathbb{E}[(v^\top z)^2]=\|z\|^2$ for every
$z\in\mathbb{R}^d$, and hence
\begin{align}\label{eq:rademacher_moments}
  \mathbb{E}\|(vv^\top-I)z\|^2
  &=\mathbb{E}\bigl[(v^\top z)^2\|v\|^2\bigr]-2\,\mathbb{E}[(v^\top z)^2]+\|z\|^2
   \nonumber\\
  &=(d-1)\|z\|^2 .
\end{align}
On $\mathcal{E}^{\rm ZO}_t$, Lemma~\ref{lem:heterogeneity} gives
$\|\Delta_n\|\le\Delta_{\max}$, hence
\begin{equation}\label{eq:gn_bound}
  \|\tilde g_n\|\le\|\tilde g_t\|+\Delta_{\max},\qquad n\in[M].
\end{equation}

\emph{Step 1 (dropout term).} Condition on $\{\tilde g_n\}$, which
determine $\mathcal{E}^{\rm ZO}_t$. The vectors
$Y_n:=(I_n-q_{\rm act})\Delta_n$, $n\in[M]$, are independent with
$\mathbb{E}[Y_n]=0$, $\|Y_n\|\le\Delta_{\max}$, and
$\sum_n\mathbb{E}\|Y_n\|^2\le Mq_{\rm act}(1-q_{\rm act})\Delta_{\max}^2$.
Lemma~\ref{lem:bernstein} with $\delta/(3T)$ yields, with probability at
least $1-\delta/(3T)$,
\begin{equation}\label{eq:edrop_bound}
  \|e^{\rm drop}_t\|\le\frac{1}{|\mathcal{A}_t|}\Bigl[
  \sqrt{2Mq_{\rm act}(1-q_{\rm act})\lambda_T}+\tfrac{2}{3}\lambda_T\Bigr]\Delta_{\max}.
\end{equation}

\emph{Step 2 (projection term).} Condition on $\{\tilde g_n\}$ and
$\{\chi_{n,s}\}$; then $|\mathcal{A}_t|$ and $N^{\rm rec}_n$ are fixed,
and $\mathcal{E}^{\rm ZO}_t\cap\mathcal{E}^{\rm rec}_t$ is determined.
Let $\mathcal{I}:=\{(n,s):I_n\chi_{n,s}=1\}$ and, for
$(n,s)\in\mathcal{I}$,
$X_{n,s}:=\frac{1}{|\mathcal{A}_t|N^{\rm rec}_n}(v_{n,s}v_{n,s}^\top-I)\tilde g_n$,
so that $e^{\rm proj}_t=\sum_{(n,s)\in\mathcal{I}}X_{n,s}$. The
$X_{n,s}$ are independent with mean zero. On
$\mathcal{E}^{\rm ZO}_t\cap\mathcal{E}^{\rm rec}_t$,
\[
  \|X_{n,s}\|\le\frac{(d-1)(\|\tilde g_t\|+\Delta_{\max})}{|\mathcal{A}_t|\bar\nu},
\]
and, by~\eqref{eq:rademacher_moments},
\[
  \!\!\sum_{(n,s)\in\mathcal{I}
}\!\!\!\!\!\mathbb{E}\|X_{n,s}\|^2
  =\!\!\!\sum_{n\in\mathcal{A}_t}\!\!\!\frac{(d-1)\|\tilde g_n\|^2}{|\mathcal{A}_t|^2N^{\rm rec}_n}
  \!\le\!\frac{(d-1)(\|\tilde g_t\|+\Delta_{\max})^2}{|\mathcal{A}_t|\bar\nu}.
\]
Lemma~\ref{lem:bernstein} with $\delta/(3T)$ yields, with probability
at least $1-\delta/(3T)$,
\begin{equation}\label{eq:eproj_bound}
  \|e^{\rm proj}_t\|\le\Bigl[\sqrt{\tfrac{2(d-1)\lambda_T}{|\mathcal{A}_t|\bar\nu}}
  +\tfrac{2(d-1)\lambda_T}{3|\mathcal{A}_t|\bar\nu}\Bigr]\bigl(\|\tilde g_t\|+\Delta_{\max}\bigr).
\end{equation}

\emph{Step 3 (channel-noise term).} Condition on $\{\tilde g_n\}$,
$\{\chi_{n,s}\}$, and $\{v_{n,s}\}$; the noises $w_{n,s}$ remain
independent, zero-mean, and sub-Gaussian with variance proxy
$\sigma_{\rm ch}^2$. Then
$e^{\rm ch}_t=\sum\nolimits_{(n,s)\in\mathcal{I}}c_{n,s}w_{n,s}$ with
$c_{n,s}:=\frac{v_{n,s}}{|\mathcal{A}_t|N^{\rm rec}_n}$, and,
by~\eqref{eq:fixed_norm}, on $\mathcal{E}^{\rm rec}_t$,
\[
  \sum\nolimits_{(n,s)\in\mathcal{I}}\|c_{n,s}\|^2
  =\frac{d}{|\mathcal{A}_t|^2}\sum\nolimits_{n\in\mathcal{A}_t}\frac{1}{N^{\rm rec}_n}
  \le\frac{d}{|\mathcal{A}_t|\bar\nu}.
\]
Lemma~\ref{lem:subgauss} with $\delta/(3T)$ yields, with probability at
least $1-\delta/(3T)$,
\begin{equation}\label{eq:ech_bound}
  \|e^{\rm ch}_t\|\le\sigma_{\rm ch}\sqrt{\tfrac{2d\lambda_T}{|\mathcal{A}_t|\bar\nu}} .
\end{equation}

\emph{Step 4 (combination).} Each of~\eqref{eq:edrop_bound},
\eqref{eq:eproj_bound}, and~\eqref{eq:ech_bound} holds with conditional
probability at least $1-\delta/(3T)$ given its conditioning variables,
hence unconditionally. By~\eqref{eq:erec_split} and a union bound, they
hold simultaneously with probability at least $1-\delta/T$. Using
$1/\bar\nu\le2/\nu$, $d-1\le d$, $q_{\rm act}\le1$, $\sqrt2\le2$, and
$\tfrac43\le2$,
\begin{align*}
  \|e^{\rm rec}_t\|
  &\le\Bigl[2\sqrt{\tfrac{d\lambda_T}{|\mathcal{A}_t|\nu}}
     +\tfrac{4d\lambda_T}{3|\mathcal{A}_t|\nu}\Bigr]
     \bigl(\|\tilde g_t\|+\Delta_{\max}\bigr)
    +2\sigma_{\rm ch}\sqrt{\tfrac{d\lambda_T}{|\mathcal{A}_t|\nu}}\\
  &\quad+\Bigl[\sqrt{\tfrac{2M(1-q_{\rm act})\lambda_T}{|\mathcal{A}_t|^2}}
    +\tfrac{2\lambda_T}{3|\mathcal{A}_t|}\Bigr]\Delta_{\max},
\end{align*}
which is~\eqref{eq:hp_erec_resifed} with $C=2$.

\emph{Probability of $\mathcal{E}^{\rm rec}_t$.} If $\nu\le2$, then
$\bar\nu=1$ and $\mathcal{E}^{\rm rec}_t$ holds trivially. If $\nu>2$,
then $N^{\rm rec}_n\sim\mathrm{Bin}(N_{\rm sc},q_{\rm sc})$ has mean
$\nu$, and the multiplicative Chernoff bound gives
$\mathbb{P}(N^{\rm rec}_n<\nu/2)\le e^{-\nu/8}$; a union bound over
$n\in[M]$ gives $\mathbb{P}(\mathcal{E}^{\rm rec}_t)\ge1-Me^{-\nu/8}$.
\hfill$\blacksquare$

\subsection{Proof of Corollary~\ref{cor:uniform}}

By the Chernoff lower tail,
$\mathbb{P}(|\mathcal{A}_t|<Mq_{\rm act}/2)\le e^{-Mq_{\rm act}/8}$.
On $\{|\mathcal{A}_t|\ge Mq_{\rm act}/2\}$, substituting
$1/|\mathcal{A}_t|\le2/(Mq_{\rm act})$ into~\eqref{eq:hp_erec_resifed}
bounds its first bracket by $\sqrt2\theta+2\theta^2\le4\theta$, using
$\theta\le1$, and its second by
$2\sqrt{(1-q_{\rm act})\lambda_T/(Mq_{\rm act}^2)}+2\lambda_T/(Mq_{\rm act})$.
Then $\|\tilde g_t\|\le\|g_t\|+\varepsilon$ and
$\|e_t^{\rm tot}\|\le\|e_t^{\rm rec}\|+\varepsilon$ give~\eqref{eq:rel_add}.
For each round, the bound~\eqref{eq:hp_erec_resifed}, the Chernoff
event, $\mathcal{E}^{\rm rec}_t$, and $\mathcal{E}^{\rm ZO}_t$ fail with
probabilities at most $\delta/T$, $e^{-Mq_{\rm act}/8}$, $Me^{-\nu/8}$,
and $\delta_{\rm ZO}$, respectively, the last by~\eqref{eq:zo_event}; a
union bound over the $T$ rounds completes the proof.
\hfill$\blacksquare$

\subsection{Proof of Lemma~\ref{lem:federated_descent}}

Since $\|K^+-K\|\le r_c$, the segment $[K,K^+]$ lies in
$\mathcal{S}_c^+$, on which $\|\nabla^2J_{\rm avg}\|_2\le L_c$; hence
$J_{\rm avg}(K^+)\le J_{\rm avg}(K)-\eta\langle g,\widehat g\rangle
+\frac{L_c\eta^2}{2}\|\widehat g\|^2$. Using
$\langle g,e\rangle\ge-a\|g\|^2-b\|g\|$ and
$\|\widehat g\|^2\le\bigl((1+a)\|g\|+b\bigr)^2\le2(1+a)^2\|g\|^2+2b^2$,
\begin{gather*}
  J_{\rm avg}(K^+)\le J_{\rm avg}(K)
  -\eta\bigl[(1-a)-L_c\eta(1+a)^2\bigr]\|g\|^2\\
  +\eta b\|g\|+L_c\eta^2b^2 .
\end{gather*}
For the stated $\eta$, the bracket is at least $(1-a)/2$. Young's
inequality gives
$\eta b\|g\|\le\frac{\eta(1-a)}{4}\|g\|^2+\frac{\eta b^2}{1-a}$, and
$L_c\eta^2b^2\le\frac{\eta(1-a)b^2}{2(1+a)^2}\le\frac{\eta b^2}{2(1-a)}$.
\hfill$\blacksquare$

\subsection{Proof of Theorem~\ref{thm:main}}

Work on $\mathcal{E}_T\cap\mathcal{E}^{\rm ZO}_T$ of
Corollary~\ref{cor:uniform} and argue by induction on $t$ that
$F_t\le F_0$, which, since $c=J_{\rm avg}(K_0)$
and by~\eqref{eq:finite_iff_stable}, is equivalent to
$K_t\in\mathcal{S}_c$. The claim holds at $t=0$. If round $t$ is
inactive, $K_{t+1}=K_t$. Otherwise,~\eqref{eq:rel_add} holds, so
$\|\widehat g_t\|\le(1+\bar a)G_c+\bar b$ and
$\eta\|\widehat g_t\|\le r_c$ by~\eqref{eq:stepsize}.
Lemma~\ref{lem:federated_descent} with $(a,b)=(\bar a,\bar b)$ and
Lemma~\ref{lem:collective_pl}, $\|g_t\|^2\ge2\mu_cF_t-2b_{\rm het}$,~give
\begin{multline*}
  F_{t+1}\le\Bigl(1-\frac{\eta\mu_c(1-\bar a)}{2}\Bigr)F_t
  +\frac{\eta(1-\bar a)}{2}b_{\rm het}\\
  +\frac{3\eta\bar b^{\,2}}{2(1-\bar a)}
  =\rho F_t+(1-\rho)\bar F .
\end{multline*}
Since $\eta\le\frac{1-\bar a}{2L_c(1+\bar a)^2}$ and $\mu_c\le L_c$,
$1-\rho\le\frac14$. Hence $F_{t+1}\le\max\{F_t,\bar F\}\le F_0$
by~\eqref{eq:certif}, which proves (i). For (ii),
$F_{t+1}-\bar F\le\rho\,(F_t-\bar F)$ on active rounds and
$F_{t+1}-\bar F=F_t-\bar F$ on inactive rounds; since $\rho>0$,
induction gives $F_t-\bar F\le\rho^{N^{\rm up}_t}(F_0-\bar F)$, which
rearranges to (ii).
\hfill$\blacksquare$

\subsection{Proof of Proposition~\ref{prop:fedlqr_channel}}

Under full-gradient transmission, agent $n$ transmits $N_{\rm F}$
packets, each carrying $\tilde g_n$; the $s$th packet, if delivered, is
received as $\tilde g_n+w_{n,s}$ with $w_{n,s}\in\mathbb{R}^d$ having
independent sub-Gaussian entries; and the server computes
$\widehat g^{\rm F}_t=\frac{1}{|\mathcal{A}_t|}\sum_{n}\frac{I_n}{N^{\rm rec}_n}
\sum_{s=1}^{N_{\rm F}}\chi_{n,s}(\tilde g_n+w_{n,s})$, where now
$N^{\rm rec}_n\sim\mathrm{Bin}(N_{\rm F},q_{\rm sc})$ and
$\mathbb{P}(I_n=1)=q^{\rm F}_{\rm act}$. Let $\mathcal{E}^{\rm rec}_t$ be
defined as in Lemma~\ref{lem:hp_erec_resifed} with
$\bar\nu_{\rm F}:=\max\{1,\nu_{\rm F}/2\}$ in place of $\bar\nu$. The
manipulation leading to~\eqref{eq:erec_split} gives
$\widehat g^{\rm F}_t-\tilde g_t=e^{\rm drop}_t+e^{{\rm ch},{\rm F}}_t$
with $e^{\rm drop}_t$ as in~\eqref{eq:edrop_pf} and
$e^{{\rm ch},{\rm F}}_t=\sum_{(n,s)\in\mathcal{I}}\frac{1}{|\mathcal{A}_t|N^{\rm rec}_n}w_{n,s}$; there is no projection term. Step~1 of the proof of
Lemma~\ref{lem:hp_erec_resifed} applies
verbatim and gives~\eqref{eq:edrop_bound} with $q^{\rm F}_{\rm act}$ in place of $q_{\rm act}$. For the noise term, condition on $\{\tilde g_n\}$
and $\{\chi_{n,s}\}$ and apply the second part of Lemma~\ref{lem:subgauss}
with the scalar coefficients $c_{n,s}=\frac{1}{|\mathcal{A}_t|N^{\rm rec}_n}$,
for which
$\sum_{(n,s)\in\mathcal{I}}c_{n,s}^2
=\frac{1}{|\mathcal{A}_t|^2}\sum_{n\in\mathcal{A}_t}\frac{1}{N^{\rm rec}_n}
\le\frac{1}{|\mathcal{A}_t|\bar\nu_{\rm F}}$ on $\mathcal{E}^{\rm rec}_t$.
This gives, with probability at least $1-\delta/(3T)$,
$\|e^{{\rm ch},{\rm F}}_t\|\le\sigma_{\rm ch}\sqrt{2d\lambda_T/(|\mathcal{A}_t|\bar\nu_{\rm F})}$,
which is~\eqref{eq:ech_bound} with $\bar\nu_{\rm F}$ in place of
$\bar\nu$. Hence, on $\mathcal{E}^{\rm ZO}_t\cap\mathcal{E}^{\rm rec}_t$
and with probability at least $1-\delta/T$,
\[
  \|\widehat g^{\rm F}_t-\tilde g_t\|
  \le2\sigma_{\rm ch}\sqrt{\tfrac{d\lambda_T}{|\mathcal{A}_t|\nu_{\rm F}}}
  +\Bigl[\sqrt{\tfrac{2M(1-q^{\rm F}_{\rm act})\lambda_T}{|\mathcal{A}_t|^2}}
  +\tfrac{2\lambda_T}{3|\mathcal{A}_t|}\Bigr]\Delta_{\max},
\]
and $\mathbb{P}(\mathcal{E}^{\rm rec}_t)\ge1-Me^{-\nu_{\rm F}/8}$ by the
same Chernoff argument. Repeating the proof of
Corollary~\ref{cor:uniform}, i.e., substituting
$1/|\mathcal{A}_t|\le2/(Mq^{\rm F}_{\rm act})$ on the event
$\{|\mathcal{A}_t|\ge Mq^{\rm F}_{\rm act}/2\}$, adding
$\|e^{\rm ZO}_t\|\le\varepsilon$, and taking a union bound over the
rounds, yields~\eqref{eq:rel_add} with $\bar a^{\rm F}=0$ and
$\bar b^{\rm F}$ as in~\eqref{eq:ab_fedlqr} with $C=2$, with probability
at least
$1-\delta-T(e^{-Mq^{\rm F}_{\rm act}/8}+Me^{-\nu_{\rm F}/8}+\delta_{\rm ZO})$.
Since $\bar a^{\rm F}=0$, the step
$\sqrt2\theta+2\theta^2\le4\theta$ is not used, and no condition on
$\theta_{\rm F}$ is needed; the noise bracket satisfies
$2\sqrt{d\lambda_T/(|\mathcal A_t|\nu_{\rm F})}\le2\sqrt2\,\theta_{\rm F}\le4C\theta_{\rm F}$
directly.
Theorem~\ref{thm:main} uses only~\eqref{eq:rel_add} and the properties
of $J_{\rm avg}$ established in Lemmas~\ref{lem:fleet}
and~\ref{lem:collective_pl}, so it applies with
$(\bar a^{\rm F},\bar b^{\rm F})$.
\hfill$\blacksquare$

\subsection{Proof of Proposition~\ref{prop:elementwise}}

(a) Conditional on the received set,
$\widehat g_{t,n}-\tilde g=\frac1k\sum_s\bigl[(v_sv_s^\top-I)\tilde g+w_sv_s\bigr]$,
a sum of $k$ independent mean-zero terms with second moments
$(d-1)\|\tilde g\|^2$ by~\eqref{eq:encoding_variance} and
$d\sigma_{\rm ch}^2$ by~\eqref{eq:fixed_norm}; the cross terms vanish
because $w_s$ is independent of $v_s$ and has zero mean.
(b) Coordinate $i$ has error $\frac1{k_i}\sum w$ with second moment
$\sigma_{\rm ch}^2/k_i$ if $k_i\ge1$, and error $-\tilde g_i$, which
occurs with probability $p_{\rm drop}^{N_{\rm F}}$, otherwise. Summing
over $i$ gives~\eqref{eq:mse_elem}. For bundled packets, all coordinates
share the same $k$, which does not change the expectation.
(c) By (a) and the zero decoder, the mean-square error of GL is
$\kappa_{\rm sc}\bigl[(d-1)\|\tilde g\|^2+d\sigma_{\rm ch}^2\bigr]
+p_{\rm drop}^{dN_{\rm F}}\|\tilde g\|^2$; comparing
with~\eqref{eq:mse_elem} gives~\eqref{eq:elem_crossover}.
\hfill$\blacksquare$

\subsection{Proof of Corollary~\ref{cor:ol}}

Let $\widehat g^{\,\circ}_{t,n}$ denote the decoded direction without
channel noise. Then
\begin{align*}
  \widehat g_t-g_t
  ={}&\tfrac{1}{|\mathcal A_t|}\textstyle\sum_{n\in\mathcal A_t}
     \bigl(\widehat g^{\,\circ}_{t,n}-\nabla J^{(n)}(K_t)\bigr)\\
  &+\tfrac{1}{|\mathcal A_t|}\textstyle\sum_{n\in\mathcal A_t}
     \bigl(\nabla J^{(n)}(K_t)-g_t\bigr)+e^{\rm ch}_t .
\end{align*}
On $\mathcal E^{\rm rec}_t$, every active agent has at least
$\max\{1,\nu_{\rm O}/2\}$ received rollouts, so the first term is at
most $\varepsilon_{\rm O}$ on the corresponding oracle event. The second
term is bounded as in Step~1 of the proof of
Lemma~\ref{lem:hp_erec_resifed}, with $\Delta_{t,n}$ replaced by
$\nabla J^{(n)}(K_t)-g_t$, whose norm is at most $2\sqrt{b_{\rm het}}$
by~\eqref{eq:grad_het}. The third term is bounded as in Step~3, with
the coefficients $c_{n,s}$ multiplied by $\sqrt d/\mathsf r$. The
remaining steps follow the proof of Corollary~\ref{cor:uniform}.
\hfill$\blacksquare$

\end{document}